\documentclass[12pt,english,oneside,openany,floatfix]{book}
\usepackage{bookman}
\usepackage{mathrsfs}
\usepackage[T1]{fontenc}
\usepackage[latin1]{inputenc}
\usepackage{fancyhdr}
\usepackage[Sonny]{fncychap}

\usepackage{babel}
\usepackage{psfrag}
\usepackage{graphicx}
\usepackage{amsmath}
\usepackage{amssymb}
\usepackage{float}
\usepackage{mathtools}
\usepackage{setspace}
\usepackage[nottoc]{tocbibind}
\usepackage[sectionbib]{chapterbib}
\usepackage[toc,titletoc]{appendix}

\makeatletter

\usepackage{verbatim}
\usepackage{subfigure}
\graphicspath{{./graphics/}}

\begin{document}
	\pagestyle{fancy}
	\pagenumbering{roman}
	\clearpage
	\begin{center}
{\bf{\huge Strongly Interacting Matter Under Extreme Conditions  }}
\end{center}
\vskip 2in
\begin{center}
Thesis Submitted to\\
 The University of Calcutta\\
     for the Degree of\\
    Doctor of Philosophy (Science)
\end{center}
\vskip 1.8in
\begin{center}
{\it {By}}\\
{\large{\bf{ Somenath Pal}}}\\
Department of Physics\\
University of Calcutta\\
92, Acharya Prafulla Chandra Road\\
Kolkata-700009, India\\

\vskip 10pt
2022
\end{center}

	\newpage
	\newpage
\begin{center}
\textbf{DECLARATION}
\end{center}

I, hereby declare that this dissertation represents my works in my own words
and where others' ideas have been included, I have adequately cited and referred
the original sources. The work is original and has not been submitted earlier as a
whole or in part of a degree/diploma at this or any other Institution/University.\\
\\
\\
\\
\\
\\
\\
Somenath Pal\\
Kolkata, India
	\newpage
	\begin{figure*}[p]
\begin{center}
\includegraphics[width=3.5in]{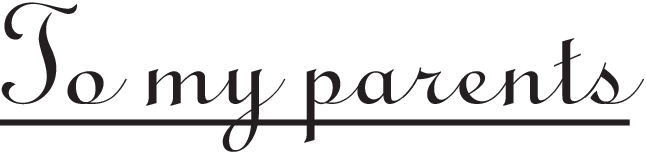}
\end{center}
\end{figure*}

	\newpage 
	{\centering \Large\textit{Acknowledgements}\vspace{0.3in}  \par}

{\it {

I gratefully acknowledge the invaluable guidance and constant support of my thesis 
supervisor Abhijit Bhattacharyya. From the very first day he has encouraged me 
to carry out this work and provided me all the assisstance during the whole research period.
Without his dedicated supervision it would
have never been possible to complete the thesis. I find myself really fortunate to have 
had the opportunity to work with him and I express my sincere gratitude to him. 
\vskip 0.2in
I am grateful to my collaborators Guruprasad Kadam, Rajarshi Ray, Hiranmaya Mishra, Anton Motornenko, Jan Steinheimer and Horst Stoecker. The collaboration with them was really fruitful and enjoyable.
\vskip 0.2in
I have always got the support from the faculty and staff members of my department.  
Their help is gratefully acknowledged. 
Also, I am grateful to all the research scholars of our department for their immense support and 
encouragement to complete the research work.
I will always remember the time that I have spent in this department. 
\vskip 0.2in
I am thankful to the Univerity Grants Commission (UGC) for financial support.
\vskip 0.2in
Last but not the least I am grateful to my Parents, Friends and 
relatives who accorded full support to my research work by encouraging me 
to approach towards my destination. I dedicate my research work to my parents.}}

	\newpage
	\vspace{-2.5in}
{\centering \Large\textit{Abstract}  \par}

In this thesis, different variants of the Hadron Resonance Gas (HRG) model have been used to describe the hadronic phase of strongly interacting matter. HRG model is a statistical thermal model which considers the hadronic matter as a dilute gas of point-like particles. The attractive interactions are mimicked in this model by taking the unstable particles as stable particles. Since repulsive interactions among the hadrons are also important, HRG model is improved by including  repulsive interaction through the inclusion of excluded volume in Excluded Volume Hadron Resonance Gas (EVHRG) model. Here, the EVHRG model is further improved by incorporating unequal radii of hadrons which has been named as Modified Excluded Volume Hadron Resonance Gas (MEVHRG) model and also by taking their Lorentz contraction in Lorentz contracted MEVHRG model, namely, LMEVHRG model. Another way of incorporating the repulsive interactions in the HRG model is through the introduction of a mean-field potential. This is done in the HRG mean-field (HRGMF) model where the single particle energies are modified by a density dependent term. It has been found that the thermodynamic quantities and the susceptibilities are significantly influenced by the inclusion of repulsive interaction. 
Both the inclusion of unequal radii and the Lorentz contraction have significant effects on  the thermodynamic quantities and susceptibilities of conserved charges.        Inclusion of repulsive interaction in HRG model through mean-field approach results in a satisfactory reproduction of the lattice data for the susceptibilities.  The centre of mass energy dependence of some thermodynamic quantities in presence of repulsive interaction has been studied. Our study clearly indicates that proper modelling of repulsive interaction among hadrons is very important to explain the thermodynamic quantities and susceptibilities of conserved charges. The effect of static magnetic field in HRG and EVHRG models by the means of Landau levels has also been investigated.  The vacuum part of the pressure, in presence of magnetic field, has been properly renormalised. The magnetic field is found to effect the thermodynamic quantities and the susceptibilities significantly. The electric charge susceptibility is influenced more strongly by the magnetic field than the baryon susceptibility.
 The total magnetization of hadronic matter is found to be positive. 
	\newpage
	\centerline{\underline{\large\sc List of Publications} }

\vskip 0.3in

\noindent{ A) Included in this thesis:}
\vskip 0.15in
\begin{enumerate}
\item {\it "Hadron resonance gas model with repulsive mean-field interactions: specific heat, isothermal compressibility and speed of sound"},
{\bf Somenath Pal}, G. Kadam and A. Bhattacharyya, Nucl. Phys. A {\bf 1023}, 122464 (2022).

\item {\it "Effects of hadronic repulsive interactions on the fluctuations of conserved charges"},
{\bf Somenath Pal}, G. Kadam, H. Mishra and A. Bhattacharyya. 
Phys. Rev. D {\bf 103}, 054015 (2021).

\item {\it "Modified Excluded Volume Hadron Resonance Gas Model with Lorentz Contraction"}, 
{\bf Somenath Pal}, A. Bhattacharyya and R. Ray, 
Nucl. Phys. A {\bf 1010}, 122177 (2021). 

\item {\it "Interacting hadron resonance gas model in magnetic field and the fluctuations of conserved charges"},
G. Kadam, {\bf Somenath Pal} and A. Bhattacharyya;
J. Phys. G {\bf 47}, 125106 (2020).
\end{enumerate}
\noindent{ B) Not included in this thesis:}
\begin{enumerate}
	\item {\it "Repulsive properties of hadrons in lattice QCD data and neutron stars"}, A.Motornenko, {\bf Somenath Pal}, A. Bhattacharyya, J. Steinheimer and H. Stoecker; Phys. Rev. C {\bf 103}, 054908 (2021).
\end{enumerate}


	\newpage
	\tableofcontents
	\newpage
	\rhead{List of Figures}
{\centering \Large\textit{List of Figures}\vspace{0.2in}  \par}
\begin{itemize}
\item [1.1]  Phase diagram in temperature and baryonic 
 chemical potential plane.

\item [1.2]  Different stages of relativistic heavy-ion collision experiment.

\item [3.1]  Scaled pressure, energy density and entropy density in HRG, EVHRG, MEVHRG, LMEVHRG models at $\mu_B=\mu_Q=\mu_S=0$ for $R_b=0.35 fm, R_\pi=0.2 fm, R_m=0.3 fm$.

\item [3.2]  Scaled pressure, energy density and entropy density at $\mu_B=\mu_Q=\mu_S=0$ in HRGMF model.

\item [3.3]  Scaled specific heat as a function of temperature in HRGMF model.

\item [3.4] Isothermal compressibility as a function of temperature in HRGMF model.

\item [3.5] Speed of sound squared ($C_s^2$) as a function of temperature in HRGMF model.

\item [3.6] Scaled specific heat $(C_V/T^3)$, isothermal compressibility $(\kappa_T)$ and speed of sound squared $(c_S^2)$ as a function of temperature in HRG, EVHRG, MEVHRG and LMEVHRG models for $R_b=0.35 fm, R_\pi=0.2 fm, R_m=0.3 fm$.

\item [3.7] Collision energy dependence of $C_V$, $\kappa_T$ and $C_s^2$ in HRGMF model.

\item [4.1]  Baryon number susceptibilities of different orders in HRG, EVHRG, MEVHRG and LMEVHRG models.

\item [4.2] Electric charge susceptibilities of different orders in HRG, EVHRG, MEVHRG and LMEVHRG models. 

\item [4.3] Strangeness susceptibilities of different orders in HRG, EVHRG, MEVHRG and LMEVHRG models.
 
\item [4.4] Baryon number susceptibilities of different orders in HRGMF model.

\item [4.5] Electric charge susceptibilities of different orders in HRGMF model.
 
\item [4.6] Strangeness susceptibilities of different orders in HRGMF model.

\item [4.7] Ratios of fourth order and second order susceptibilities in HRGMF model.

\item [4.8] Differences of second order and fourth order susceptibilities in HRGMF model.

\item [4.9] $R_B^{12}$, $R_B^{31}$ and $R_B^{42}$ as 
functions of $\mu_B/T$ in HRGMF model.

\item [5.1] Vacuum pressure of charged particles computed using MFIR scheme plotted against dimensionless variable $x=\frac{m^2}{2eB}$ and the total pressure as a function of temperature in presence of magnetic field for $\mu_B=0$.

\item [5.2] Energy density and  entropy density calculated in HRG and EVHRG models with and without  magnetic field for $\mu_B=0$.

\item [5.3] Magnetization of hadronic matter estimated in HRG and EVHRG models in presence of magnetic field at  $\mu_B=0$.

\item [5.4] Thermodynamic variables at $\mu_B=300 MeV$ in   HRG and EVHRG models with and without  magnetic field.

\item [5.5] Baryon susceptibilities of the hadronic matter estimated in HRG and EVHRG models with and without magnetic field.

\item [5.6] Electric charge susceptibilities of the hadronic matter estimated in HRG and EVHRG models with and without magnetic field.

\item [5.7] Ratios of baryon susceptibilities of the hadronic matter estimated in HRG and EVHRG models with and without magnetic field.

\item [5.8] Ratios of electric charge susceptibilities of the hadronic matter estimated in HRG and EVHRG models with and without magnetic field.

\end{itemize}

	\newpage
	\rhead{List of Tables} 
{\centering \Large\textit{List of Tables}\vspace{0.2in}  \par}
\begin{itemize}
\item [5.1] Hadrons and resonances included in the hadron resonance gas model with magnetic field.

\end{itemize}

	\newpage
	\pagenumbering{arabic}
	\setcounter{page}{1}
	\newpage
\rhead{Introduction}
\chapter{\label{chap:Introduction} Introduction}
One of the objectives of modern science is to understand the origin and evolution of the universe. On the microscopic level, this requires a sound understanding of the fundamental particles and the way they interact among themselves. To develop a theory for the interactions among these particles, the fundamental properties of the forces acting among these particles should be identified. There exist four fundamental forces in nature. Among these, strong force, which acts at a range of $\sim$ 1 fm, is responsible for the tight bonding of the nucleons in the nucleus of atoms. Quantum Chromodynamics (QCD)~\cite{QCD} is the theory of strong interactions, which is the quantum field theory for color charged quarks and gluons. In QCD, quarks are the fundamental particles and massless gluons are the mediators of strong force among them. Every quark can have one of the three colors: red, green and blue or their respective anti-colors. A color-charged particle has never been detected experimentally. Only color-singlet hadrons, which are made up of quarks and gluons, are observed in nature. The strong interaction among the quarks are mediated by exchange of gluons, which, are themselves color-charged. The color charge of gluons is a result of the non-abelian nature of QCD. It predicts two important natures of strong interaction: asymptotic freedom and confinement. The gluons interact strongly among themselves which finally leads to confinement when the distance among the quarks become large. It is argued that, if an attempt to separate two color-charged particles are made, then the gluon fields between the two particles form a narrow tube of color charge with high energy density. Such a tube is called color flux tube or color string. Thus, beyond a certain point, the energy between the two color-charged particles becomes so large that new color charged particles are created out of the energy available. These newly created particles interact strongly with the older particles and form new hadrons which are color-neutral. Evidences of jet fragmentation in heavy-ion experiments, in which highly focused bunches of color-neutral particles are generated, support this phenomenon. Thus, at low energy regime, the QCD coupling becomes large and quarks are in confined states which are called hadrons.  The scenario is opposite in the case of high energy regime, where the QCD coupling is small. This leads to asymptotic freedom~\cite{Politzer:1974fr} at small distances between quarks.
\section{QCD phase transition}
The asymptotic freedom of QCD at high energy densities suggests that at some critical energy density, the hadronic matter should undergo a confinement-deconfinement phase transition. Above this critical energy density, the QCD matter is made of deconfined quasi-free quarks and gluons, which is called Quark Gluon Plasma (QGP). On the other hand, below this critical energy density, the QCD matter is hadronic~\cite{yagi}. Normal hadronic matter is believed to undergo a phase transition to QGP phase at high temperature and/or density. Study of such matter is of significant interest as hot strongly interacting matter may have existed microseconds after the Big Bang. On the other hand, dense strongly interacting matter is of astrophysical importance as it is expected to be found inside cores of neutron stars.

There are significant indications that such a state of strongly interacting matter can be created in relativistic heavy-ion collision experiments. Relativistic Heavy-Ion Collider (RHIC) at Brookhaven and Large Hadron Collider (LHC) at CERN have found signals which suggest that QGP has actually been formed in these colliders. These two colliders have provided us a lot of data which have given us useful insights about the properties of the fireball created in such collisions. In future, Facility for Anti-proton and Ion Research (FAIR) at GSI will be used to probe the matter with high net baryon density and low temperatures.
\begin{figure}[h]
	\begin{center}
	\includegraphics[scale=0.35]{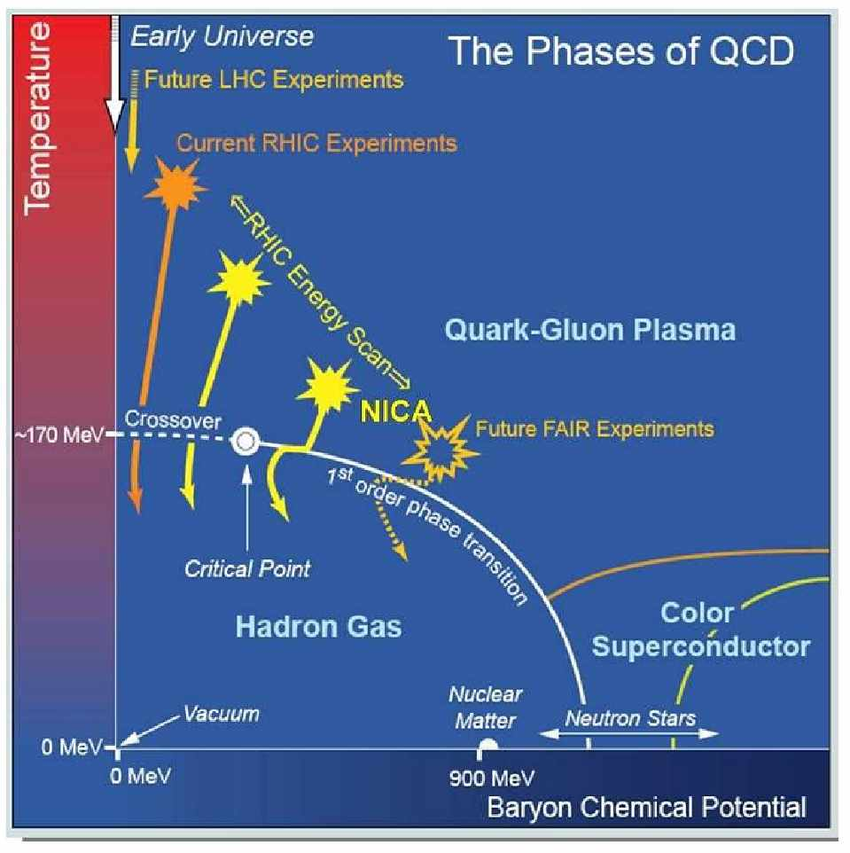}
	\caption{QCD phase diagram in the temperature-baryon chemical potential plane.}
	\end{center}
	\label{phase_diagram}
\end{figure}

Phase diagram is a diagrammatic way of representing the various phases of matter. The dimension and axes of a phase diagram depends on the independent parameters chosen to describe the matter. A schematic phase diagram is shown in Fig.~\ref{phase_diagram} with temperature and baryon chemical potential as the independent parameters. Other chemical potentials have not been shown in the phase diagram because those are related to the baryon chemical potential for a relativistic heavy-ion collision experiment. The region at vanishing chemical potential and high temperature shows the situation which is similar to the early universe scenario. The LHC is operated in this regime. On the other hand, the region at high baryon chemical potential and low temperature reflects the scenario which is expected to exist in neutron stars. Such a scenario will be investigated in the Compressed Baryonic Matter (CBM) experiment at the FAIR. As one moves from the high baryon chemical potential domain to low baryon chemical potential domain, along the baryon chemical potential axis, one crosses the quark-hadron phase boundary. Various model studies predict that this phase transition at high density is of first order. This first order phase transition line bends gradually to the temperature axis as the chemical potential is decreased and eventually ends at the Critical End Point (CEP). If one goes to higher temperatures, the transition from hadronic matter to QGP is a smooth crossover, as predicted by lattice QCD. At very high density, cooper pairs  (each made up of two quarks) are formed  due to the attractive interactions between the quarks. The condensate of cooper pairs breaks the local color gauge symmetry. This phenomenon is called the Color Superconductivity.

When one talks about different phases of matter, a criterion to distinguish between two different phases should be stated. A phase transition is characterized by the non-analyticity of some observables. If that observable shows some non-analytic behaviour near some line in the phase space, one says that the matter undergoes phase transition along that line, as one can distinguish the two different phases present on either sides of that line. This line is called the phase boundary and the observable which shows non-analytic behaviour is called the order parameter of phase transition. It should be mentioned that there can be multiple order parameters that contain the same non-analyticity. Hence, the order parameter is not unique.

\section{Theoretical approaches to QCD}
QCD is not currently exactly solvable due to its non-perturbative character. However, some other theoretical approaches have been developed till date which give significant insight about strongly interacting matter. In the high momentum transfer or small distance regime, where the QCD coupling $\alpha_s$ is small, perturbative QCD~\cite{Politzer:1973fx,Lepage:1980fj} can be successfully used. However, perturbative method is not applicable when the QCD coupling constant is large. There are two major alternatives to the direct use of QCD: Lattice QCD and the effective model approach.
\subsection{Lattice QCD}
Lattice QCD (LQCD) is a first-principle method to study the properties of strongly interacting matter. In this non-perturbative approach, the QCD lagrangian is treated on a grid which is discretized in space-time~\cite{Boyd:1996bx,Boyd:1995zg,Fodor:2001au,Fodor:2002km}. The lattice spacing between two lattice points serves as an ultraviolet regulator (cut-off). LQCD has been very useful to describe the thermodynamic properties of strongly interacting matter at high temperatures in last three decades or so. In lattice QCD, the infinite dimensional integrals for thermal traces are approximated by finite number of integrals using Monte Carlo method. The functional integrals in the partition function are performed by discretizing the space-time (Eucledian), the fields being defined at those discrete space-time points.

 Lattice QCD predicts that the deconfinement transition between quarks and hadrons is a crossover for small quark masses. Over the years, 
 as the computational power has increased,  results from LQCD have improved. Earlier results provided by lattice QCD suggests that the QGP to hadron phase transition for vanishing baryon chemical potential occurs at a temperature of approximately 170 MeV~\cite{Sharma:2013hsa}. However, recently, this temperature has been found to be 155 MeV~\cite{HotQCD:2018pds} as predicted by the continuum extrapolated calculations.
 
Lattice QCD has some inherent problems. One has to compromise with the magnitude of the lattice spacing in terms of the computational power, although, in recent years, there has been huge progress in the available computational powers. The second problem is much more difficult to overcome. At finite density, the action becomes imaginary. This problem is called the sign problem of lattice QCD. Few methods have been developed to circumvent this problem to some extent. Taylor series expansion and reweighting techniques have been introduced to get approximate results at finite density. But unfortunately, all these methods have limited applicability as they fail at large densities. So, the physics along the chemical potential axis in the QCD phase diagram remains largely inaccessible by LQCD.
\subsection{Effective Models}
In the low energy regime, where the quarks are confined in the hadrons, $\alpha_s$ becomes large. As a result, the perturbative method fails. This is the region where effective models become useful. These models have been constructed to reproduce low energy properties of strongly interacting matter. Some of the popular effective models are Hadron Resonance Gas (HRG) Model~\cite{HRG1,HRG2,HRG3,HRG4,HRG5,HRG6,HRG7,HRG8,HRG9,HRG10,HRG11}, Polyakov loop enhanced Nambu-Jona-Lasinio (PNJL) model~\cite{PNJL1,PNJL2,PNJL4,
	PNJL10,PNJL11,PNJL12,
	PNJL15}, Polyakov-Quark-Meson (PQM) 
model~\cite{PQM1,PQM2,PQM3,PQM4}, Chiral Perturbation Theory~\cite{chiral1}, etc. These models have been
successfully used to describe some of the properties of the strongly interacting matter at high temperature and density. The results  reproduce important lattice data at vanishing chemical potentials quite successfully~\cite{PNJL11,PNJL12}. There have been attempts to describe both the confined and deconfined phases of QCD matter by the use of hybrid models. Two such models are the Polyakov loop enhanced Chiral mean-field (CMF) model~\cite{Steinheimer:2011ea,Mukherjee:2016nhb} and HRG-PNJL hybrid model~\cite{Bhattacharyya:2017gwt}. The thermodynamic quantities like pressure, energy density, entropy density, number density, susceptibilities of conserved charges, etc. have been well reproduced within these models. In this work, the HRG model and its variants will be used to look at different properties of hot/dense hadronic matter.

HRG model is used to describe a system of dilute gas consisting of hadrons and is based on Dashen, Ma, Bernstein theorem~\cite{Dashen:1969ep}. In this model attractive interactions among hadrons are taken care of by considering the unstable resonance particles as stable particles. HRG model successfully describes some of the experimental data of a system at freeze out~\cite{HRG_freezeout1}. It was soon realized that the non-interacting pure HRG model is not sufficient to describe hadronic matter,
especially, at higher temperatures.
The reliability of the results obtained from HRG model calculations can only be validated by confronting those with LQCD simulations. Various studies
have confronted ideal HRG EoS with lattice results and found reasonable agreement all the way up to the crossover temperature except 
for interaction measure~\cite{Karsch:2003vd}. Later studies including continuous Hagedorn states in HRG model, found good agreement with the lattice data even for interaction measure~\cite{NoronhaHostler:2008ju,Kadam:2014cua}. However, 
recent studies have shown that commonly performed  comparisons of ideal HRG model with LQCD simulations and heavy-ion data may 
lead to misconceptions which might further render wrong conclusions and it is necessary to take into account short range 
repulsive interactions  among the hadrons as they play  a crucial role in the thermodynamics 
of hadron gas~\cite{Vovchenko:2016rkn}. The repulsive interaction among hadronic particles becomes increasingly important as one approaches the critical temperature. This is taken into account in the modified version of HRG model, namely, Excluded Volume Hadron Resonance Gas (EVHRG) model where repulsive interaction comes into play due to finite excluded hardcore volume of the particles~\cite{EVHRG1,EVHRG2,EVHRG3,EVHRG4,EVHRG5,EVHRG6,EVHRG7,EVHRG8}. Thus, the hadron gas created from the fireball in the heavy-ion collisions, at a given temperature and chemical potentials, cannot accomodate more than a certain number of particles due to finite excluded volumes of the particles. Consideration of excluded volume also restricts the motion of the particles in the fireball, which, has a significant impact on the thermodynamic quantities of the system. For example, the pressure gets significantly reduced if excluded volumes of the hadrons are considered.

Recently, several works have been done on EVHRG model using different sizes for different hadrons. 
The effect of excluded volume on the equation of state has been studied by some authors~\cite{alba}. In particular, they have looked into the pressure and the trace 
of the energy momentum tensor and compared their results with the lattice data. They have found that the best fits are obtained when the excluded volume is inversely 
proportional to the mass of the particle. There is a renewed interest in studying multiplicity data~\cite{diptak1,diptak2}. However, a more prudent approach has also been followed 
by studying the HRG model in the light of both multiplicity and fluctuation data~\cite{gupta}. The authors have treated temperature and chemical potential as parameters 
and tried to fit those by fitting the data using non-interacting HRG model. Fluctuations of baryon number and strangeness within HRG model with repulsive mean field approach with the effect of missing resonances have been studied by some authors~\cite{Huovinen:2018ziu}. HRG model with parity-doubled baryons having temperature dependent mass has also been used to calculate charge susceptibilities~\cite{Morita:2017hgr}.
 It was found that the reduction in effective mass at high temperatures makes the model results to overshoot the lattice data. The contribution of heavy resonances through exponential Hagedorn mass spectrum to fluctuations of conserved charges has also been investigated~\cite{Lo:2015cca}. A comparison between models with EVHRG hardcore repulsion and interaction based on S-matrix in HRG framework found that for a $\pi N \Delta$ system, there is good agreement between the two approaches with excluded volume radius $R = 0.3$ fm for pions and nucleons~\cite{Lo:2017ldt}. It has been shown that the mid-rapidity data for hadron yield ratios at AGS, SPS and two highest RHIC energies, are best fit when $R_\pi = 0, R_K = 0.35$ fm, $R_{mesons} = 0.35$ fm, $R_{baryon} = 0.5$ fm~\cite{Bugaev:2012wp}. A choice of 2nd virial coefficient for nucleons  to be $3.42\ \text{fm}^3$ has been found to generate the ground state nuclear properties well~\cite{Vovchenko:2017zpj}. Some other choices can be found in Refs.~\cite{Samanta:2017yhh,Vovchenko:2018cnf,Vovchenko:2020lju}. This indicates that there is no strict consensus about the value of excluded volume of hadrons.
\section{Experimental studies}
QGP can be formed in laboratory by colliding two heavy nuclei at very high energies. A schematic evolution of the matter created in heavy-ion collisions is shown in Fig. 1.2. In such relativistic high energies, the heavy nuclei experience Lorentz contraction. At this time, the self-interaction among gluons leads to a large number of low-x gluons (x is the fraction of momentum of the hadron carried by the gluon). The gluon densities inside the colliding hadrons increase and particles with high transverse momentum become suppressed at this stage. The colliding nuclei in off-central collisions have a participant part and a spectator part. The spectator parts do not participate directly in the formation of QGP. Just after the collisions, the colliding nuclei enter the partonic stage. At this stage, the partons interact by large momentum transfer and high transverse momentum jets
\begin{figure}[h]
	\begin{center}
		\includegraphics[scale=0.51]{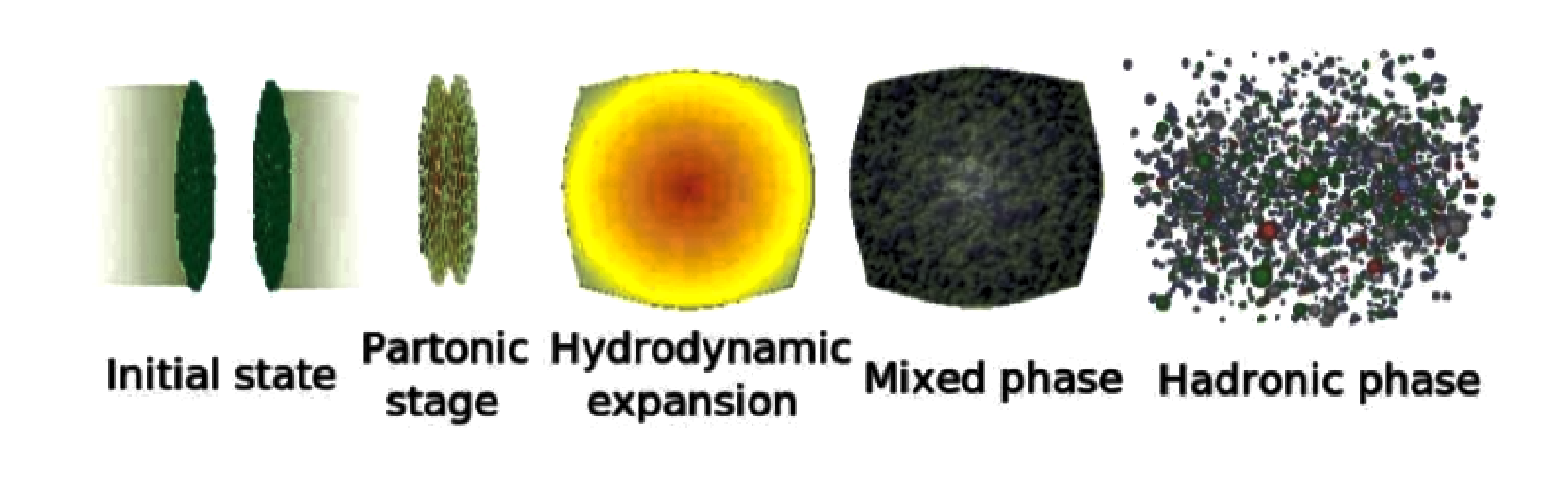}
		\caption{Different stages of relativistic heavy-ion collision experiment.}
	\end{center}
	\label{evolution1}
\end{figure}
 and heavy quarks are produced. A large number of particles are produced out of the energy available and they form a fireball of hot and dense strongly interacting matter. The fireball then thermalizes after a very small time and starts to expand in size. As it expands, the energy density decreases and the system cools gradually and finally, the system hadronizes. At this time, the hadrons in the system are mainly resonances. This state is called resonance-gas state. At the end of this state, the rate of expansion of the state is far greater than the rate of the inelastic collisions. The system cools rapidly and the density of higher mass resonances decrease sharply. At a temperature close to pion mass, the system is dominated by lighter and stable hadrons. This is a dilute-gas state. Particles interact via elastic collisions in this state. Chemical equilibrium is maintained in this state and the particle composition of the system remains almost same. If the centre of mass energy per nucleon $(\sqrt{s_{NN}})$ of the colliding nuclei is large, which is achievable in top RHIC or LHC energies, the two colliding nuclei tend to pass through each other. Thus the produced system has very high temperature and energy density in the central region with low net baryon density. If the centre of mass energy per nucleon is between a few GeV to few tens of GeV, the colliding nuclei tend to stay with each other. In this case, matter with high net baryon density but relatively low temperature is produced. This type of matter has been created in Super Proton Synchrotron (SPS) at CERN and Alternating Gradient Synchrotron (AGS) at BNL. The Beam Energy Scan (BES) program at RHIC is aimed at creating much wider range of net baryon density. Future projects like Compressed Baryonic Matter (CBM) at Facility for Anti-proton and Ion Research (FAIR) at GSI and Nuclotron-based Ion Collider fAcility (NICA) at JINR will shed more light on our understanding of such matter.
\section{Heavy-ion collision and magnetic field}
A huge magnetic field, $B\sim m_{\pi}^2$ ($\sim 10^{18}$G) is expected to be created in the off-central HICs due to relativistic motion of charged  particles~\cite{Skokov:2009qp,Bzdak:2011yy,Deng:2012pc}. The presence of magnetic field $(B)$ can affect the equation of state and hence it can impact significantly the overall structure of the phase diagram. For instance, magnetic field may induce interesting phenomena on QCD matter, $viz.$ chiral magnetic effect~\cite{Fukushima:2008xe}, magnetic catalysis~\cite{Shovkovy:2012zn}, inverse magnetic catalysis~\cite{Bali:2011qj,Preis:2012fh} etc. It is important to mention that, strong magnetic fields are also expected to be present in dense neutron stars~\cite{Duncan:1992hi,dey}. Such magnetic fields might have been present during the electroweak transition in the early universe~\cite{Vachaspati:1991nm,Bhatt:2015ewa}. The way the magnetic field influences transport phenomena are also very important in the context of heavy-ion collisions as well as in the context of neutron stars~\cite{Kadam:2014xka}. Thus it is very important to study the effect of magnetic field on the hot and dense strongly interacting matter.
\section{Present work}
In the present work, a study of the hadronic matter with different variants of the HRG model is planned. In chapter 2, brief descriptions of the ideal HRG model and its different variants, which include repulsive interactions, are presented. The effect of unequal excluded volume of different hadronic species has been introduced and the  model has been named as Modified Excluded Volume Hadron Resonance Gas (MEVHRG) model. Then the effect of Lorentz contraction of the excluded volumes in the rest frame of the hadronic matter has been incorporated. Another method of introducing the repulsive interactions in the HRG model, namely, the HRG mean-field model has also been discussed in this chapter. In chapter 3, the thermodynamic properties like pressure, energy density, entropy density, specific heat, isothermal compressibility, speed of sound, etc have been investigated, in the ambit of those variants of HRG model. Chapter 4 has been devoted to the study of the susceptibilities of conserved charges. The effect of repulsive interactions on the susceptibilities have been studied in certain amount of detail and the results have been compared with the LQCD data. In chapter 5, a brief description of the EVHRG model with external magnetic field has been presented. Then  the results for thermodynamic quantities and susceptibilities of conserved charges in presence of magnetic field have been presented and those have been compared with the case where magnetic field is absent. Finally, chapter 6 contains a brief summary and concluding remarks.
	\newpage
\rhead{Models of hadronic matter}
\chapter{\label{chap:model} Models of hadronic matter}
Effective models are useful to obtain reliable quantitative results in the non-perturbative domain of strong interaction. These models make some approximations regarding the interactions in strongly interacting matter and simplify the calculations to a great extent. Also, these models can be confidently used in the non zero baryon chemical potential region, unlike lattice QCD. In this work, the different variants of Hadron Resonance Gas (HRG) model have been used to look at the hadronic matter at high temperatures and densities. 
\section{Hadron Resonance Gas Model}
Hadron resonance gas model is a low temperature statistical thermal model describing hadronic phase of QCD. This model 
is based on S-matrix formulation of statistical mechanics~\cite{Dashen:1969ep}. In the relativistic virial expansion of 
the partition function the interactions are manifested in the form of  phase shifts in the two particle scattering. 
If such scattering occurs through the exchange of a  narrow resonance state then the interacting partition function becomes  a non-interacting partition function with the additional contributions from the exchanged 
resonances~\cite{Dashen:1974yy,Welke:1990za,Venugopalan:1992hy}. 
In the HRG model, the thermodynamic quantities 
of low temperature hadronic matter  can be obtained from the partition function which contains all relevant degrees of freedom 
of the confined QCD phase and  implicitly includes the attractive interaction that results in resonance formation.

Thermodynamic properties of ideal hadron resonance gas model can be derived from the grand canonical partition function given by
\begin{eqnarray}
\text{ln}\mathcal{Z}(T,\mu_B,\mu_Q,\mu_S, V)= 
\sum_{i}\text{ln}\mathcal{Z}_i(T,\mu_B,\mu_Q,\mu_{S},V)
\end{eqnarray}
where $\mu_B, \mu_Q$ and $\mu_S$ are the chemical potentials corresponding to baryon number, electric charge and strangeness respectively. There is a summation over all the hadronic species '$\it{i}$'.

The grand canonical partition function for an ideal gas $\mathcal{Z}^{id}_i$ is given by~\cite{pathria}:
\begin{equation}
\ln{\mathcal{Z}^{id}_i}=\pm\frac{Vg_i}{2\pi^2}\int_{0}^{\infty}p^2\ dp\ln\{1\pm exp[-(E_i-\mu_i)/T]\}
\label{HRG_pressure}
\end{equation}
Here $V$ is the volume of the system, 
$g_i$ is the degeneracy factor, 
$p$ is the momentum of a particle, 
$E_i=\sqrt{p^2+m^2_i}$ is the energy of a single particle, 
$m_i$ is the mass of particle species $\it{i}$, 
$T$ is the temperature and 
$\mu_i=B_i\mu_B+S_i\mu_S+Q_i\mu_Q$ is the chemical potential of particle species $i$.  
For the case of strong interactions, baryon number, electric charge and strangeness are conserved. Hence,  $\mu_B, \mu_Q$ and $\mu_S$ are the three chemical potentials associated with these three conserved charges.
In the above equation, the (+) sign is for fermions and the (-) sign is for bosons.

Various thermodynamic quantities like pressure $(P)$, number density $(n)$, energy density $(\epsilon)$, entropy density $(s)$, etc. can be derived from the expression of the partition function.
The pressure $(P_i)$ of the species $i$ can be obtained from the partition function as
\begin{eqnarray}
P_i^{id}(T,\mu_B,\mu_Q,\mu_S,V)=
\lim_{V\rightarrow\infty}\frac{ T}{V}\  \text{ln}\mathcal{Z}^{id}_i(T,\mu_B,\mu_Q,\mu_S,V)
\end{eqnarray}
Number density $n_i$ is given by
\begin{equation}
n_i^{id}=\frac{T}{V}\Big(\frac{\partial \ln \mathcal{Z}^{id}_i}{\partial \mu_i}\Big)_{V,T}=\frac{g_i}{2\pi^2}\int_{0}^{\infty}\frac{p^2\ dp}{\exp[(E_i-\mu_i)/T]\pm1}
\label{ndensity_HRG}
\end{equation}
The energy density is given by
\begin{equation}
\begin{split}
\varepsilon_i^{id}&=\frac{E_i^{id}}{V}=-\frac{1}{V} \left(\frac{\partial \ln\mathcal{ Z}_i^{id}}{\partial\frac{1}{T}}\right)_{\frac{\mu}{T}}
=\frac{g_i}{2\pi^2}\int_0^\infty\frac{p^2\,dp}{\exp[(E_i-\mu_i)/T]\pm1}E_i,
\end{split}
\end{equation}

The entropy density is given by
\begin{eqnarray}
s_i^{id}&=&\frac{S_i^{id}}{V}=\frac{1}{V}\left(\frac{\partial\left({T \ln Z_i^{id}}\right)}{\partial T}\right)_{V,\mu}\\ \nonumber
&=&\pm\frac{g_i}{2\pi^2}\int_0^\infty p^2\,dp \Bigg [ \ln\left(1\pm\exp(-\frac{(E_i-\mu_i)}{T})\right).\\ \nonumber
&\pm&\frac{(E_i-\mu_i)}{T(\exp((E_i-\mu_i)/T)\pm1)}\Bigg ]
\end{eqnarray}
The above equations are valid for a system in thermal equilibrium. Hence, when one calculates the contribution to these quantities from a particular hadronic species, one basically assumes that the particle species is stable at that temperature. Thus, one sums up contribution from all the particle species in HRG model: whether it is a stable particle or a resonance particle. Consideration of resonance particles as stable ones takes care of the attractive interaction in the system.
\section{EVHRG Model}
In EVHRG Model, a short range hardcore repulsive hadron-hadron interaction is taken into account by considering excluded volume~\cite{EVHRG1,EVHRG2,EVHRG3,EVHRG4,EVHRG5,EVHRG6,EVHRG7,EVHRG8} of the hadrons.

One knows that $\mathcal{Z}$ can be defined as
\begin{equation}
\mathcal{Z}(T,\mu,V)=\sum_0^{\infty}e^{\frac{\mu N}{T}}Z(T,N,V)
\end{equation}
where $Z(T,N,V)$ is the canonical partition function.
The excluded volume $V_{ev}$ of a particle can be incorporated by modifying the $Z$ as
\begin{equation}
Z^{ev}(T,N,V)=Z(T,N,V-V_{ev}N)\ \theta(V-V_{ev}N)
\end{equation}
the $\theta(V-V_{ev}N)$ function ensures that the maximum number of particles which can be accommodated in the system should not have total excluded volume greater than the system volume $V.$
\begin{equation}
\mathcal{Z}^{ev}(T,\mu,V)=\sum_0^{\infty}e^{\frac{\mu N}{T}}Z(T,N,V-V_{ev}N)\ \theta(V-V_{ev}N)
\end{equation}
A Laplace transformation of the above equation gives
\begin{eqnarray}
\hat{\mathcal{Z}}^{ev}(T,\mu,\xi)&=\int_0^{\infty}dV\ e^{-\xi V}\ \mathcal{Z}^{ev}(T,\mu,V)\\ \nonumber
&=\int_0^{\infty}dx\ e^{-\xi x}\ \mathcal{Z}^{ev}(T,\hat{\mu},x)
\end{eqnarray}
where $\hat{\mu}=\mu-V_{ev}T\xi$ and the integration variable has been changed from $V$ to $x=V-V_{ev}N$.

From the properties of Laplace transform one can write,
\begin{equation}
P^{ev}(T,\mu)=\lim_{V\rightarrow\infty}\frac{T}{V}\mathcal{Z}^{ev}(T,\mu,V)=T\xi^*(T,\mu)
\label{pexcl}
\end{equation}
Here $\xi^*(T,\mu)$ is the extreme right singularity of the function $\mathcal{Z}^{ev}$ in the variable $\xi$. Here, $\mathcal{Z}^{ev}$ is singular only at $x\rightarrow\infty$. At that point,
\begin{equation}
\xi^*=\lim_{x\rightarrow\infty} \frac{ln\ \mathcal{Z}(T,\tilde{\mu},x)}{x}
\end{equation}
with $\tilde{\mu}=\mu-V_{ev}T\xi^*$.

From equation~(\ref{pexcl}) one has
\begin{equation}
P^{ev}(T,\mu)=P^{id}(T,\tilde{\mu});\ \ \ \ \tilde{\mu}=\mu-V_{ev}P^{ev}(T,\mu)
\end{equation}
It has been found that the inclusion of repulsive interactions by introduction of excluded volume corrections in ideal HRG model can have significant effect on various thermodynamic observables, especially on higher order fluctuations~\cite{zeeb,bugaev,vovchenko1,Albright:2015uua,Garg:2013ata,ab2}. It is also important in the context of statistical hadronization~\cite{BraunMunzinger:1999qy}. Study of the transport coefficients of hadronic matter in HRG model with repulsive interactions has also been done by some authors ~\cite{Kadam:2015xsa,Kadam:2017iaz,Kadam:2019peo}. Thus, it is necessary to include the repulsive interactions in the ideal HRG model to understand the properties of hadronic matter in the context of HICs.

In this thesis, two different types of excluded volume corrections have been considered:\\
\begin{itemize}
	\item Type 1 EVHRG model
	\item Type 2 EVHRG model
\end{itemize}
\subsection{Type 1 EVHRG model}
In this type of EVHRG model, only meson-meson, baryon-baryon and antibaryon-antibaryon repulsive interactions are considered. There is no repulsive interaction among any other pairs.
In type 1 EVHRG model pressure is given as
\begin{eqnarray}
P(T,\mu_1,\mu_2,...)&=&P_{(m)}(T,\mu_1,\mu_2,...)+P_{(b)}(T,\mu_1,\mu_2,...)\nonumber \\
&+&P_{(\bar{b})}(T,\mu_1,\mu_2,...)
\label{type1_pressure}
\end{eqnarray}
\begin{equation*}
P_{(m)}(T,\mu_1,\mu_2,...)=\sum_{p}P_{(m)p}^{id}(T,\hat{\mu}_{(m)1},\hat{\mu}_{(m)2},...)
\end{equation*}
\begin{equation*}
P_{(b)}(T,\mu_1,\mu_2,...)=\sum_{q}P_{(b)q}^{id}(T,\hat{\mu}_{(b)1},\hat{\mu}_{(b)2},...)
\end{equation*}
\begin{equation*}
P_{(\bar{b})}(T,\mu_1,\mu_2,...)=\sum_{r}P_{(\bar{b})r}^{id}(T,\hat{\mu}_{(\bar{b})1},\hat{\mu}_{(\bar{b})2},...)
\end{equation*}
where $P_{(m)}$, $P_{(b)}$ and $P_{(\bar{b})}$ are mesonic, baryonic and anti-baryonic contribution to pressure with $\hat{\mu}_{(m)p}$, $\hat{\mu}_{(b)q}$ and $\hat{\mu}_{\bar{(b)}r}$ as the effective chemical potential for $\it{p}$-th meson, $\it{q}$-th baryon and $\it{r}$-th anti-baryon respectively which can be written as
\begin{equation*}
\hat{\mu}_{(m)p}=\mu_p-V_{ev,p}P_{(m)}(T,\mu_1,\mu_2,...)
\end{equation*}
\begin{equation*}
\hat{\mu}_{(b)q}=\mu_q-V_{ev,q}P_{(b)}(T,\mu_1,\mu_2,...)
\end{equation*}
\begin{equation}
\hat{\mu}_{(\bar{b})r}=\mu_r-V_{ev,r}P_{\bar{(b)}}(T,\mu_1,\mu_2,...)
\label{type1_mu}
\end{equation}
where $V_{ev,p} =4\frac{4}{3}\pi R_p^3$ is the excluded volume for $\it{p}$-th meson having hard-core radius $R_p$ and same treatment holds for baryons and anti-baryons.

Equations (\ref{type1_pressure}) and (\ref{type1_mu}) are iteratively solved to get the pressure.
Since effective chemical potential $\hat{\mu}_{(m)p}$ is smaller than chemical potential $\mu_p$, pressure $P_{(m)}(T,\mu_1,\mu_2,...)$ is smaller than ideal mesonic pressure $P_{(m)}^{id}$ and same is true for baryons and anti-baryons.
One can calculate various thermodynamic quantities like number density of $\it{p}$-th meson in the EVHRG model, $n_{(m)p}$ as
\begin{equation}
n_{(m)p}=n_{(m)p}(T,\mu_1,\mu_2,...)=\frac{\partial P_{(m)}}{\partial\mu_p}=\frac{n_{(m)p}^{id}(T,\hat{\mu}_p)}{1+\sum_pV_{ev,p}n_{(m)p}^{id}(T,\hat{\mu}_k)}
\end{equation}
Similar relations hold for baryons and anti-baryons. In what follows,  the notification $(m)$, $(b)$ and $(\bar{b})$ shall be dropped and the equations introduced will be valid for mesons, baryons and anti-baryons separately.
\subsection{Type 2 EVHRG model}
In this type of EVHRG model, repulsive interactions among all the hadronic species are considered. In type 2 EVHRG model pressure is given as
\begin{equation}
P(T,\mu_1,\mu_2,...)=\sum_{p}P_p^{id}(T,\hat{\mu}_{1},\hat{\mu}_{2},...)
\end{equation}
with the effective chemical potential given 
\begin{equation*}
\hat{\mu}_{p}=\mu_p-V_{ev,p}P_(T,\mu_1,\mu_2,...)
\end{equation*}
The thermodynamic quantities like pressure can be obtained in a similar way as the type 1 EVHRG model.
\section{Modified Excluded Volume Hadron Resonance Gas (MEVHRG) model}
To analyse the effect of unequal size of different hadron species, one 
takes recourse to virial expansion method. Such an analysis is done in
Refs.~\cite{st1,st2}. In this analysis, excluded volume for a single
particle is taken to be half the volume excluded by two touching spheres
of unequal radii instead of equal radii as taken in the previous EVHRG
analysis. This Model is named as Modified Excluded Volume Hadron
Resonance Gas Model (MEVHRG model).

%
When many particle species are present, the second virial coefficients of particle $"k"$ and $"n"$ are
\begin{equation}
a_{kn}=\frac{2}{3}\pi(R_k+R_n)^3=\frac{2}{3}\pi(R_k^3+3R_k^2R_n+3R_kR_n^2+R_n^3)
\end{equation}
Then the pressure of the system becomes~\cite{st2}
\begin{eqnarray}
P^M&=&T\sum_{k=1}^{N}\phi_ke^{\frac{\mu_k}{T}}\Big[1-\frac{4}{3}\pi R_k^3\sum_{n=1}^{N}\phi_ne^{\frac{\mu_n}{T}}\nonumber \\
&-&2\pi R_k^2\sum_{n=1}^{N}R_n\phi_ne^{\frac{\mu_n}{T}}-2\pi R_k\sum_{n=1}^{N}R_n^2\phi_ne^{\frac{\mu_n}{T}}\Big]
\end{eqnarray}
where $\phi_k=g_k\gamma_S^{|S_K|}\int\frac{d^3p}{2\pi^3}\exp[-\frac{\sqrt{p^2+m^2}}{T}]$ is the thermal density of particles and $'M'$ stands for MEVHRG model. Here $\gamma_S$ is the strangeness suppression factor and $|S_K|$ is the number of valence strange quarks and antiquarks in that particular hadron.
Here an approximation has been made that surface and curvature terms in the above expression which are proportional to $R_k^2$ and $R_k$ respectively, are equal and then one gets
\begin{equation}
P^M\simeq T\sum_{k=1}^{N}\phi_ke^{\frac{\mu_k}{T}}\Big[1-\frac{4}{3}\pi R_k^3\sum_{n=1}^{N}\phi_ne^{\frac{\mu_n}{T}}-4\pi R_k^2\sum_{n=1}^{N}R_n\phi_ne^{\frac{\mu_n}{T}}\Big]
\end{equation}
To facilitate an iterative algorithm for obtaining $P^M$, one replaces $\phi_ne^{\frac{\mu_n}{T}}\simeq\frac{P_n}{T}$
\begin{equation}
P^M\simeq T\sum_{k=1}^{N}\phi_ke^{\frac{\mu_k}{T}}\Big[1-\frac{4}{3}\pi R_k^3\frac{P^M}{T}-4\pi R_k^2\sum_{n=1}^{N}\frac{R_nP^M_n}{T}\Big]
\end{equation}
For $\mu_k/T<1$ one further approximates 
\begin{equation}
P^M\simeq T\sum_{k=1}^{N}\phi_k\exp\Big[\frac{\mu_k}{T}-\frac{4}{3}\pi R_k^3\frac{P^M}{T}-4\pi R_k^2\sum_{n=1}^{N}\frac{R_nP^M_n}{T}\Big]
\end{equation}
The above equation can be written as
\begin{equation}
P^M=\sum_{k=1}^{N}P_k(T,\hat{\mu}_k^M)
\label{eq_PM}
\end{equation}
where
\begin{equation}
\hat{\mu}_k^M=\mu_k-\frac{4}{3}\pi R_k^3P^M-4\pi R_k^2\sum_{n=1}^{N}R_nP^M_n
\label{mu_k}
\end{equation}
Eqn.~(\ref{eq_PM}) and (\ref{mu_k}) can be solved iteratively to get pressure.

Number density of $i$-th hadron is given as
\begin{equation}
n_i^M=\frac{[(1+\sum_k4\pi R_k^3n_k^{id}(T,\hat{\mu}_k^M))n_i^{id}(T,\hat{\mu}_i^M)-\sum_j4\pi R_j^2n_j^{id}(T,\hat{\mu}_j^M)R_in_i^{id}(T,\hat{\mu}_i^M)]}{\splitfrac{[(1+\sum_k4\pi R_k^3n_k^{id}(T,\hat{\mu}_k^M))(1+\sum_j\frac{4}{3}\pi R_j^3n_j^{id}(T,\hat{\mu}_j^M))}{-\sum_j4\pi R_j^2n_j^{id}(T,\hat{\mu}_j^M)\sum_k\frac{4}{3}\pi R_k^4n_k^{id}(T,\hat{\mu}_k^M)]}}
\end{equation}

\section{Effect of Lorentz Contraction (LMEVHRG Model)} 
In a hadron gas, the constituting particles can have large kinetic energies and hence they can have large velocities. So the excluded volume of a particular particle in the rest frame of the thermal medium is Lorentz contracted.  This weakens the excluded volume repulsive interaction compared to the case where the effect of Lorentz contraction is not taken into account. Generalisation of cluster and virial expansions for momentum dependent inter-particle potentials to incorporate Lorentz contraction has been studied by some authors~\cite{st1}. The effect of Lorentz contraction has been introduced through a Lorentz contracted radius dependent potential. Calculating the Lorentz contracted excluded volume of two particles is very complex because the shape of the particles become ellipsoidal. Therefore, the distance between two touching ellipsoids depends not only on the size at the particles' rest frame but also on the angle of their relative velocities. The relevant area which one is  looking for is when the mean energy per particle is higher compared to the masses of the particles. The collisions with collinear velocities in the system are more common than collisions with non-collinear velocities since in the later case the effective excluded volume is larger and hence such collisions are suppressed. At reasonable densities, the orientation of different particles become correlated and hence the system tends to adjust itself in the configuration with minimum excluded volume. Also there is possibility of rotation of Lorentz contracted ellipsoids. Effect of rotation is negligible in a dilute system because the average number of collision per second among the particles of the system is very small. On the other hand, in a highly dense system the compact arrangement of particles hinders the ellipsoids from rotating. These effects will be ignored in this work. Here, one takes into account the effect of Lorentz contraction on the excluded volume of the particles in a simple phenomenological way in the rest frame of the system and compare the results obtained for the thermodynamic quantities and susceptibilities with those of Lattice QCD. If a particle has momentum $p$, its size is contracted in the direction of motion according to the formula $R^{\prime}=R\sqrt{1-v^2}$ where $v$ is the velocity of that particle. Now, $p=mv=\frac{m_0}{\sqrt{1-v^2}}v$ where $m_o$ is the rest mass of the particle. From this one gets $1-v^2=\frac{m_0^2}{p^2+m_o^2}$.

So, the effective chemical potential with Lorentz contraction becomes
\begin{eqnarray}
\hat{\mu}_k^{LM}=\mu_k-\frac{4}{3}\pi R_k^3\frac{m
	_0}{\sqrt{p^2+m_0^2}}P^{LM}-4\pi R_k^2\frac{m
	_0}{\sqrt{p^2+m_0^2}}\sum_{n=1}^{N}R_nP^{LM}_n
\label{mu_LMEVHRG}
\end{eqnarray}
Here 'LM' in the superscript stands for Lorentz Contracted Modified EVHRG Model. Eqn. (\ref{mu_LMEVHRG}) is inserted into the momentum integral of eqn. (\ref{HRG_pressure}) to calculate $P^{LM}$.

The average velocity of lighter particles are larger than that of heavier particles. As a result, lighter particles experience larger Lorentz contraction than heavier particles. Different particles of the same species with different velocities should have different degree of Lorentz contraction of excluded volumes and the two particle excluded volume matrix elements are different for different combinations of particles belonging to the same species but with different velocities. Particle species with larger excluded volumes in their rest frames, will experience larger amount of volume contraction as compared to particle species with smaller excluded volumes in their rest frames. If lighter particles are assigned larger excluded volumes than heavier particles then difference between the results of MEVHRG and LMEVHRG models will be greater than the case when lighter particles are assigned smaller excluded volumes.

There is a problem in consideration of the Lorentz contraction to the excluded volume. As the average excluded volume per particle decreases with increasing temperature, there is more available space in the system for new particles and hence phase transition is delayed when this model is used in combination with a model incorporating deconfinement. If one considers the effects of orientations and rotations of particles then this problem can be circumvented up to some extent as they tend to decrease the particle density.

\section{Mean-field HRG (HRGMF) model}
Ideal HRG model can be extended by including short range repulsive interactions between hadrons. These 
repulsive interactions can be treated in mean field approach where the single particle energies 
$\epsilon_a$ get shifted by the mean field repulsive interaction 
as ~\cite{ Kapusta:1982qd,Olive:1980dy}
\begin{equation}
\varepsilon_{a}=\sqrt{p^2+m_{a}^2}+U(n)=E_{a}+U(n)
\label{dispersion}
\end{equation}
where $n$ is the total hadron number density. The potential energy $U$ represents repulsive interaction between hadrons 
and it is taken to be function of total hadron density $n$. For an arbitrary inter-hadron potential, 
the potential energy $U$  is given by

\begin{equation}
U(n)=Kn
\label{poten}
\end{equation}
where $K$ is a  phenomenological parameter given by the spatial integration of the inter hadron potential.

In the present investigation,  different repulsive interaction parameter for baryons  and mesons have been assumed. 
One denotes the mean field parameter for baryons ($B$) and anti-baryons ($\bar{B}$) by $K_B$, while for mesons 
one denotes it by  $K_M$. Thus,  for baryons (antibaryons)
\begin{equation}
U(n_{B\{\bar{B}\}})=K_Bn_{B\{\bar{B}\}}
\label{potenbar}
\end{equation}

and for mesons
\begin{equation}
U(n_M)=K_Mn_M
\label{potenmes}
\end{equation}

The total hadron number density is

\begin{equation}
n(T,\mu)=\sum_{a}n_{a}=n_B+n_{\bar{B}}+n_M
\end{equation}
where $n_{a}$ is the number density of $a$-th hadronic species. Also, $n_B$, $n_{\bar{B}}$  and $n_M$ are total baryon, anti-baryon and meson number densities respectively. For baryons,

\begin{equation}
n_{{B}}=\sum_{a\in B}\int d\Gamma_{a}\:\frac{1}{e^{\frac{(E_{a}-{\mu_{\text{eff},B}})}{T}}+1}
\label{numdenbaryon}
\end{equation}
where $\mu_{\text{eff},{B}}=c_i\mu_i-K_Bn_{B}$ and $c_i =(B_i,Q_i,S_i), {\mu_i}=(\mu_B,\mu_Q ,\mu_S)$ and $d\Gamma_{a}\equiv\frac{g_{a}d^{3}p}{(2\pi)^3}$.
The sum is over all the baryons. Similarly, the number density of antibaryons is

\begin{equation}
n_{{\bar{B}}}=\sum_{a\in \bar{B}}\int d\Gamma_{a}\:\frac{1}{e^{\frac{(E_{a}-{\mu_{\text{eff},\bar{B}}})}{T}}+1}
\label{numdenantibaryon}
\end{equation}
where $\mu_{\text{eff},{\bar B}}=\bar{c_i}{\mu_i}-K_Bn_{\bar{B}}$. Note that repulsive mean-field parameter is same for baryons as well as anti-baryons.
For mesons,

\begin{equation}
n_{{M}}=\sum_{a\in M} \int d\Gamma_{a}\:\frac{1}{e^{\frac{(E_{a}-{\mu_{\text{eff},M}})}{T}}-1}
\label{numdenmeson}
\end{equation}
where $\mu_{\text{eff},{M}}={c_i}\mu_i-K_Mn_{M}$ and the sum is over all the mesons. 

Eqs. (\ref{numdenbaryon})-(\ref{numdenmeson})  are actually self consistent equations for number densities which can be solved numerically.
The expressions for the pressures  of baryons and mesons are then given, respectively, as
\begin{eqnarray}
P_{B\{\bar{B}\}}(T,\mu)&=&T\sum \limits_{a\in B \{\bar B \}} \int d\Gamma_{a}
\text{ln}\bigg[1+ e^{-(\frac{E_a-\mu_{\text{eff}}\{\bar{\mu}_{\text{eff}}\}}{T})}\bigg]\nonumber \\
&-&\phi_{B\{\bar{B}\}}(n_{B\{\bar{B}\}})
\label{pbbar}
\end{eqnarray}
\begin{eqnarray}
P_M(T,\mu)&=&-T \sum_{a\in M} \int d\Gamma_{a}\text{ln}\bigg[1- {e^{\frac{(E_{a}-{\mu_{\text{eff},M}})}{T}}}\bigg] \nonumber \\
&-&\phi_M(n_M)
\end{eqnarray}
where,

\begin{equation}
\phi_B(n_{B\{\bar{B}\}})=-\frac{1}{2}K_Bn_{B\{\bar{B}\}}^2
\end{equation}
and
\begin{equation}
\phi_M(n_M)=-\frac{1}{2}K_Mn_M^2
\end{equation}

Thermodynamic quantities can be readily calculated by taking appropriate derivatives of the partition function 
or equivalently of the pressure.
	\newpage
\rhead{Thermodynamics}
\chapter{\label{chap:thermodynamics} Thermodynamics of Hadronic Matter}
Behaviour of thermodynamic quantities like pressure, energy density, entropy density, specific heat at constant volume, isothermal compressibility and speed of sound are important to understand the properties of any physical system. Pressure ($P$) of a system as functions of temperature and chemical potential represents the equation of state of the system. Energy density ($\epsilon$) represents the energy stored in the system. Entropy density (s) is given by the derivative of pressure with respect to the temperature. These quantities depend on the degrees of freedom and the interaction prevailing in the system. Certain thermodynamic observables like specific heat ($C_V$), isothermal compressibility ($\kappa_T$) and speed of sound ($C_s^2$) are sensitive to the phase transition of the system. Thus their behaviour provide useful information about the nature of the phase transition. Specific heat $C_V$ is defined as the change in internal energy density for unit change in temperature. This quantity is important in the context of heavy-ion collision phenomenology as it is related to the temperature fluctuations of the system produced in heavy-ion collision by means of event-by-event analysis~\cite{Stodolsky:1995ds,Shuryak:1997yj,Basu:2016ibk}. The $C_V$ is expected to show a jump near a first order phase transition and a singularity for a second order phase transition. Thus, the behavior of $C_V$ can provide useful indication about the nature of phase transition. Isothermal compressibility represents the rate of change of volume of the system with respect to pressure at constant temperature. It is given by the second order derivative of Gibbs free energy and is expected to diverge at CEP. Fluctuations in a system causes deviation from equilibrium and, for a brief time, a non-equilibrium condition is reached~\cite{Khuntia:2018non}.  
Speed of sound reflects the small perturbations produced in the system in its local rest frame. Speed of sound depends on the degrees of freedom of the system and the interactions among them. Hence it shows critical behaviour near phase transition i.e. a minimum near phase transition~\cite{Tiwari:2011km}. In this section the results for thermodynamic quantities calculated within different variants of the HRG model have been presented.

The system with large volume in the ambit of some variants of HRG model namely EVHRG, MEVHRG, LMEVHRG, HRGMF models has been analyzed. In Ref.~\cite{Venugopalan:1992hy}, analysis of hadron-hadron scattering was performed. It was found that there is little evidence for hardcore repulsive interaction in hadron pairs other than in nucleons. HRG model description improved by including excluded volume correction with repulsive interaction among baryons only was considered in Ref.~\cite{Vovchenko:2016rkn}. With this type of repulsive interactions, the baryon susceptibilities calculated within this framework were in good agreement with LQCD data. Consideration of mesonic eigenvolume comparable to that of baryons leads to suppression of some thermodynamic quantities and degrades the agreement with Lattice data~\cite{EVHRG7,Vovchenko:2014pka}. Short range baryon-antibaryon interactions, are expected to be dominated by annihilation processes. With this consideration, excluded volume corrections lead to good qualitative agreement of pressure and baryon susceptibilities with LQCD data~\cite{Vovchenko:2017xad}. Here, only, meson-meson, baryon-baryon and antibaryon-antibaryon repulsive interactions have been considered. To include quantum correction to excluded volumes, one has to know the scattering phase shifts of different interaction channels, which is 
an involved procedure. Such correction in EVHRG model was considered  in Refs.~\cite{Vovchenko:2017drx,typel}. In Ref.~\cite{Vovchenko:2017drx}, taking such a correction, a monotonically decreasing second virial coefficient  with increasing temperature was obtained. This suggests that effects of quantum correction decreases with the increase in temperature.
\section{Pressure, energy density and entropy density}
\begin{figure}[h]
	\begin{center}
	\hspace{-1.2cm}
	\begin{tabular}{c c}
		\includegraphics[scale=0.55]{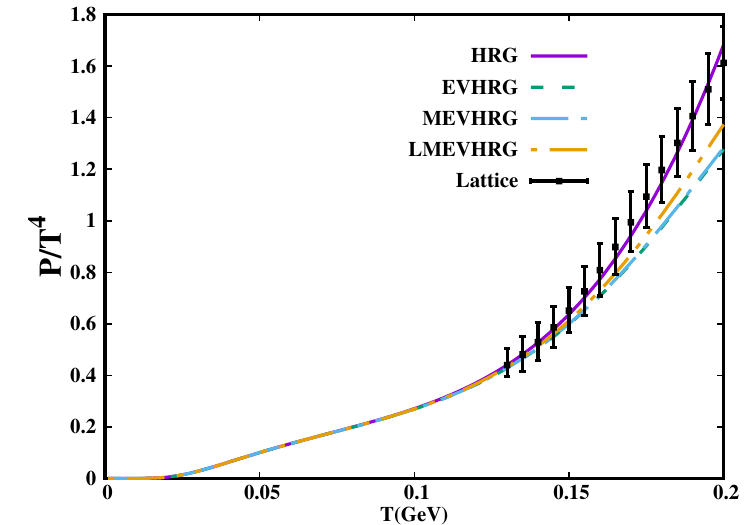}  
		\hspace{-0.4cm}
		\includegraphics[scale=0.55]{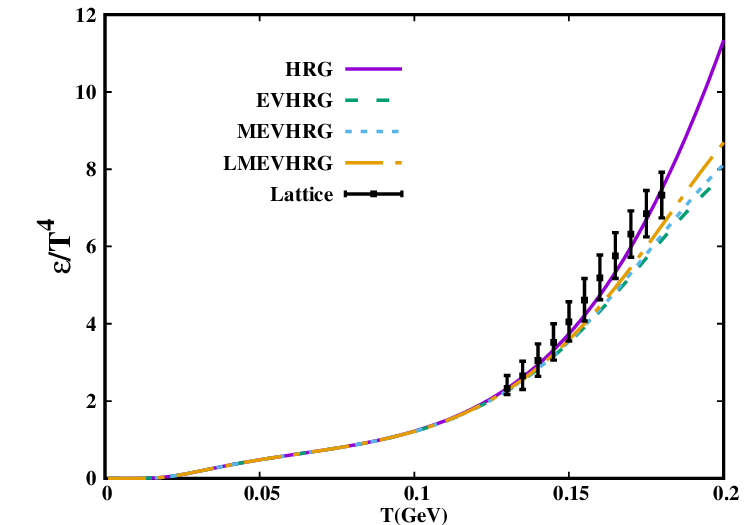}\\
		\includegraphics[scale=0.55]{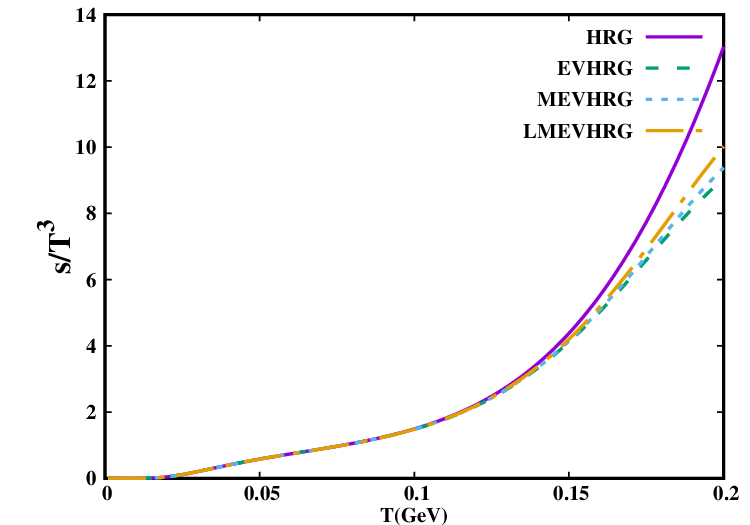}
	\end{tabular}
	\caption{Scaled pressure, energy density and entropy density in HRG, EVHRG, MEVHRG, LMEVHRG models at $\mu_B=\mu_Q=\mu_S=0$ for $R_b=0.35 fm, R_\pi=0.2 fm, R_m=0.3 fm$.}
	\label{fig 1}
	\end{center}
\end{figure}

In Fig.~\ref{fig 1} the scaled pressure ($P/T^4$), energy density ($\epsilon/T^4$) and entropy density ($s/T^3$) as functions of temperature ($T$) for three variants of HRG model namely, EVHRG, MEVHRG, LMEVHRG models along with the original HRG model have been shown. The findings have been compared with the available lattice results~\cite{Bazavov:2014pvz}. In this case, baryon radius $R_b = 0.35 fm$, pion radius $R_\pi = 0.2 fm$ and radii of other mesons 
$R_m = 0.3 fm$ have been considered. All particles listed in particle data book up to $3 \ GeV$ mass have been considered. The pressure corresponding to pure HRG is maximum. When hardcore repulsive interaction is included, there is a reduction in scaled pressure. Scaled pressure for MEVHRG model is greater than that for EVHRG model since the effective chemical potential in MEVHRG case is larger than the EVHRG case. Scaled pressure for LMEVHRG model is even higher because in this case repulsive interaction is weaker. Scaled pressure for EVHRG and MEVHRG differ only slightly from each other. It can be argued that this difference will be greater if one considers different excluded volumes for different hadron species instead of only three types of excluded volumes. This effect can be expected from eqn.~(\ref{mu_k}) where the last term in effective chemical potential deviates more from the EVHRG value as greater variety of radii is used. It is seen that Lorentz contraction of excluded volume of the particles has significant effect on the pressure of the system. Fig.~\ref{fig 1} also shows the scaled energy density and entropy density for HRG, EVHRG, MEVHRG and LMEVHRG models. The behaviour of energy density and entropy density are similar to that of the pressure. It is to be noted that the effect of excluded volume is larger in energy density and entropy density than that in pressure. Also, the effect of unequal radii is more prominent in energy density and entropy density. In all these three quantities, the EVHRG results are the smallest for the specific choice of excluded volumes.
\begin{figure}[h]
	\begin{center}
	\hspace{-1.2cm}
	\begin{tabular}{c c}
		\includegraphics[scale=0.55]{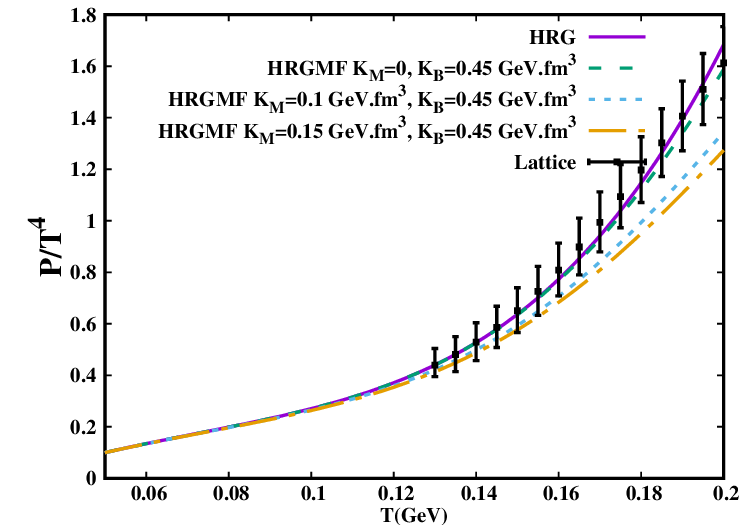}  
		\hspace{-0.4cm}
		\includegraphics[scale=0.55]{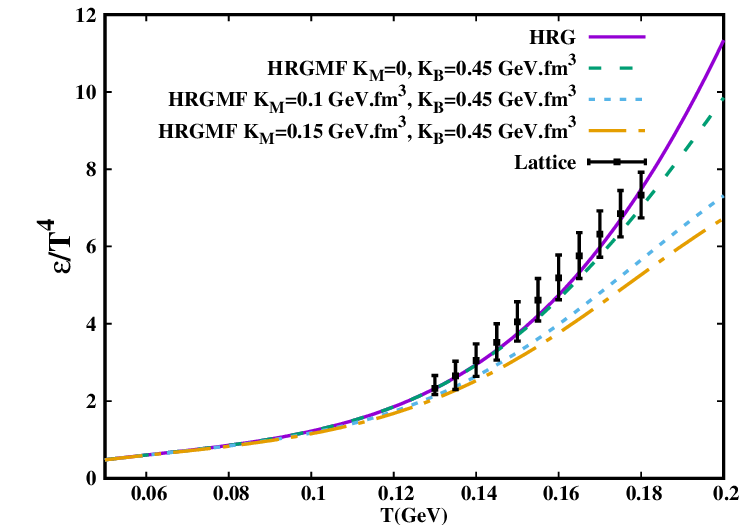}\\
		\includegraphics[scale=0.55]{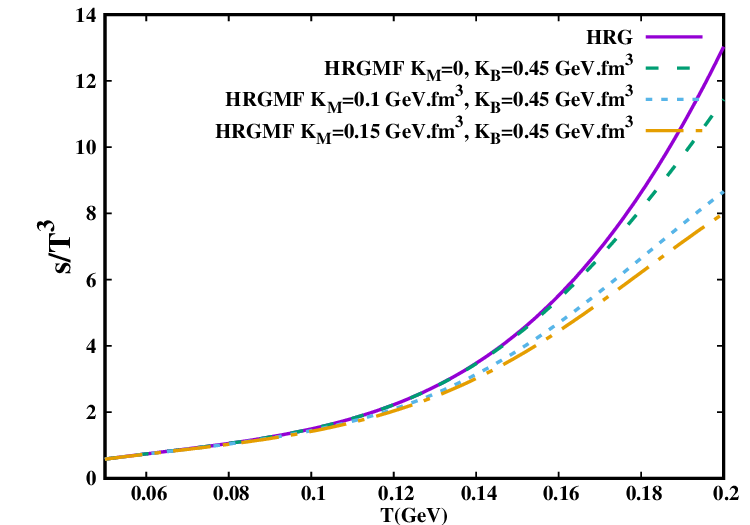}
		
	\end{tabular}
	\caption{Scaled pressure, energy density and entropy density at $\mu_B=\mu_Q=\mu_S=0$ in HRGMF model.}
	\label{fig 2_}
	\end{center}
\end{figure}

Fig.~\ref{fig 2_} shows the scaled pressure, energy density and entropy density as functions of temperature for HRGMF model. The only parameters in this model are $K_M$ and $K_B$. Three different representative values for meson mean field parameter, $viz.$, $K_M=0, 0.1$ and $0.15$ GeV$\cdot$fm$^{3}$ have been chosen,  while the baryon mean-field parameter has been fixed at $K_B=0.45$ GeV$\cdot$fm$^{3}$. It has been found that a value of $K_B$ close to 0.45 GeV$\cdot$fm$^{3}$ leads to good agreement with LQCD data for pressure and some of the cumulants of baryon number susceptibilities~\cite{Kadam:2019peo,Huovinen:2017ogf,Pal:2020ucy}. It can be seen that all the three quantities, considered, increase with temperature for HRGMF model, although, the numerical values differ significantly from each other with the variation of $K_M$. Mesons are the dominant contributors to these quantities at any temperature and hence, the effect of changing $K_M$ should have larger effect on these quantities than the effect of changing $K_B$. It is seen that these quantities decrease as $K_M$ is increased.
\section{Specific heat, compressibility and speed of sound}
Let us now discuss the results  for specific heat ($C_V$),  isothermal compressibility ($\kappa_T$) 
and speed of sound ($C_s^2$) as a function of temperature ($T$) at fixed values of baryon chemical potential ($\mu_B$).
It is interesting to explore what results are obtained, for these quantities, when  repulsive interaction is included. Also, the results have been compared with those obtained from ideal HRG model.

Two representative values of baryon chemical potential:  $\mu_B=0$ and $0.3$ GeV have been considered in the HRGMF model.  These values of $\mu_B$ have been chosen because the critical end point is expected to be near $\mu_B$\ = 0.3 GeV~\cite{Parotto:2018pwx,Stephanov:1998dy}. For comparison with excluded-volume HRG model (EVHRG),  the radii of hadrons have been taken as $R_M=0$ (for mesons) and  $R_B=0.3 \:\text{fm}$ (for baryons and anti-baryons). This choice is in line with the study of Ref.~\cite{Vovchenko:2017xad}. Here repulsive interaction among baryon-baryon and antibaryon-antibaryon pairs only has been considered i.e., type 1 EVHRG model has been considered here. No repulsive interaction among other pairs has been considered. If one does not consider any repulsive interaction among meson pairs and if only baryon-baryon and antibaryon-antibaryon repulsive interactions are considered, it leads to a better agreement of baryon number susceptibilities with Lattice data. This assumption is consistent with Refs.~\cite{Vovchenko:2016rkn,Satarov:2016peb}.

In case of ideal HRG model, the expression for the specific heat is written as
\begin{eqnarray}
C_{V,\mu_B}(T)&=&\bigg(\frac{\partial \epsilon}{\partial T}\bigg)_{V,\mu_{B}}\\\nonumber
&=&\sum_i \int_{0}^{\infty}d\Gamma_i \frac{ E_i(E_i-B_i\mu_B)\exp[(E_i-B_i\mu_B)/T]}{T^2\{\exp[(E_i-B_i\mu_B)/T]\pm 1\}^2}\\\nonumber
&=&\sum_i\int_{0}^{\infty} d\Gamma_i \: f_{\pm}(1\mp f_{\pm})\frac{E_i(E_i-B_i\mu_B)}{T^2}
\end{eqnarray}
where $f_{\pm}$ are the distribution functions defined as
\begin{equation}
f_{\pm}=\frac{1}{e^{(x/T)}\pm1}
\label{df}
\end{equation}
where $x=E_i-B_i\mu_B$ for ideal HRG model.

In case of HRGMF model, the expression for the specific heat picks up mean-field dependent terms.

For baryons, with $x=E_i-B_i\mu_B+K_Bn_B$ one gets
\begin{eqnarray}
C_{V,\mu_B}(T)&=&\sum_{i \in B}\int d\Gamma_i\: f^{\text{mf}}_{+}(1- f^{\text{mf}}_{+})(E_i+K_Bn_B)\nonumber\\
&\times& \Bigg[\frac{(E_i+K_Bn_B-B_i\mu_B)}{T^2}
-\frac{K_B}{T}\Bigg(\frac{\partial n_B}{\partial T}\Bigg )_{V,\mu_B}\Bigg]
\end{eqnarray}

where 

\begin{equation}
\frac{\partial n_{B}}{\partial T}=\sum_{i\in B}\frac{\int d\:\Gamma_i\:\bigg[\frac{(E_i+K_Bn_B-B_i\mu_B)}{T^2}\bigg]f_{+}^{\text{mf}}(1- f_{+}^{\text{mf}})}{1+\frac{K}{T}\int d\Gamma_{i}\: f^{\text{mf}}_{+}(1- f_{+}^{\text{mf}})}
\end{equation}
Similar expressions can be written for mesons but with $f_{-}, \ K_M$ and $n_M$. The $f^{\text{mf}}_{\pm}$ are the distribution functions given by Eq.(\ref{df}) but with $x=E_i-\tilde\mu_{\text{eff}}$.

\begin{figure}[h]
	\begin{center}
	\hspace{-1.2cm}
	\begin{tabular}{c c}
		\includegraphics[scale=0.55]{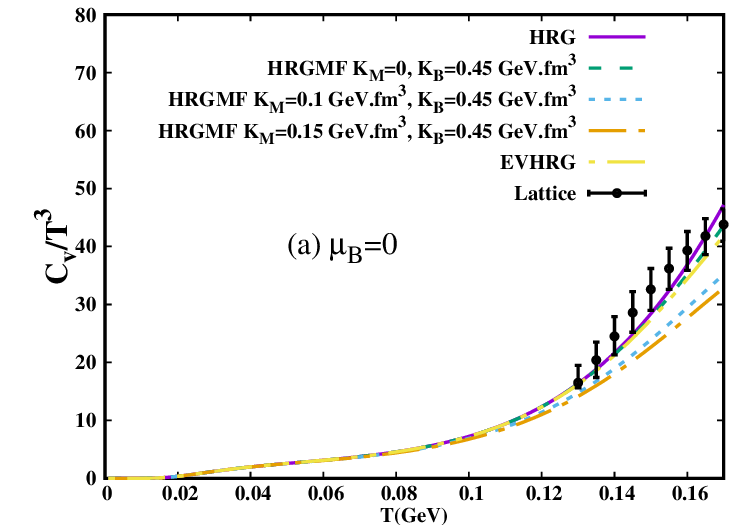}  
		\hspace{-0.4cm}
		\includegraphics[scale=0.55]{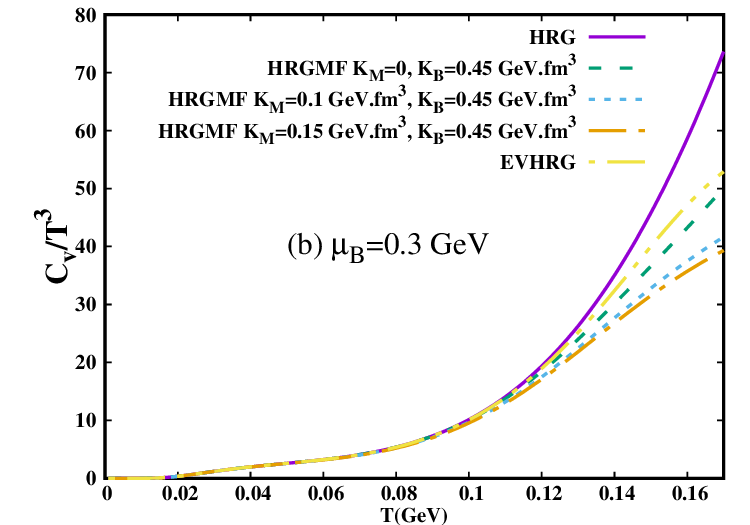} \\
		\hspace{-0.4cm}
		\includegraphics[scale=0.55]{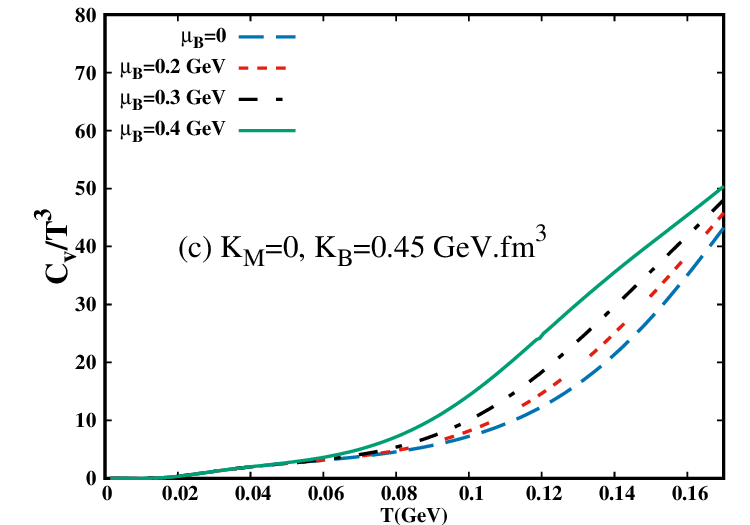}
		
	\end{tabular}
	\caption{Scaled specific heat as a function of temperature in HRGMF model. The first two panels are for $\mu_B = 0$ and 0.3 GeV respectively. Third panel shows scaled specific heat for $K_m=0$ at $\mu_B = 0,$ 0.2, 0.3 and 0.4 GeV.}
	\label{cv}
	\end{center}
\end{figure}

Figs.~\ref{cv}(a)  and \ref{cv}(b) show $C_V/T^3$ as a function of temperature for two different values of $\mu_B$ i.e. 0 and 0.3 GeV for ideal HRG, EVHRG and HRG mean field model (HRGMF). It can be seen that scaled specific heat ($C_V$) increases monotonically with temperature for ideal HRG. It rises rapidly near $T \sim 0.15$ GeV which is close to critical temperature ($T_c$) where the phase transition from hadronic phase to Quark-gluon plasma (QGP) phase is expected to occur as suggested by lattice QCD simulations. In fact, the specific heat is expected to show a power law behaviour $C_V\sim |t|^{-\alpha}$ near the QCD critical point, where $t=\frac{T-T_c}{T_c}$ and $\alpha$ is the critical exponent.  However,  in the hadron resonance gas model only a rapid rise as temperature reaches $T_c$ is observed.  In case of HRG model, with discrete mass spectrum and with only spin degree of freedom $J_i$, the spin degeneracy $(2J_i+1)$ results in hadron resonance abundances which goes like $g_i \propto m_i^2 $.  As a result, the partition function and all of its higher order derivatives are continuous. Thus, the specific heat does not show any divergence as $T\longrightarrow T_c$. The repulsive  interactions suppress the rapid rise of $C_V$ at high temperature as can be noted in Fig.\ref{cv}. This suppression is expected. By definition, $C_V$ is the amount of energy required to raise the temperature by unit value. If there is a repulsive interaction between hadrons, then it requires more energy to create a given species of hadron in a medium.  Hence the corresponding yield of hadrons is reduced.  Thus, the energy poured into the system is utilised to raise the temperature instead of creating massive hadrons.  As a result, the contribution of the degrees of freedom, which are hadrons in case of hadronic matter, to the  scaled specific heat $C_V/T^3$,  will be suppressed in the models with repulsive interactions. 

For $\mu_B = 0$, the results have been confronted with that obtained in Lattice QCD~\cite{Bazavov:2014pvz}. It is seen that HRGMF results are in good agreement with the lattice data at temperatures higher than $0.15 $ GeV. The inclusion of mesonic mean-field interactions slightly underestimates the lattice data although it reproduces the overall behaviour. The EVHRG result lies close to HRGMF result for $K_M=0$ as  repulsive interaction among meson pairs has not been considered.  It should be mentioned that confronting the HRG model and its extension beyond $T_c$ may not be reasonable. But this study clearly  indicates that the repulsive interactions play an important role near $T_c$.

Fig.~\ref{cv}(c)  shows specific heat estimations of HRGMF model with $K_M=0$ and $K_B=0.45$ GeV $\text{fm}^3$  at various values of $\mu_B$. It is to be noted that the scaled ratio $C_V/T^3$ is higher in magnitude at higher $\mu_B$. 

Isothermal compressibility ($\kappa_T$) is defined as
\begin{equation}
\kappa_T|_{T,\langle N_i\rangle}=-\frac{1}{V}\bigg(\frac{\partial V}{\partial P}\bigg)_{T,\langle N_i\rangle}
\label{kt}
\end{equation}
Considering the pressure as a function of temperature and individual chemical potentials $\mu_i$ of the particle species,
\begin{equation}
dP=\Big( \frac{\partial P}{\partial T}\Big)dT+\sum_i \Big(\frac{\partial P}{\partial \mu_i}\Big)d\mu_i
\end{equation}
From which one gets
\begin{equation}
\Big( \frac{\partial P}{\partial V}\Big)_{T,\{\langle N_i\rangle\}}=\sum_i \Big(\frac{\partial P}{\partial \mu_i}\Big)\Big( \frac{\partial \mu_i}{\partial V}\Big )\Big \vert_{T,\{\langle N_i\rangle\}}
\label{delpv}
\end{equation}
The change in the number of particle species $N_i$ is given by
\begin{equation}
dN_i=\Big( \frac{\partial N_i}{\partial T}\Big)dT+\Big( \frac{\partial N_i}{\partial \mu_i}\Big)d\mu_i+\Big( \frac{\partial N_i}{\partial V}\Big)dV
\end{equation}
For constant $N_i$ and $T$,
\begin{equation}
\Big( \frac{\partial \mu_i}{\partial V}\Big )\Big \vert_{T,\{\langle N_i\rangle\}}=-\Big( \frac{\partial N_i}{\partial V}\Big)\Big/\Big( \frac{\partial N_i}{\partial \mu_i}\Big)
\end{equation}
Using $\frac{\partial N}{\partial V}=\frac{\partial P}{\partial \mu}$ one has from eq.~(\ref{delpv})
\begin{equation}
\Big( \frac{\partial P}{\partial V}\Big)_{T,\{\langle N_i\rangle\}}=-\sum_i\frac{\Big(\frac{\partial P}{\partial \mu_i}\Big)^2}{\Big( \frac{\partial N_i}{\partial \mu_i}\Big)}
\label{kt1}
\end{equation}
Using equations~(\ref{kt}) and~(\ref{kt1}) one gets,
\begin{equation}
\kappa_T|_{T,\langle N_i\rangle}=\frac{1}{\sum_i[(\frac{\partial P}{\partial\mu_i})^2/{(\frac{\partial n_i}{\partial\mu_i})]}}
\end{equation}
Here, $\mu_i$ is the chemical potential linked to the $i$'th particle species. The $\mu_i$ is related to $\mu_B$, $\mu_Q$, and $\mu_S$ by the relation $\mu_i=B_i\mu_B+Q_i\mu_Q+S_i\mu_S$. The $n_i$ appearing in this expression is the density of $i^{th}$ species which is not zero. So to evaluate the sum, one first writes the expression for $P$ and $n_i$ as a function of $\mu_i$ (without explicitly writing $\mu_i$ in terms of $\mu_B,\ \mu_Q$ and $\mu_S$ at this time) and takes the derivative with respect to $\mu_i$ defined for that species. Note that the derivatives are non-zero even at $\mu_i$ = 0. Hence $\kappa_T$ gets contribution from all the species: mesons as well as baryons. The term  $\frac{\partial n_i}{\partial \mu_i}$  for a given baryon or meson species in HRG is given by

\begin{equation}
\frac{\partial n_i}{\partial \mu_i}=\int_{0}^{\infty}d\Gamma_{i}\:  \frac{1}{T}\:f_{\pm}[1\mp f_{\pm}]	
\end{equation}

while, in the HRGMF model it is given by,
\begin{equation}
\frac{\partial n_i}{\partial \mu_i}=\int_{0}^{\infty}d\Gamma_{i}\:  \frac{1}{T}\:f^{\text{mf}}_{\pm}[1\mp f^{\text{mf}}_{\pm}]	\bigg(1-K_{B\{M\}}\frac{\partial n_{B\{M\}}}{\partial \mu_i}\bigg)
\end{equation}
Here the upper sign is for baryons and lower sign is for mesons. The $K_B$ and $K_M$ are used for baryons and mesons respectively. (Here 
electric charge chemical potential ($\mu_Q$) and strangeness chemical potential ($\mu_S$) have been taken be zero). 

Figs.~\ref{kt}(a) and ~\ref{kt}(b)  show plots of isothermal compressibility $\kappa_T$ as a function of temperature at baryon chemical potential 0 and 0.3 GeV respectively. At $\mu_B=0$ (Fig.\ref{kt}(a)), for all the three variants of HRG model considered, $\kappa_T$ decreases with increasing temperature. In case of ideal HRG, this decreasing behaviour of $\kappa_T$ as a function of $T$ is expected as $\kappa_T\propto \text{pressure}^{-1}$.  As the temperature is increased, more and more hadrons populate the system. As a result, the pressure also increases which render the decreasing behaviour of $\kappa_T$. In case of HRGMF,  the scaled pressure ($P/T^4$) is smaller
\begin{figure}[h]
	\begin{center}
		\hspace{-1.2cm}
		\begin{tabular}{c c}
			\includegraphics[scale=0.55]{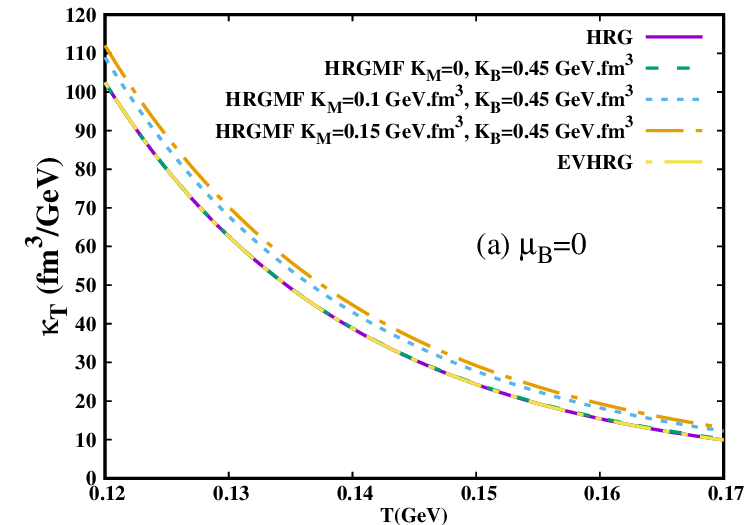} 
			\includegraphics[scale=0.55]{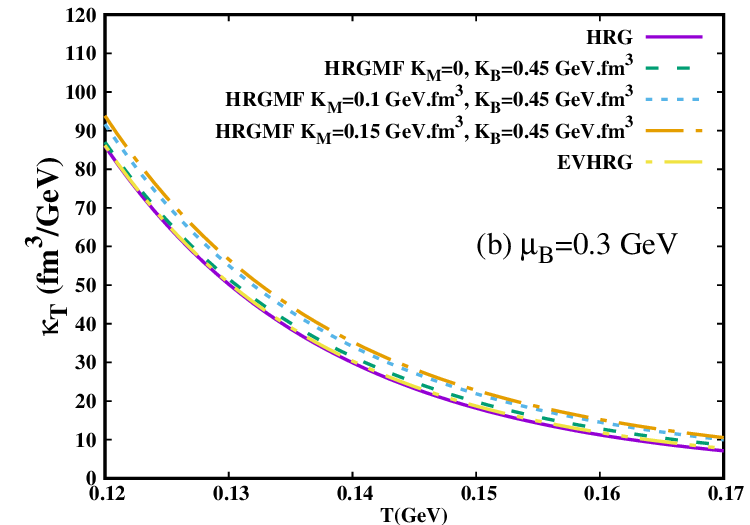} \\
			\includegraphics[scale=0.55]{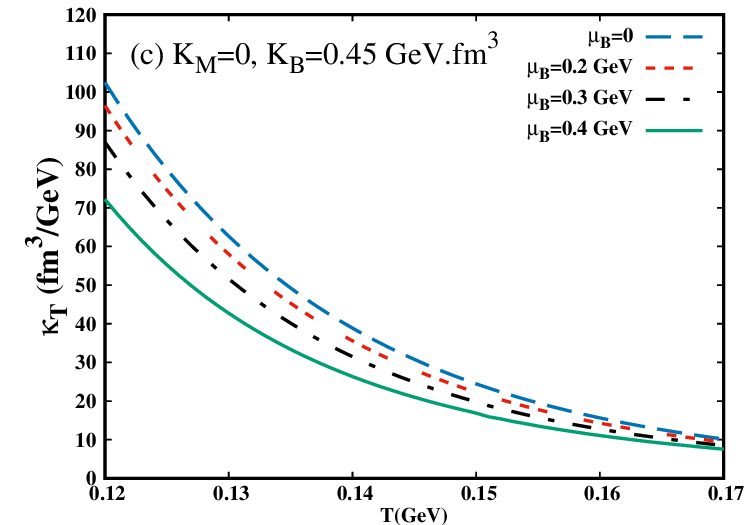}
		\end{tabular}
		
		\caption{Isothermal compressibility as a function of temperature in HRGMF model. The first two panels are for $\mu_B = 0$ and 0.3 GeV respectively. Third panel shows compressibility for $K_m=0$ at $\mu_B = 0,$ 0.2, 0.3 and 0.4 GeV.}
		\label{kt}
	\end{center}
\end{figure}
 compared to ideal HRG pressure. Hence the magnitude of $\kappa_T$ is larger in HRGMF and EVHRG at a given temperature. The choice of excluded volumes here ($R_M=0$ for mesons and $R_B$=0.3 fm for baryons and anti-baryons) lead to only small deviation in compressibility as compared to HRG results, which, for $\mu_B=0$, is negligible. It can be seen that the isothermal compressibility is more sensitive to the sizes of the mesons rather than those of the baryons. This is corroborated by the fact that in the HRGMF model, for $K_M=0$, the result is almost same as ideal HRG model. So the impact of repulsion in the mesonic sector seems to be more compared to that in the baryonic sector. At finite $\mu_B$ (Fig.\ref{kt}(b)), while the overall behaviour of $\kappa_T$ as a function of $T$ is similar to the $\mu_B=0$ case,  the magnitude of $\kappa_T$ is smaller than that at $\mu_B=0$ case. This behaviour is the reflection of that fact that the scaled pressure ($P/T^4$) is higher at finite $\mu_B$ as compared to that at $\mu_B=0$. Fig.~\ref{kt}(c) shows $\kappa_T$ as a function of temperature for various values of $\mu_B$ for $K_M=0$. It is seen that $\kappa_T$ decreases with the increase of $\mu_B$. The difference of $\kappa_T$ between two values of $\mu_B$ is larger at lower temperatures as compared to that at higher temperatures. The reason is that, at low temperatures, the number density of the particles is low, and hence, the hadronic matter is more compressible.

The (adiabatic) speed of sound is defined as 
\begin{equation}
C_s^2=\bigg(\frac{\partial P}{\partial \epsilon}\bigg)_{s/n}
\end{equation}

Physical waves propagate at constant $s/n$. In order to discuss the implications of the results obtained in the context of HIC, one needs to find explicit expression for adiabatic speed of sound which is also valid at finite $\mu_B$. It can be easily shown that at constant $s/n$ the speed of sound is~\cite{Albright:2015fpa}

\begin{equation}
C_s^2(T,\mu_B)=\frac{v_n^2 Ts+v_s^2\mu_Bn_B}{\epsilon+P}
\end{equation}

where,

\begin{equation}
v_s^2(T,\mu_B)=\frac{n_B\chi_{TT}-s\chi_{\mu_B T}}{\mu_{B}(\chi_{TT}\chi_{\mu_B\mu_B}-\chi_{\mu_BT})}
\end{equation}
is the speed of sound at constant $s$ and,

\begin{equation}
v_n^2(T,\mu_B)=\frac{s\chi_{\mu_B\mu_B}-n_B\chi_{\mu_B T}}{T(\chi_{TT}\chi_{\mu_B\mu_B}-\chi_{\mu_BT})}
\end{equation}
is the speed of sound at constant $n$. The susceptibilities ($\chi$'s) are defined as, $\chi_{ab}=\partial^2 P/\partial a \partial b $, with $a,b=T,\mu_B$.

\begin{figure}[h]
	\begin{center}
	\hspace{-1.2cm}
	\begin{tabular}{c c}
		\includegraphics[scale=0.55]{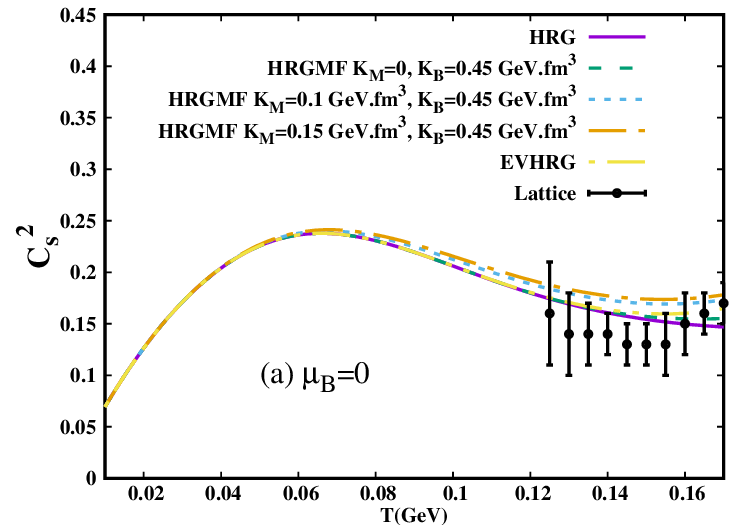}
		\hspace{-0.4cm}
		\includegraphics[scale=0.55]{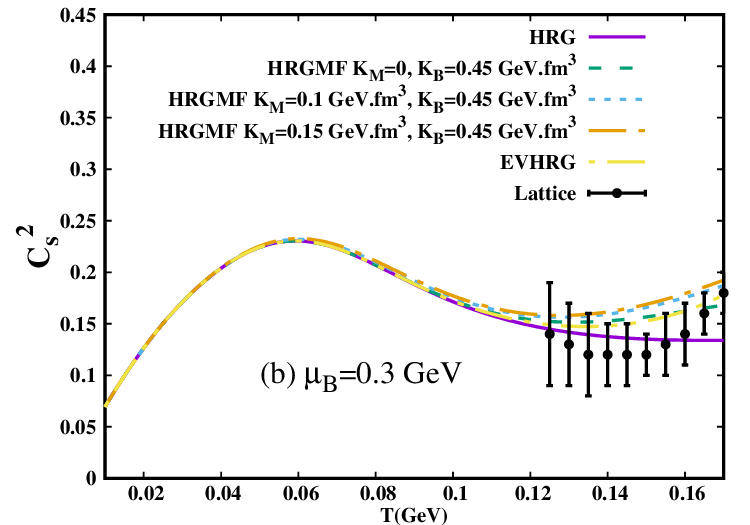} \\
		\hspace{-0.4cm}
		\includegraphics[scale=0.55]{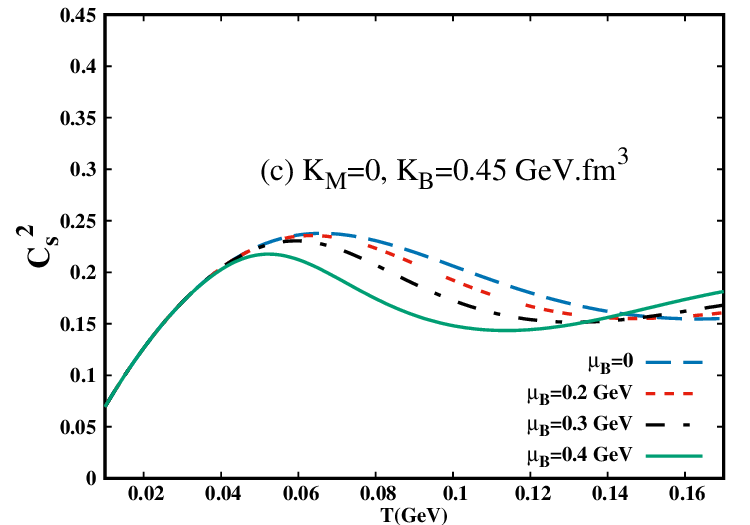}
	\end{tabular}
	
	\caption{Speed of sound squared ($C_s^2$) as a function of temperature in HRGMF model. The first two panels are for $\mu_B = 0$ and 0.3 GeV respectively. Third panel shows speed of sound squared ($C_s^2$) for $K_m=0$ at $\mu_B = 0,$ 0.2, 0.3 and 0.4 GeV.}
	\label{cs}
		\end{center}
\end{figure}

In Fig.~\ref{cs}  the variation of speed of sound squared ($C_s^2$)  with temperature has been plotted for some values of baryon chemical potential.  The $C_s^2$ depends on the degrees of freedom of the system, the equation of state and hence the interaction among the hadrons.  At very low temperatures $T\sim m_\pi$ the matter is dominated mostly by lighter pions. The energy pumped into the system is utilised to 
increase the temperature of the matter in this temperature regime. Hence the change in pressure is larger than the
change in energy density of the system resulting in monotonic rise in speed of sound.   This is true also when the 
repulsive interactions are considered at $\mu_B\neq 0$. As the temperature increases more massive hadrons and hadronic resonances start
to populate the system.  The energy pumped into the system is utilised to create heavier hadrons in this regime. As a result, the increase
in pressure lags behind the corresponding increase in the energy density.  Thus, the rapid rise of $C_s^2$ slows down and 
shows a maximum around $T\sim 0.08$ GeV.  After this temperature, it decreases slowly and saturates at a constant value around $T\sim 
0.16$ GeV.  The effect of repulsive interaction becomes visible after the system achieves an energy density sufficient to 
create baryons.  Again,  because in presence of repulsive interactions more energy is required to create a baryon,  increase in 
energy density lags behind increase in pressure.  This can be seen in Fig.\ref{cs} where one notes that the magnitude of $C_s^2$
is greater in HRGMF and EVHRG models as compared to HRG model.  Also, in presence of repulsive interactions, $C_s^2$
increases with temperature. An interesting observation with regard to the minimum in $C_s^2$ can be noted from Fig.(\ref{cs}). The minimum of the speed of sound, in the presence of repulsive interactions, is shifted to lower temperature as compared to ideal gas case. The results are compared with those of lattice QCD~\cite{Borsanyi:2012cr}. The results with repulsive interaction at temperatures higher than $0.15$ \ GeV approaches the lattice results. For all the values of $\mu_B$ there is an excellent fit with the lattice results. Furthermore, as the temperature increases, stronger repulsion seems to be preferable.  Fig.~\ref{cs}(c) shows speed of sound at $K_M=0$ for various values of $\mu_B$. There is almost no difference in $C_s^2$ among various $\mu_B$ cases at low temperatures, since, number of baryons is smaller at low temperatures. At mid-range temperatures, $C_s^2$ decreases as $\mu_B$ is increased. This is because the increase in pressure lags behind the increase in energy density  more for higher $\mu_B$. The trend becomes opposite at high temperatures when the increase in energy density lags behind the increase in pressure again.
\begin{figure}[h]
	\hspace{-1.2cm}
	\begin{tabular}{c c c}
		\includegraphics[scale=0.5]{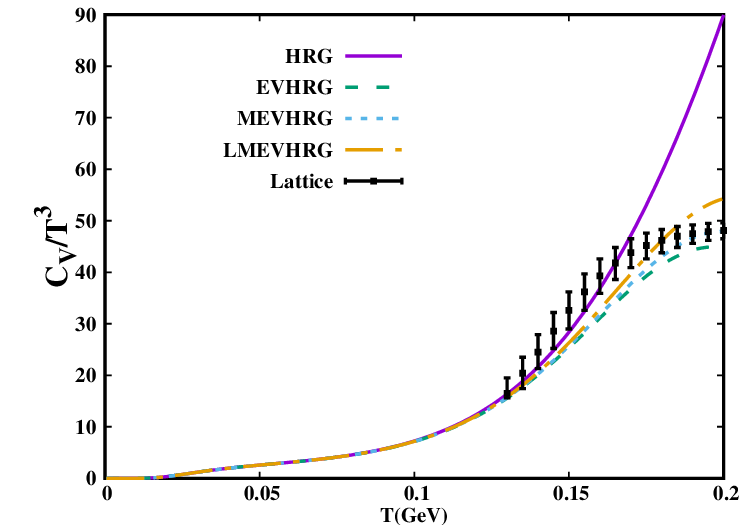}  
		\hspace{-0.4cm}
		\includegraphics[scale=0.5]{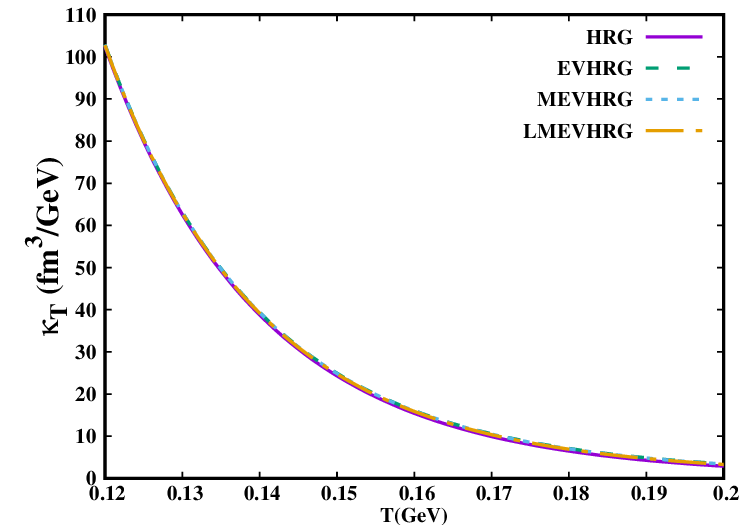} \\
		\hspace{-0.4cm}
		\includegraphics[scale=0.5]{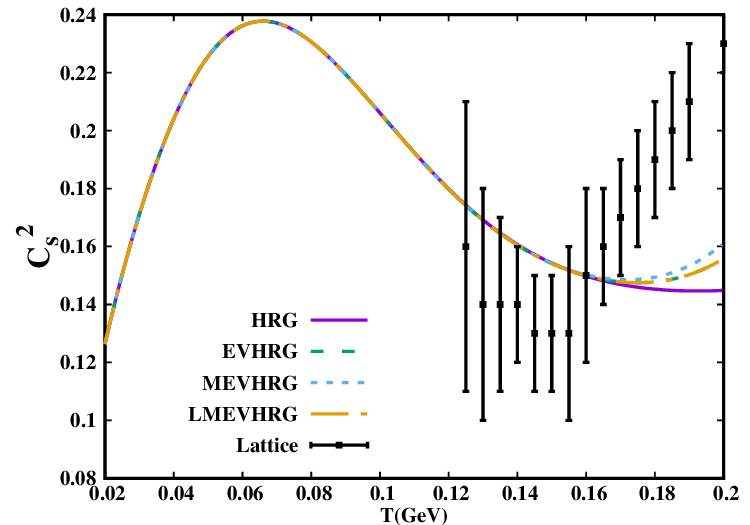}
		
	\end{tabular}
	\caption{Scaled specific heat $(C_V/T^3)$, isothermal compressibility $(\kappa_T)$ and speed of sound squared $(c_S^2)$ as a function of temperature in HRG, EVHRG, MEVHRG and LMEVHRG models. Here, $R_b=0.35 fm, R_\pi=0.2 fm, R_m=0.3 fm$.}
	\label{LMEVHRG_thermo}
\end{figure}

For the sake of completeness, the scaled specific heat $(C_V/T^3)$, isothermal compressibility $(\kappa_T)$ and speed of sound squared $(C_S^2)$, as functions of temperature, have been studied in MEVHRG and LMEVHRG models. The results have been compared with HRG and EVHRG models and are plotted in Fig.~\ref{LMEVHRG_thermo}. It can be seen that among all these three quantities, $(C_V/T^3)$ show the most distinct separation among the results of the four variants of HRG model. For $(\kappa_T)$, the difference among the results of all the four variants are very small. There is no significant difference in speed of sound at low and medium temperatures. This difference is prominent only at high temperatures,
\section{Connection with heavy-ion collisions}
Let us now discuss the effect of repulsive interactions on specific heat, isothermal compressibility and speed of sound in the light of heavy-ion collision experiments.  Connection of any physical quantity with the heavy-ion collision experiments can be made by finding the beam energy ($\sqrt{s}$) dependence of the temperature and chemical potential. This is extracted from a statistical thermal model description of the particle yield at various $\sqrt{s}$~\cite{Cleymans:2005xv}. The freeze-out curve is parametrised by 

\begin{equation}
T(\mu)=a-b\mu^2-c\mu^4;\hspace{0.1 cm} \mu=d/(1+e\sqrt{s})
\label{FO}
\end{equation}	
where $a=0.140$ GeV, $b=139\pm 0.016\:\text{GeV}^{-1}$, $c=0.053\pm0.021\:\text{GeV}^{-3}$, $d=1.308\pm 0.028$ GeV and $e=0.273 \pm 0.008\:\text{GeV}^{-1}$. It is to be noted that this parametrization is based on ideal HRG model.

The event-by-event  temperature fluctuation is controlled by  heat capacity for a system in equilibrium
\begin{equation}
P(T)\sim \text{exp}\bigg (-\frac{\tilde{C}}{2}\frac{(\Delta T)^2}{\langle T \rangle^2}\bigg)
\end{equation}
where $\Delta T=T-\langle T \rangle$ and $\langle T \rangle$ is the mean temperature. This yields the relation between  heat capacity and temperature fluctuation~\cite{LL}  as 
\begin{equation}
\frac{1}{\tilde{C}_{V,N}}=\frac{\langle T^2 \rangle - \langle T \rangle^2}{\langle T \rangle^2}
\label{sh}
\end{equation}
In the above equation, the heat capacity is calculated at constant volume and number of particles. On the other hand, here the chemical potential, has been taken as a parameter in the calculations, instead of number of particles. However, it is straightforward to deduce a relation connecting the two specific heats. The heat capacity at constant volume and number of
particles is given by
\begin{equation}
\tilde{C}_{V,N}=T\Big (\frac{\partial S}{\partial T}\Big)_{N,V}
\end{equation}
while the heat capacity at constant volume and chemical potential is
\begin{equation}
\tilde{C}_{V,\mu}=T\Big (\frac{\partial S}{\partial T}\Big)_{V,\mu}
\end{equation}
If one takes the entropy as a function of temperature $T$ and chemical potential $\mu$ then, one can write at constant volume $V$,
\begin{equation}
dS|_{\mu}=\Big(\frac{\partial S}{\partial T}\Big)_{\mu,V}dT+\Big(\frac{\partial S}{\partial {\mu}}\Big)_{T,V}d\mu
\label{ds}
\end{equation}
Writing the differential $d\mu$ as a function of $T$ and $N$,
\begin{equation}
d\mu|_V=\Big(\frac{\partial \mu}{\partial T}\Big)_{N,V}dT+\Big(\frac{\partial \mu}{\partial N}\Big)_{T,V}dN
\end{equation}
For constant $N$, the above equation gives
\begin{equation}
d\mu|_V=\Big(\frac{\partial \mu}{\partial T}\Big)_{N,V}dT
\label{dmu}
\end{equation}
Putting equation~(\ref{dmu}) into~(\ref{ds}) one gets,
\begin{equation}
\tilde{C}_{V,N}=\tilde{C}_{V,\mu}+T\Big(\frac{\partial S}{\partial {\mu}}\Big)_{T,V}\Big(\frac{\partial \mu}{\partial T}\Big)_{N,V}
\label{cvnvmu}
\end{equation}
One can write
\begin{equation}
dN|_V=\Big(\frac{\partial N}{\partial T}\Big)_{\mu,V}dT+\Big(\frac{\partial N}{\partial \mu}\Big)_{T,V}d\mu
\end{equation}
Putting $dN|_V=0$, one has
\begin{equation}
\Big(\frac{\partial \mu}{\partial T}\Big)_{N,V}=-\Big(\frac{\partial N}{\partial T}\Big)_{\mu,V}\Big/\Big(\frac{\partial N}{\partial \mu}\Big)_{T,V}
\end{equation}
From equation~(\ref{cvnvmu}), one can get
the relationship between $C_{V,N}$ and $C_{V,\mu}$:
\begin{equation}
C_{V,N}=C_{V,\mu}-T\frac{{\big(\frac{\partial s}{\partial \mu}}\big)_{T,V}{\big(\frac{\partial n}{\partial T}\big)}_{\mu,V}}{{\big(\frac{\partial n}{\partial \mu}\big)}_{T,V}}
\end{equation}

The equation~(\ref{sh}) is valid in the presence of a large energy reservoir. This role may be played by the longitudinal degrees of freedom of the beam in heavy-ion collision~\cite{Stodolsky:1995ds}. Thus, it is possible to extract the specific heat using temperature fluctuations through  $p_T$ spectra measured on the basis of  event-by-event analysis of experimental data. While such analysis could be done, the discussion, here, has been restricted only to the beam-energy dependence of $C_{V,\mu}$ extracted using  eq. (\ref{FO}). This will give information about the magnitude of specific heat at the freeze-out.
\begin{figure}[h]
	\begin{center}
		\hspace{-1.2cm}
		\begin{tabular}{c c}
			\includegraphics[scale=0.55]{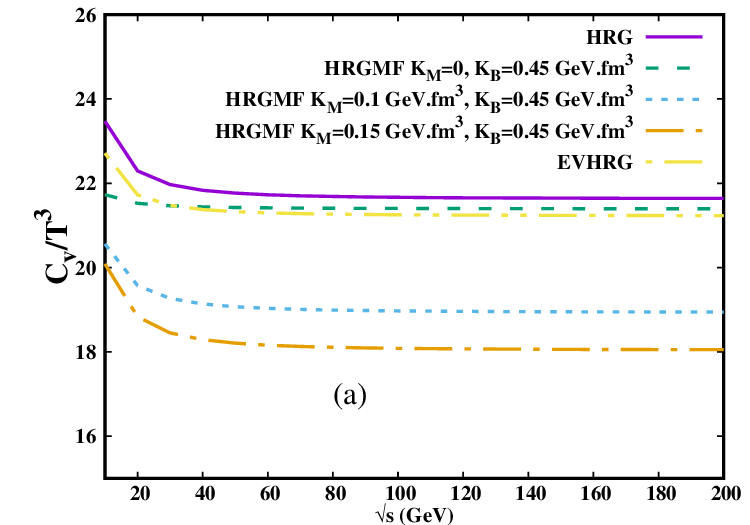}
			\hspace{-0.4cm}
			\includegraphics[scale=0.55]{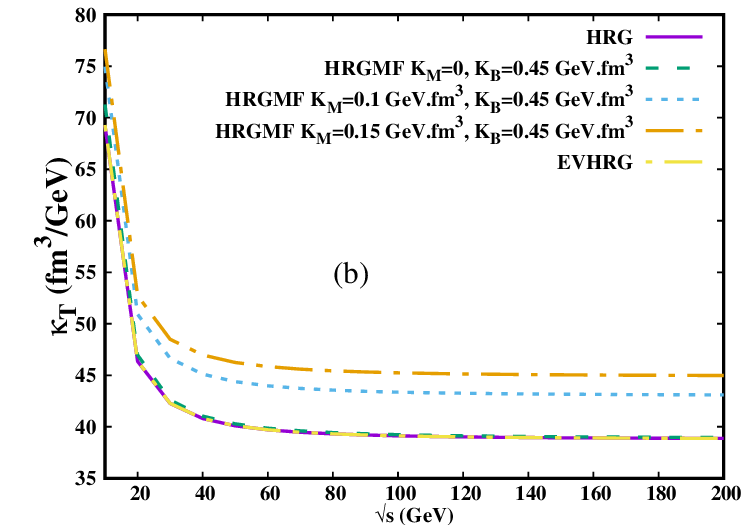} \\
			\hspace{-0.4cm}
			\includegraphics[scale=0.55]{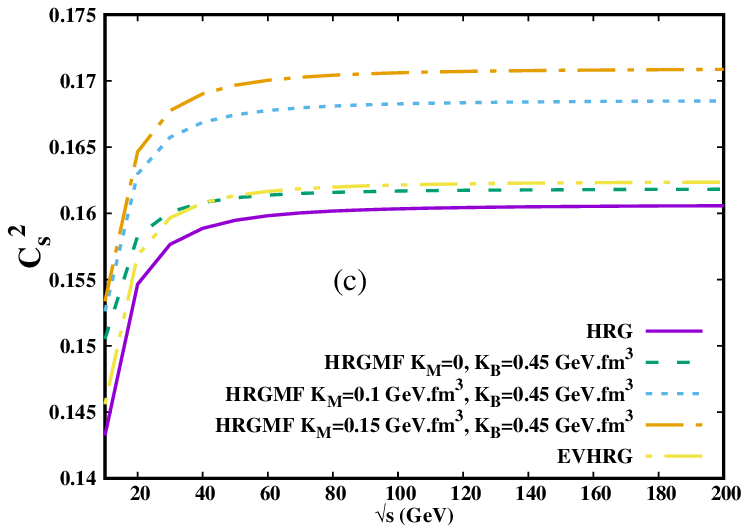}
		\end{tabular}
		
		\caption{Collision energy dependence of $C_V$, $\kappa_T$ and $C_s^2$ in HRGMF model. }
		\label{sqrts}
	\end{center}
\end{figure}

In the event-by-event analysis, temperature can be extracted from transverse momentum ($p_T$) spectra. From the plot of $C_{V,\mu}$ (Fig.~\ref{sqrts}(a)) it can be noted that the specific heat drops a bit sharply at a low value of $\sqrt{s}$ and then it almost saturates. With the increase of the centre of mass energy of heavy-ion collision, the colliding nuclei become more transparent and they create smaller particle number density. Hence the energy pumped into the system goes to increase the temperature and specific heat decreases with increasing centre of mass energy. It can also be seen that the specific heat is suppressed due to repulsive interactions. The temperature distribution of event-by-event collisions become broadly peaked in presence of repulsive interactions. However, repulsive interaction between hadrons does not change the overall qualitative behaviour of $C_{V,\mu}$. A similar finding was observed in Ref.~\cite{Basu:2016ibk} in which, the energy dependence of specific heat  of hadronic matter at freeze-out in Au+Au and Cu+Cu collisions at the  Relativistic Heavy Ion Collider (RHIC) energies within the  ambit of hadron resonance gas model was studied. These authors found that $C_{V,\mu}$ drops sharply from low collision energy till $\sqrt{s}=62.4$ GeV after which, it saturates. Here one should note that the freeze-out  parameters are different in their work and as a result the numerical values of $C_{V,\mu}$ are different.

In the context of HIC experiments, isothermal compressibility ($\kappa_T$) is related to the fluctuation in the particle multiplicities, temperature and volume of the system formed. In the grand canonical ensemble framework the particle multiplicity fluctuation $\omega$ is given by
\begin{eqnarray}
\omega=\frac{\langle N^2 \rangle- \langle N \rangle^2}{\langle N \rangle} 
=\frac{k_B T\langle N \rangle}{V}\kappa_T
\end{eqnarray}
In the above equation, temperature and volume can be extracted from the mean hadron yield and $\kappa_T$ can be extracted through the event-by-event multiplicity fluctuation measurements. 
Here, the beam-energy dependence of $\kappa_T$ has been studied, which, would provide an information about its magnitude at freeze-out.

In Fig. \ref{sqrts}(b) it can be seen that the compressibility sharply decreases till $\sqrt{s}=25$ GeV, and then it almost saturates. This suggests that the matter produced in low energy collision experiments is more compressible. The effect of repulsive interactions on the magnitude of $\kappa_T$ is smaller at higher collision energies as compared to at lower ones. The magnitude of $\kappa_T$ for low collision energies is not very different from ideal HRG results. At high collision energies its magnitude is significantly higher as compared to ideal HRG estimations. However, the repulsive interaction does not change the overall behaviour of $\kappa_T$. In 
Ref.~\cite{Khuntia:2018non,Mukherjee:2017elm} authors carried out similar studies and, qualitatively, the findings are same as obtained here.

In the relativistic hydrodynamic simulation of the matter produced in HIC experiments, the speed of sound plays a very crucial role.  The value of $C_s^2$ depends on the degrees of freedom and the interactions. Study of $C_s^2$ carried out in this work has vindicated this fact. It has been seen how the behaviour of $C_s^2$ changes as new degrees of freedom (massive resonances and baryons) enter the system.  The effect of the inclusion of repulsive interactions has also been shown. In the hydrodynamical description of matter produced in HIC,  $C_s^2$ sets the expansion time scale as  $\tau_{\text{exp}}^{-1}\sim\frac{1}{\epsilon}\frac{d\epsilon}{d\tau}=\frac{1+c_s^2}{\tau}$~\cite{Mohanty:2003va}. The thermal equilibrium is maintained only if $\tau_{\text{exp}}>\tau_{\text{coll}}$, where $\tau_{\text{coll}}$ is the collision time. 
The $C_S^2$ rises very rapidly 
at low $\sqrt{s}$ and then it saturates at higher collision energies. Repulsive interaction has quite a strong effect. The magnitude of $C_S^2$ is larger as compared to ideal HRG model and it keeps on increasing as the effect of repulsion increases. As $C_S^2$ increases with the introduction of repulsive interaction, the expansion time scale of the system decreases. Thus it becomes more difficult to maintain the thermal equilibrium.  In Ref.~\cite{Mohanty:2003va} authors discussed the sensitivity of the hadronic spectra on  EoS obtained within the ambit of hadron resonance gas model. They found that in fixing the value of freeze-out temperature at $T\sim 0.12$ GeV from the transverse momentum ($p_T$) spectra, the value $C_s^2=1/5$ gives good description of the data for Landau's hydrodynamical model. This value is smaller than ideal gas EoS limit $1/3$. Thus the presence of interactions lowers the value of speed of sound and hence for such system the maintenance of thermal equilibrium is easier. However, in the present work it has been found that that the  speed of sound is higher in magnitude as compared to ideal HRG model when repulsive interactions are taken into account. This has very non-trivial effect on thermalisation.  The repulsive interaction would not be able to maintain thermal equilibrium for a long time in the expanding plasma. It would be interesting to study the hydrodynamical evolution with EoS which includes the effect of repulsive interactions and its effect on the thermalisation.
	\newpage
\rhead{Fluctuations of Conserved Charges}
\chapter{\label{chap:fluc}Fluctuations of Conserved Charges}
Fluctuations are measures of how a function varies with respect to one of the variables it depends on with respect to the average value of the function. On the other hand, correlations are measures of how a function of two or more variables varies with combined changes in those independent variables. 
Fluctuations of various conserved charges provide important information about the phase transition from QGP to hadronic phase. It has significant importance in the sense that fluctuations in the number of particle species is accessible in heavy-ion collision experiments. On the other hand, the thermodynamic quantities like pressure, energy density etc. are not directly measurable in experiments. Fluctuations of conserved charges are directly related to various thermodynamic quantities. There are several types of fluctuations depending upon their origin. Quantum fluctuations in a observable arise when the observable under consideration does not commute with the Hamiltonian of the system. Dynamical fluctuations arise due to the dynamics of the system. This thesis is restricted to the study of dynamical fluctuations.  There are also density fluctuations that are controlled by the compressibility.

Study of fluctuation is important in characterization of phase transition. Fluctuation of the order parameter, in case of a second order phase transition, diverges with a critical exponent specific to the universality class of the phase transition~\cite{landau}. Fluctuations of conserved charges provide very useful information about the phase
transition between hadronic and quark gluon plasma phase. Divergent fluctuations can give an indication about the existence of the critical end point
(CEP). One can calculate
charge susceptibilities which are related to fluctuations via
fluctuation-dissipation theorem. If net baryon number of the system is
small then quark-hadron phase transition is continuous and
fluctuations are expected not to show singular behaviour. On the other
hand Lattice QCD calculation shows that at small chemical potentials,
susceptibilities increase rapidly near the crossover region. A measure of the intrinsic statistical fluctuations in a system close to thermal equilibrium is provided by the corresponding susceptibilities. Susceptibilities of higher order are expected to be more sensitive near phase
transition. Fluctuation of conserved charges has been studied
in Refs.~\cite{fluctuation1, fluctuation2, fluctuation4,volume2,fuku2008,roessner1,friman,roessner2,fuku3,schaefer1,
	schaefer2,schaefer3,schaefer4,LQCD12,Borsanyi:2018grb,bazavov:2020}.

Fluctuations are related to the cumulants of partition function for a system in thermal equilibrium. These cumulants or susceptibilities are expressible in terms of integrals of equal-time correlation functions, which represent the static responses of the system. Furthermore, finite size scaling arguments on the system size dependence of fluctuations can be used to distinguish between a crossover or first order or second order phase transition.

The grand canonical partition function for a system in thermal equilibrium is given by
\begin{equation}
\mathcal{Z}=Tr\Big [\exp\Big (-\frac{H-\sum_i\mu_iQ_i}{T}\Big )\Big ]
\end{equation}
where $H$ is the Hamiltonian of the system, $Q_i$ are the conserved charges and $\mu_i$ are the corresponding chemical potentials. The mean and (co)-variances are expressible in terms of derivatives of the partition function with respect to appropriate chemical potentials:
\begin{equation}
\langle Q_i\rangle=T\frac{\partial}{\partial\mu_i}\log(\mathcal{Z})
\end{equation}
\begin{equation}
\langle \delta Q_i\delta Q_j\rangle=T^2\frac{\partial ^2}{\partial\mu_i\partial\mu_j}\log(\mathcal{Z})=VT\chi_{i,j}
\end{equation}
where $\delta Q_i=Q_i-\langle Q_i\rangle$. The susceptibilities $\chi_{i,j}$ are
\begin{equation}
\chi_{i,j}=\frac{T}{V}\frac{\partial ^2}{\partial\mu_i\partial\mu_j}\log(\mathcal{Z})
\end{equation}
These susceptibilities are a measure of (co)-variances. The diagonal susceptibilities $(\chi_{i,i})$ are a measure of fluctuations of the system and the off-diagonal susceptibilities $(\chi_{i,j}$ with $i\ne j)$ are a measure of correlation between the conserved charge $Q_i$ and $Q_j$.

Susceptibilities of conserved charges can be calculated by taking Taylor series expansion of scaled pressure $(P/T^4)$ with respect to baryon, electric and strange chemical potentials~\cite{Allton:2002zi}:
\begin{equation}
P = P_0 + T^4 \sum_{i,j,k} \frac{1}{i!j!k!} \chi^{i,j,k}_{B,Q,S} \left(\frac{\mu_B}{T}\right)^i
\left(\frac{\mu_Q}{T}\right)^j
\left(\frac{\mu_S}{T}\right)^k\
\label{taylor}
\end{equation}
where $P_0$ is the pressure at zero chemical potential, $\chi^{i,j,k}_{B,Q,S}$ are the conserved charge susceptibilities given by:
\begin{equation}
\chi_{BQS}^{ijk} = \frac{\partial^i \partial^j \partial^k P(T,\mu_B,\mu_Q,\mu_S)/T^4}{\partial \left({\mu_B/T}\right)^i \partial \left({\mu_Q/T}\right)^j \partial \left({\mu_S/T}\right)^k}\
\label{chin}
\end{equation}
Charge susceptibilities are expressible in terms of integrals of density fluctuations:
\begin{equation}
\chi_{ab}^{11}=\frac{1}{VT}\int dx^3 dy^3 <\delta \rho_a(x)\delta \rho_b(y)>
\label{susceptibility}
\end{equation}
where $\delta \rho_a(x)=\rho_a(x)-\bar{\rho_a}$ is the density fluctuation of charge $a$ at location x, $\bar{\rho_a}$ denotes the spatially averaged density of charge $a$. The diagonal elements of $\chi_{ab}^{11}$ matrix are a measure of fluctuations and the off-diagonal elements are a measure of correlations. 

In order to calculate the susceptibilities, first, one has to choose the point $(T, \mu_B^0, \mu_Q^0, \mu_S^0)$ at which the susceptibility is to be calculated. One then calculates the pressure of the system at that chosen temperature $T$ at several points in the chemical potential plane around $(\mu_B^0,\mu_Q^0,\mu_S^0)$ with small regular spacing in the chemical potentials. Then this data is fitted to the Taylor series expansion~(\ref{taylor}) with the help of gnu-plot program. The magnitude of the error in this fit has been kept under limit by varying the ranges of the chemical potentials and simultaneously keeping the values of the least squares below $10^{-10}$. Thus the susceptibility at the point $(T, \mu_B^0, \mu_Q^0, \mu_S^0)$ is obtained. This exercise is repeated for each value of $T$. In case of the susceptibility at $(\mu_B^0,\mu_Q^0,\mu_S^0)$ = (0,0,0), the Taylor series expansion used has only even terms, as in that case, the odd terms vanish due to CP symmetry.
\section{Results from different variants of EVHRG model}
\begin{figure}[H]
		\begin{center}
	\hspace{-1.2cm}
		\begin{tabular}{c c}
			\includegraphics[scale=0.55]{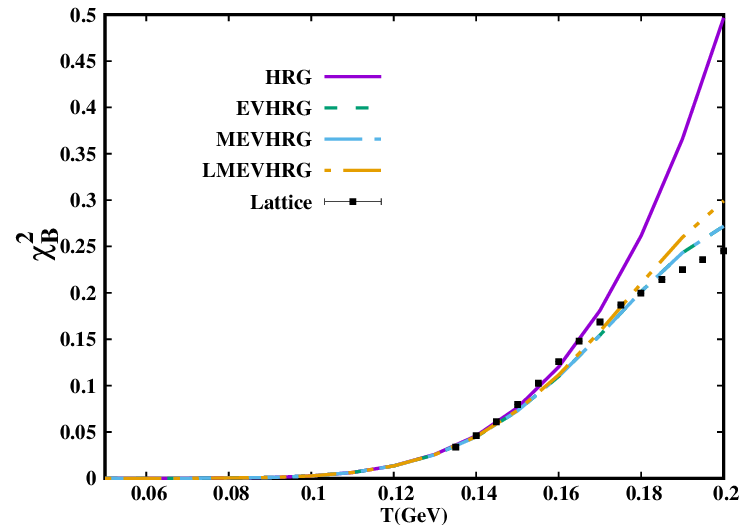}&
			\hspace{-0.4cm}
			\includegraphics[scale=0.55]{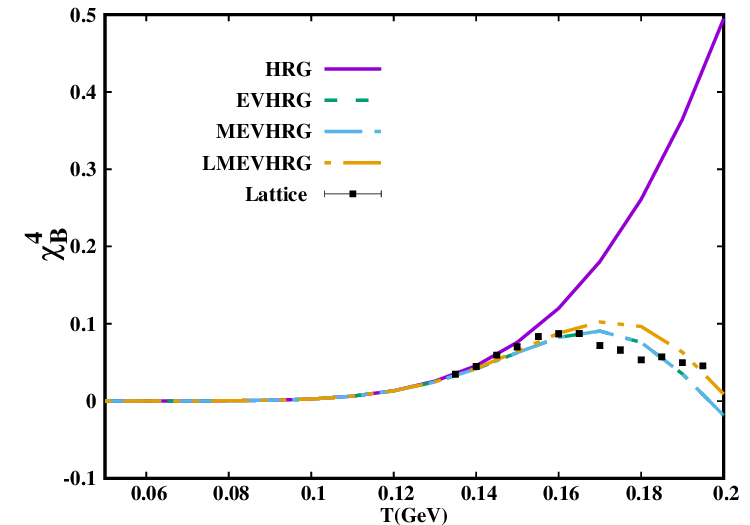}
		\end{tabular}
		
		\caption{Baryon number susceptibilities of different orders in HRG, EVHRG, MEVHRG and LMEVHRG models.}
		\label{fig 2}
	\end{center}
\end{figure}

\begin{figure}[h]
	\begin{center}
		\hspace{-1.2cm}
		\begin{tabular}{c c}
			\includegraphics[scale=0.55]{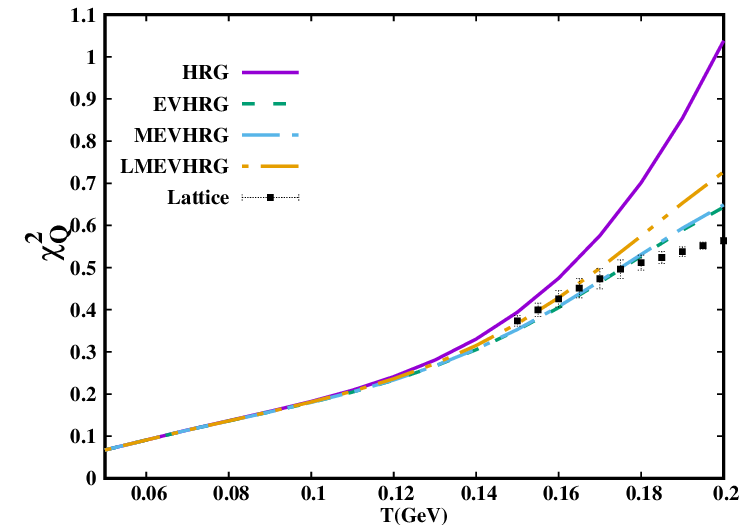}&
			\hspace{-0.4cm}
			\includegraphics[scale=0.55]{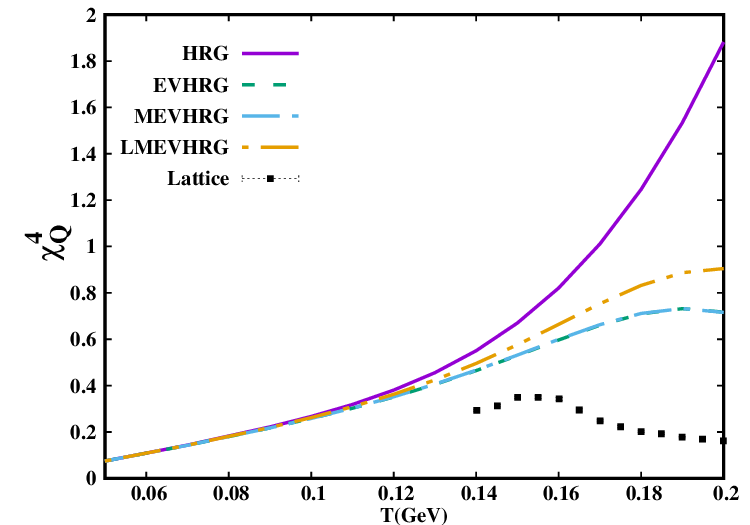}
		\end{tabular}
		
		\caption{Electric charge susceptibilities of different orders in HRG, EVHRG, MEVHRG and LMEVHRG models. }
		\label{fig 3}
	\end{center}
\end{figure}

\begin{figure}[h]
	\begin{center}
		\hspace{-1.2cm}
		\begin{tabular}{c c}
			\includegraphics[scale=0.55]{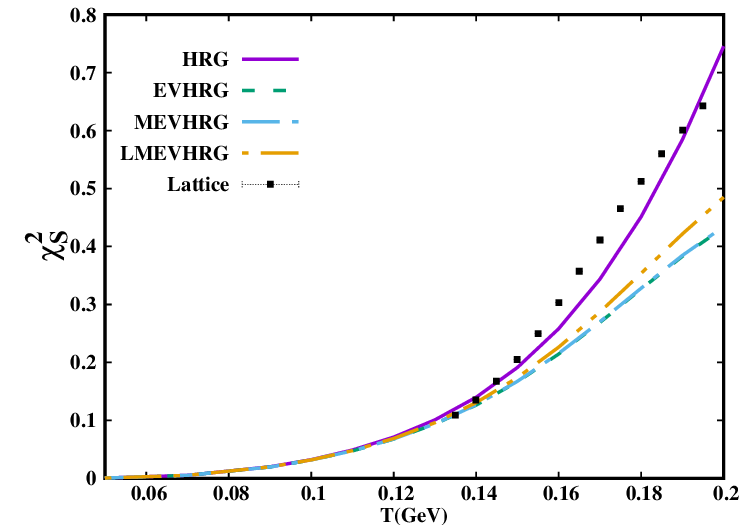}&
			\hspace{-0.4cm}
			\includegraphics[scale=0.55]{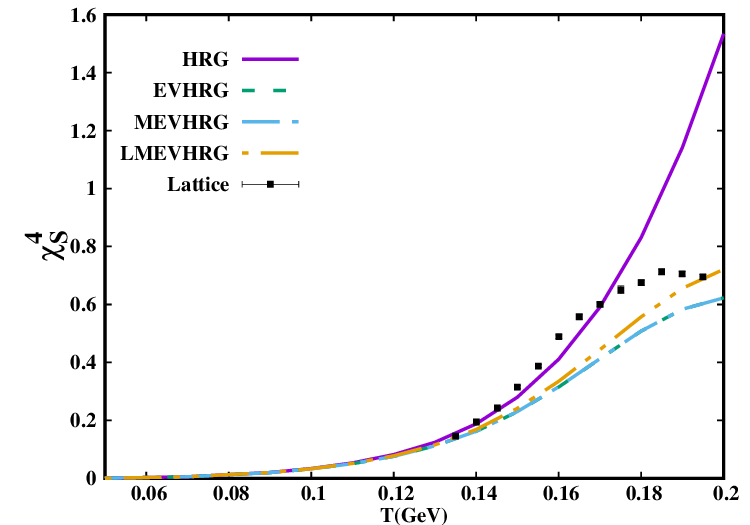}
		\end{tabular}
		
		\caption{Strangeness susceptibilities of different orders in HRG, EVHRG, MEVHRG and LMEVHRG models.}
		\label{fig 4}
	\end{center}
\end{figure}

In Figs.~\ref{fig 2}-\ref{fig 4}  the second and fourth order susceptibilities of the conserved charges at zero chemical potentials ($\mu_B=\mu_Q=\mu_S = 0$) within HRG, EVHRG, MEVHRG and LMEVHRG models have been shown.  The excluded volume radii chosen are: baryon radius $R_b = 0.35 fm$, pion radius $R_\pi = 0.2 fm$ and radii of other mesons $R_m=0.3 fm$. The results have been compared with those obtained  from LQCD. From the plots it can be seen that, at low temperatures, susceptibilities for all the four models are almost same. As temperature increases, deviations among various models become prominent. For ideal HRG model with no repulsive interaction, susceptibilities of all the conserved charges and of all orders increase rapidly with temperature. Susceptibilities  obtained in HRG model are greater than those obtained in EVHRG model (where hardcore repulsion is included). Susceptibilities for MEVHRG model are little bit higher compared to the EVHRG case for this choice of hadronic radii. Whether the absolute values of susceptibilities in EVHRG model will be larger than those calculated in MEVHRG model will depend on the choice of radii of hadrons and also on the type of conserved charge. If the dominating hadrons carrying a particular conserved charge are assigned larger excluded volume than that of dominated hadrons then the susceptibility of that conserved charge in MEVHRG model will be lower than EVHRG case. The reason is that, the last term in  eqn.~(\ref{mu_k}) will be greater in MEVHRG model than its counterpart in EVHRG model. For the particular choice of hadronic radii mentioned above, all the susceptibilities in MEVHRG model are larger than corresponding EVHRG case. It is seen that $\chi^n_Q > \chi^n_S > \chi^n_B$. This is expected because the dominating hadrons at a particular temperature govern the susceptibilities. Since pions (lightest electrically charged hadrons) are lighter than kaons (lightest strange hadrons) which are lighter than protons (lightest baryons), hence the result.

Fig. \ref{fig 2} shows second and fourth order baryon number susceptibilities at zero chemical potentials ($\mu_B=\mu_Q=\mu_S=0$). The results have been compared with LQCD results given in Refs.~\cite{LQCD12,Borsanyi:2018grb}. The left panel shows $\chi^2_B$ as a function of temperature. 
It is seen that the results of all the four models for 
$\chi^2_B$ are almost same up to $T=0.14$ \ GeV. The EVHRG, MEVHRG and LMEVHRG results don't show significant difference up to somewhat higher temperature. Up to $T=0.18$ \ GeV, for all the three models with repulsive interaction, agreement with LQCD is quite satisfactory. This clearly shows the relevance of repulsive interaction. As expected, $\chi^2_B$ does not have any dependence on the excluded volume of mesons since there is no coupling among mesons and baryons in this model and mesons don't carry baryon number. Since the radii of all baryons have been taken to be the same, there is no quantitative difference between EVHRG and MEVHRG model results in this case. As baryons need very high temperature to show significant Lorentz contraction, difference between LMEVHRG model and MEVHRG model is significant at high temperatures only. The results for $\chi^4_B$ of all the four models are shown in the right panel. The values of $\chi^4_B$ of all the models coincide up to $T=0.12$\ GeV. The results from different variants of interacting HRG model do not show significant difference up to somewhat higher temperature. For EVHRG, MEVHRG and LMEVHRG model, at low temperatures, $\chi^4_B$ increases gradually when temperature is increased. Then it reaches a maximum and then starts to decrease. The results from the interacting models  have good agreement with the LQCD data. Since  only one kind of baryonic radius has been considered, there is no difference in EVHRG and MEVHRG results in this case. Higher order susceptibilities of conserved charges are more sensitive to change in number of particles and hence the difference between the results of MEVHRG and LMEVHRG models for $\chi^4_B$ is larger than that for $\chi^2_B$ at high temperatures.

The left panel of Fig.~\ref{fig 3} shows second order electric charge susceptibility. It is seen that results of all the four models for $\chi^2_Q$ coincide up to $T=0.1$ \ GeV. Results for EVHRG and MEVHRG models deviate from each other at medium range of temperature in contrast to $\chi^2_B$. This is because the dominant contributors, here, are mesons which have been assigned two different values of radii. Here the effect of Lorentz contraction is significantly large compared to that in $\chi^2_B$ as lighter mesons are the dominant contributors. The results for LMEVHRG model has the best agreement with LQCD data at medium range of temperatures but EVHRG and MEVHRG results show better agreement at somewhat higher temperatures. Once again 
it is seen that models with repulsive interactions perform better. The right panel shows fourth order electric charge susceptibility. It is seen that, all the four model results, for $\chi^4_Q$, are almost equal up to $T=0.1$\ GeV. The pion, which is the lightest electrically charged particle, is the most significant contributor to $\chi^4_Q$. The EVHRG and the MEVHRG plots differ negligibly from each other at all temperatures. The results for $\chi^4_Q$ shows a tendency of saturation at high temperatures. Like the baryon number susceptibilities, $\chi^4_Q$ shows larger difference between MEVHRG and LMEVHRG model results than that of $\chi^2_Q$.

The left panel of Fig.~\ref{fig 4}  shows second order strangeness susceptibility. It is seen 
that results of all the four models, for $\chi^2_S$,
have almost similar results up to $T=0.11$ \ GeV. One can see that $\chi^2_S$ deviates from LQCD data much more than that of
$\chi^2_B$ and $\chi^2_Q$. The lattice data overestimates even the pure HRG model results. Kaon is the most dominant contributor to
the strangeness susceptibility. Difference between MEVHRG and EVHRG results are insignificant in this case. Effect of Lorentz contraction here is smaller than that in $\chi^2_Q$ but stronger than that in $\chi^2_B$ as kaons are more massive than pions but lighter than nucleons. The right panel shows fourth order strangeness susceptibility. The results of HRG model show good agreement with the lattice data up to a temperature of $T=0.165$\ GeV. Then they deviate from each other. The interacting HRG results approach the lattice results at high temperatures.

Though HRG result for scaled pressure shows good agreement with LQCD data, baryon and electric charge susceptibilities significantly deviate from the LQCD data. Even for the fourth order strangeness susceptibility, the non-interacting HRG scenario at high temperatures deviates significantly from the LQCD data. On the other hand, the interacting models perform much better in reproducing the lattice results. This indicates that excluded volume correction is extremely relevant in describing the hot and dense strongly interacting matter. It should be mentioned that inclusion of Lorentz contraction gives a better fit with the lattice data in some cases. For $\chi^2_Q$, $\chi^2_B$ and $\chi^2_S$
the result with Lorentz contraction scenario has a maximum deviation of $10\%$ from the Lattice data and in some cases the deviation is less than $5\%$.
\section{Results from HRGMF model}
Let us now discuss some approximate  expressions for the pressure and the number densities 
and hence some properties of the susceptibilities which will be helpful in analysing the variation of susceptibilities with temperature and/or 
chemical potentials.
For non-interacting HRG, the pressure can be expanded in powers of  fugacity. Hence, 
the baryonic pressure, in HRGMF model, as given  in 
Eq.(\ref{pbbar}) can be written as, (with $\beta=T^{-1}$)
\begin{eqnarray}
\frac{{P}_{B\{\bar B\}}}{T^4}&=&\sum\limits_{a\in B\{\bar B\}}\frac{g_a}{2\pi^2}(\beta m_a)^2\sum\limits_{l=1}^{\infty}
(-1)^{l+1} \:l^{-2} 
\times K_2(\beta l m_a)z^{l} \nonumber\\
&+&\frac{K_BT^2}{2}\left(\frac{n_{B\{\bar B\}}}{T^3}\right)^2
\label{bessell}
\end{eqnarray}
In the above, the fugacity is defined as $z=\exp ({\beta\mu_{eff}})$ and $ K_2$ is the Bessel function. It can be easily 
shown that as long
as $\beta(m_a-\mu_{eff})\gtrsim 1$, the contribution to the pressure $P^{id}_{B\{\bar B\}}$
can be approximately given by the leading term i.e. $l=1$ in the summation which, in fact, corresponds to Boltzmann approximation. In this case,
the pressure from the baryons can be written as
\begin{eqnarray}
&\frac{P_{B\{\bar B\}}}{T^4}
=\sum_{a\in B}\frac{g_a}{2\pi^2}(\beta m_a)^2K_2(\beta m_a)\nonumber\\
&\times \exp(\beta\mu_{eff}^a)+\frac{K_BT^2}{2}\left(\frac{n_{B\{\bar B\}}}{T^3}\right)^2
\end{eqnarray}

In a similar way, the number density for baryons can be approximated as
\begin{eqnarray}
\frac{n_{B}}{T^3}=\sum_{a\in{B}}\frac{g_a}{2\pi^2}(\beta m_a)^2K_2(\beta m_a)e^{\beta\mu_{eff}^a}
\label{approxnb}
\end{eqnarray}

Thus, the gas pressure in presence of interaction in the Boltzmann approximation can be written as
\begin{eqnarray}
P_{B\{\bar B\}}=T n_{B\{\bar B\}}+\frac{K_B}{2} n_{B\{\bar B\}}^2
\end{eqnarray}



For temperatures below the
QCD transition temperatures such that $n_B$ $(n_{\bar B})$ are small, one
can expand the exponential  $\exp(\mu_{eff})\simeq \exp(c_i\mu_i)(1-\beta K_B n_B)$. The number densities given in Eq.(\ref{approxnb}) can be further approximated
as
$n_B=n_B^{id}/(1+n_B^{id})$. Here $n_B^{id}$ is the number density without any repulsive interaction (Eq.(\ref{approxnbid})) i.e.
\begin{eqnarray}
\frac{n_{B}^{id}}{T^3}=\sum_{a\in{B}}\frac{g_a}{2\pi^2}(\beta m_a)^2K_2(\beta m_a)\exp(\beta c_a^i\mu_i)
\label{approxnbid}
\end{eqnarray}

.
The contribution to pressure from baryons can then be approximated as
\begin{eqnarray}
P_B=Tn_B^{id}-\frac{K_B}{2}{(n_B^{id})^2}
\end{eqnarray}
A similar expression can be obtained for anti-baryon pressure. As may be noted, the effect of the density dependent repulsive 
interaction essentially lies in 
reducing the pressure at finite densities.

The total pressure from baryons and antibaryons can then be written as

\begin{eqnarray}
&&\frac{P_B+P_{\bar B}}{T^4}\simeq \sum_{a\in B}F_a(\beta m_a)\cosh(\beta c_a^i\mu_i)\nonumber\\
&-&\frac{K_BT^2}{2}\Big (\sum_aG_a(\beta m_a,\beta\mu_Q,\beta\mu_s)\Big )^2e^{2\beta\mu_B}\nonumber\\
&-&\frac{K_BT^2}{2}\Big (\sum_aG_a(\beta m_a,-\beta\mu_Q,-\beta\mu_s)\Big )^2e^{-2\beta\mu_B}.
\label{pbapprox}
\end{eqnarray}


Here, the function, independent of all the chemical potentials, $F_a(\beta m_a)$ has been defined as
\begin{eqnarray}
F_a(\beta m_a)=\frac{g_a}{\pi^2}(\beta m_a)^2 K_2(\beta m_a)
\label{Fa}
\end{eqnarray}
and the function, which is independent of the baryon chemical potential,  is defined as
\begin{eqnarray}
&&G_a(\beta m_a,\beta\mu_Q,\beta\mu_s)=\frac{g_a}{2\pi^2}(\beta m_a)^2 K_2(\beta m_a)\nonumber\\
&\times &\exp(Q_a\beta\mu_Q+S_a\beta\mu_s)
\label{Ga}
\end{eqnarray}

%

In a similar manner, in the Boltzmann approximation, the pressure for the mesons, and with $\beta K_M n_M\le 1$, can be written as

\begin{eqnarray}
\frac{P_M}{T^4}\simeq \sum_{a\in M}\frac{n_a^{id}}{T^3}-\frac{1}{2} (K_MT^2)\bigg(\frac{n_a^{id}}{T^3}\bigg)^2
\label{pmapprox}
\end{eqnarray}
with
\begin{eqnarray}
n_a^{id}=\frac{g_a}{2\pi^2}K_2(\beta m_a) \exp(\beta \mu_a)
\end{eqnarray}

The total pressure $P=P_B+P_{\bar B}+P_M$ is thus given approximately by the sum of Eqs.(\ref{pbapprox}) and (\ref{pmapprox}).
Some interesting consequences follow from these approximate expressions for pressure. Firstly, the baryonic susceptibilities of odd order will be  small for small baryon chemical potential
and will vanish for zero baryonic chemical potential. Further, the even order baryonic susceptibilities, 
e.g. $\chi^2_B$ and $\chi^4_B$ will not be identical because of the repulsive interaction term. Indeed, 
the difference between these two is given approximately as
\begin{eqnarray}
\chi^2_B-\chi^4_B\simeq 
12\frac{K_BT^2}{2} (\beta^3n_B^{id})^2
\end{eqnarray}
Thus, while  the difference between the higher (even) order and lower (even) order baryonic 
susceptibilities vanishes for
ideal HRG, it does not vanish when there is mean field repulsive terms. In the following, the results for the susceptibilities with actual Fermi-Dirac or Bose-Einstein statistics
have been presented by solving the self consistent equations for the number densities. However, the above assertions made with the approximate
expressions for the pressure remain valid.

The only parameters in the HRGMF model are $K_M$ and $K_B$. Three different representative values for meson mean field parameter,
$viz.$, $K_M=0, 0.1$ and $0.15$ GeV$\cdot$fm$^{3}$ have been chosen, while baryon mean-field parameter has been fixed to
$K_B=0.45$ GeV$\cdot$fm$^{3}$~\cite{Kadam:2019peo,Huovinen:2017ogf}.
\begin{figure}[h]
	\begin{center}
		\begin{tabular}{c c}
			\includegraphics[scale=0.55]{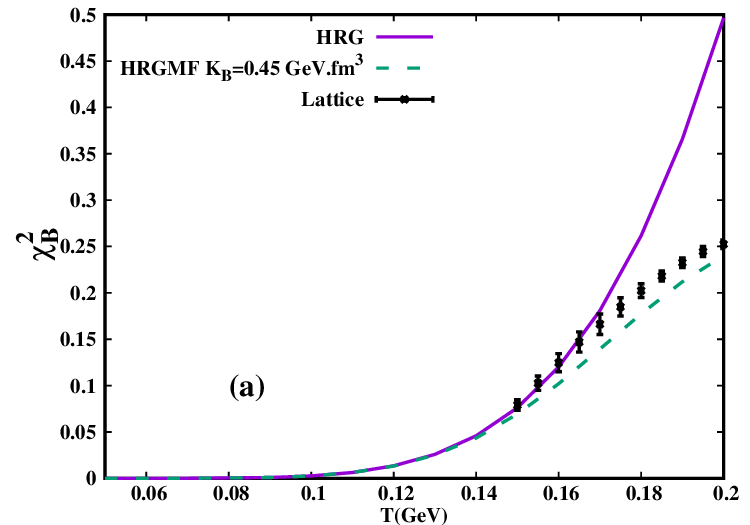}
			\includegraphics[scale=0.55]{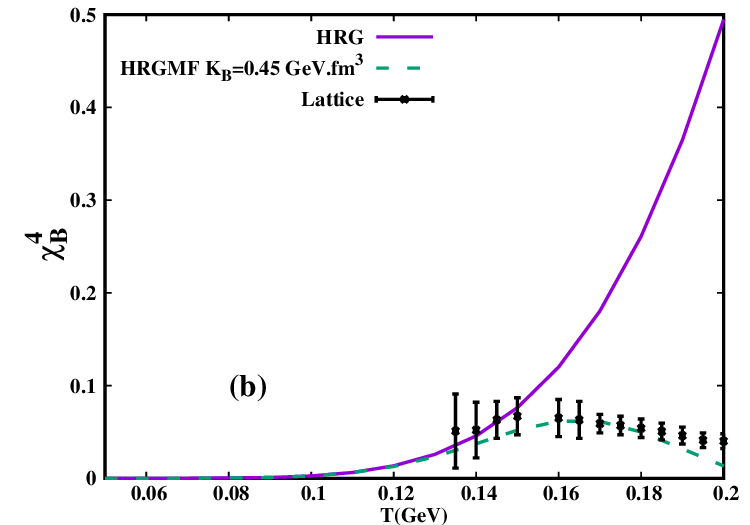}
		\end{tabular}
		\caption{Baryon number susceptibilities of different orders in HRGMF model.  This result is independent of $K_m$. The lattice data is taken from Ref.~\cite{LQCD16}.} 
		\label{chiB_HRGMF}
	\end{center}
\end{figure}

Fig.\ref{chiB_HRGMF} shows the  2nd and 4th order baryon number susceptibilities within the HRGMF model. When 
repulsion is absent, the susceptibilities, calculated in the ideal HRG model, increase monotonically with temperature. As the interaction among the baryons is turned on through assigning a non zero value to $K_B$, the susceptibilities (especially $\chi_B^4$) 
show non-monotonic behaviour. It is to be noted that the HRG model reproduces the LQCD results up to a temperature 
of  $T=0.16$ GeV, after which, it deviates. On the other hand, HRGMF provides a very good qualitative and quantitative description. 
The broad bump in $\chi_B^4$ (Fig.\ref{chiB_HRGMF}(b)) in the lattice simulations, 
is very well reproduced in HRGMF model. The introduction of repulsive interaction reduces the  baryonic susceptibility at
higher temperatures which can be explained from the approximate expressions for pressure in Eq. (\ref{pbapprox}). This also
explains that the reduction is larger as one goes to higher order susceptibility.
It is to be noted that mesons do not contribute to $\chi_{B}$ and hence the results are independent of the value of $K_M$. 
Since all the baryons have baryon number one, the numerical values of $\chi_B^2$ and $\chi_B^4$ are same for HRG.
However, for the interacting (HRGMF) scenario, this is not true. Agreement of HRGMF  with LQCD indicates that the 
role of repulsive interaction in the thermodynamics of hadron gas is important, especially at higher temperature. 
Recently, in similar studies, it has been observed that if the repulsive interactions between mesons is switched off
and if van-der Waals repulsive and attractive interaction between baryons are included,  the
resulting model turns out to be in better agreement with the lattice data~\cite{Vovchenko:2016rkn}.  Thus, ideal HRG is insufficient to describe higher order susceptibilities. The agreement with the results from mean-field HRG, indicates that repulsive interactions cannot be neglected in the studies which are being carried out to probe quark-hadron phase transition as well as QCD critical point.

\begin{figure}[H]
	\begin{center}
		\begin{tabular}{c c}
			\includegraphics[scale=0.55]{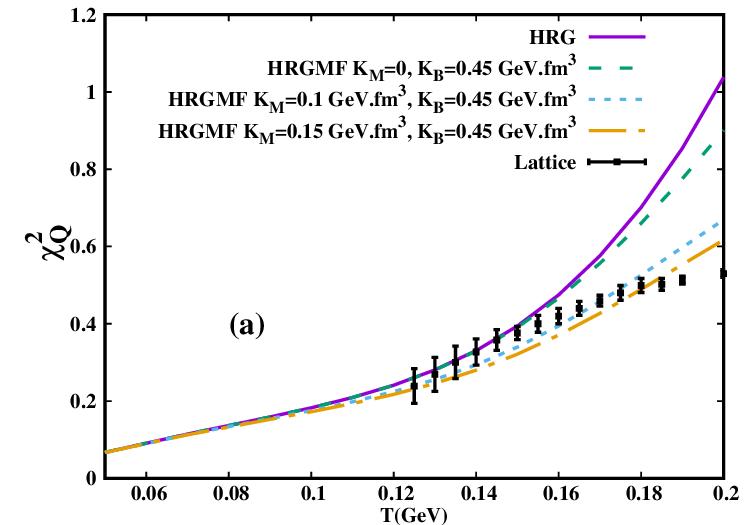}
			\includegraphics[scale=0.55]{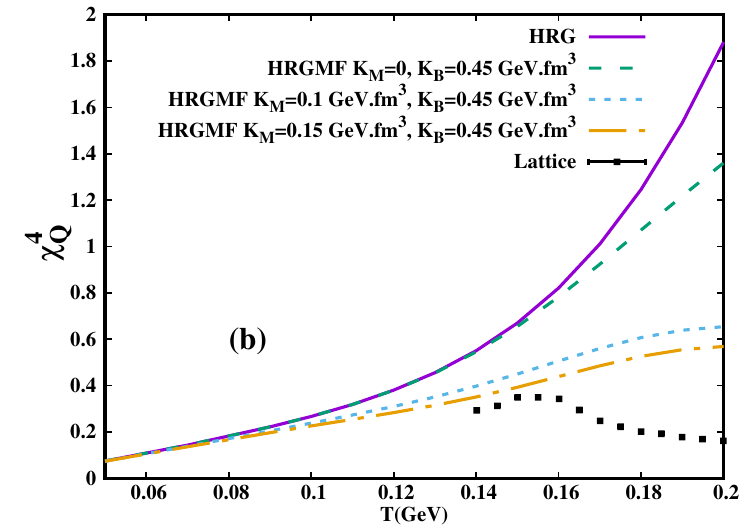}
		\end{tabular}
		\caption{Electric charge susceptibilities of different orders in HRGMF model.   The lattice data is taken from Ref.~\cite{LQCD16}.}
		\label{chiQ_HRGMF}
	\end{center}
\end{figure}
Fig.\ref{chiQ_HRGMF} shows  2nd and 4th order electric charge susceptibilities. Similar to the case of baryon number 
susceptibilities, the electric charge susceptibilities increase monotonically with temperature for HRG. However, unlike baryon number susceptibilities,
the numerical values of $\chi_Q^2$ and $\chi_Q^4$ are not same for HRG. 
It is to be noted that there is no qualitative and almost no quantitative difference between the results obtained from 
HRG and HRGMF  model for $K_M=0$  except at high temperature. The reason is that, charged mesons, which are the dominant contributors
to $\chi_Q^{n}$, have no repulsive interactions for $K_M=0$. But when mean-field interactions are switched on for mesons,  HRGMF model reproduces LQCD results for $\chi_Q^{2}$. Good
quantitative agreement is achieved for $K_M=0.1$ GeV$\cdot$fm$^{3}$. In case of $\chi_Q^{4}$, 
HRGMF results for susceptibility for all the three choices of $K_M$ are larger than that for LQCD. 
Nonetheless,  qualitative agreement can be seen for higher values of $K_M$ which also emphasizes the 
important role of repulsive interactions among mesons.

\begin{figure}[H]
	\begin{center}
		\begin{tabular}{c c}
			\includegraphics[scale=0.55]{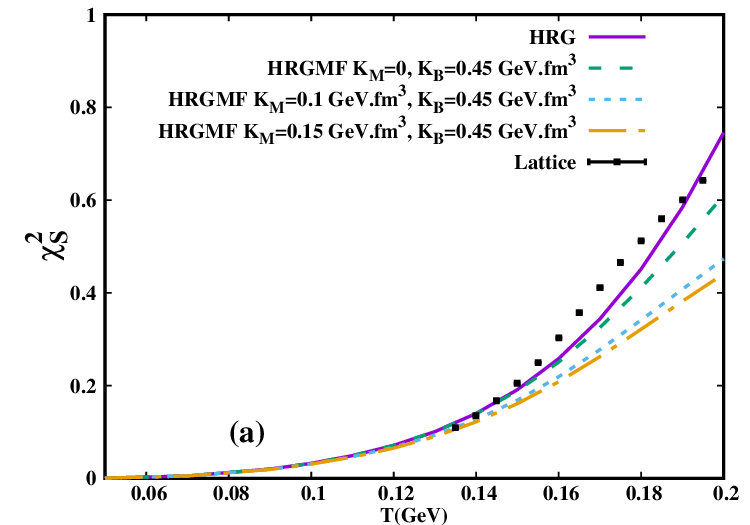}
			\includegraphics[scale=0.55]{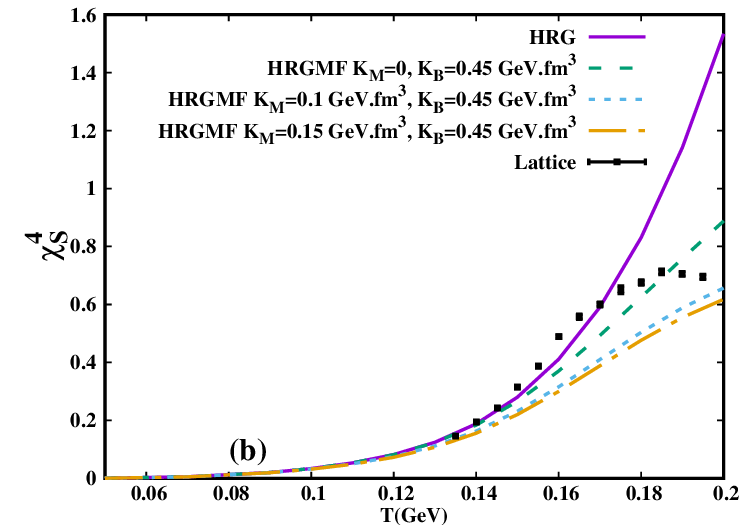}
		\end{tabular}
		\caption{Strangeness susceptibilities of different orders in HRGMF model.   The lattice data is taken from Ref.~\cite{LQCD16}. } 
		\label{chiS_HRGMF}
	\end{center}
\end{figure}

Fig.\ref{chiS_HRGMF} shows 2nd and 4th order strangeness susceptibilities. It is to be noted that the lattice 
results of $\chi_{S}^2$ is reproduced by ideal HRG model. The results of $\chi_{S}^4$ is 
also well described by HRG model up to a temperature of 0.16 GeV. 
At higher temperatures, the results from the interacting scenario, modelled by  
HRGMF model, is closer to the lattice results. It is to be noted that the repulsive interactions are found to underestimate strangeness susceptibilities. 
Previous studies have also 
observed similar behaviour of strangeness susceptibilities~\cite{bazavov_strange}. This observation can be attributed to the unknown strange 
hadronic states not included into the hadronic mass spectrum. In fact, inclusion of these unknown states has 
been found to improve HRG model estimations~\cite{lo}.

Ratios of susceptibilities of  baryon, charge and strangeness are related to the moments of conserved charge fluctuations and hence they are important in 
the context of heavy-ion collision experiments. The reason is that, heavy-ion experiments cannot measure charge susceptibilities directly. Instead, the moments of conserved charges like the mean (M), standard deviation $(\sigma)$, skewness (S) and kurtosis $(\kappa)$ are accessible in experiments. The standard deviation corresponds to the width
of the distribution, the skewness is a measure of the asymmetry
of the distribution and kurtosis describes the peakedness of the
distribution. The moments are related to the susceptibilities by the relations:
\begin{equation}
\frac{\chi_x^2}{\chi_x^1}=\frac{\sigma_x^2}{M_x},\ \ \ \frac{\chi_x^3}{\chi_x^2}=S_x\sigma_x,\ \ \ \frac{\chi_x^4}{\chi_x^2}=\kappa_x\sigma_x^2
\end{equation}
These ratios of susceptibilities are expected to be independent of the volume of the system.
\begin{figure}[h]
	\begin{center}
		\begin{tabular}{c c}
			\includegraphics[scale=0.55]{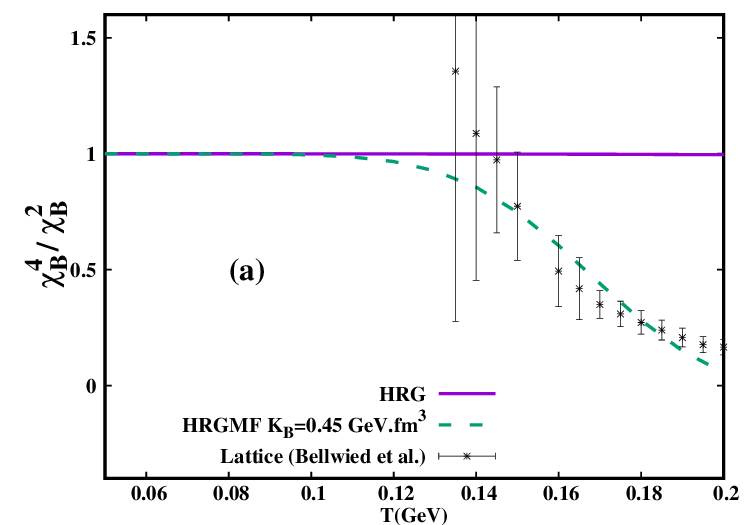}
			\includegraphics[scale=0.55]{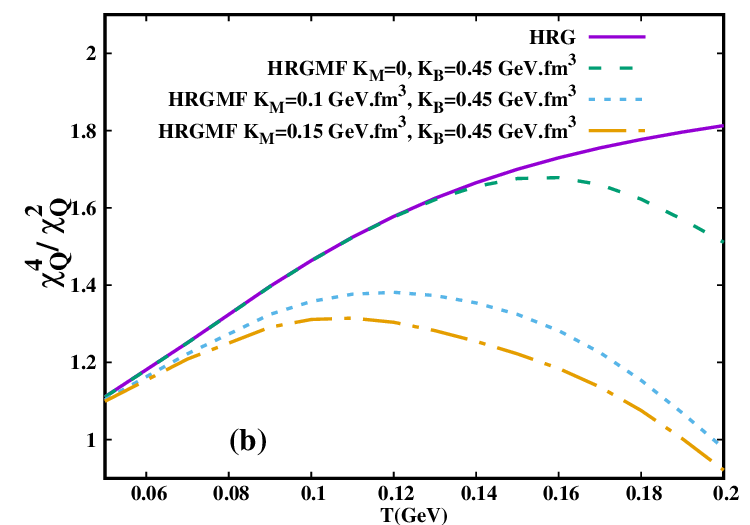}\\
			\includegraphics[scale=0.55]{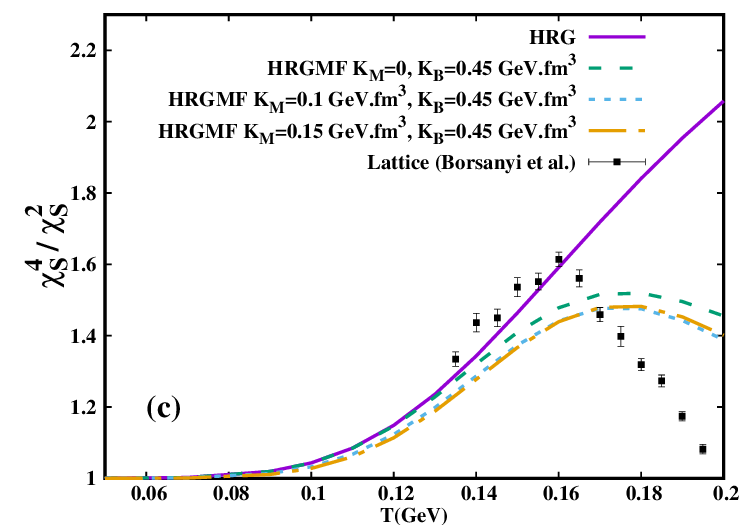}
		\end{tabular}
		\caption{Ratios of fourth order and second order susceptibilities in HRGMF model.} 
		\label{chiBQ_ratios}
	\end{center}
\end{figure}

Fig. \ref{chiBQ_ratios} shows ratios of susceptibilities of  baryon, charge and strangeness of 2nd and 4th order.
In the HRGMF model, in the Boltzmann approximation,  (and in the limit of $\beta U\ll 1$),
the susceptibility $\chi_{B}^{n}$ can be approximated as
\begin{equation}
\chi_{B}^{n} =(\chi_{B}^{id})^{n} -2^n\beta^4K_{B}(n_{B}^{id})^2
\label{chiBoltz}
\end{equation}
where $(\chi_{B}^{id})^{n}$ is the n$^{\text{th}}$ order non-interacting susceptibility.
From the above  equation, it is seen that when repulsive interactions are switched off 
($K_B=0$), then $\frac{\chi_{B}^4}{\chi_B^2}=1$. In HRG, 
and in the Boltzmann approximation, net-baryon kurtosis  shows the expected Skellam 
behavior~\cite{Vovchenko:2016rkn,Asakawa:2015ybt}. 
The repulsive interactions (2nd term in Eq.(\ref{chiBoltz})) makes this ratio to decrease. This is because the magnitude of this decrease of the susceptibility compared to ideal HRG increases with the order of the susceptibility when repulsive interaction is present.
This behaviour is also consistent 
with  the lattice QCD data obtained in Ref.~\cite{bazavov_prd95}. 
In Ref.~\cite{bazavov_prd95}, it was shown  that, at low temperature the ratio has 
a value of unity and, at high 
temperatures, it decreases with temperature to reach the free quark limit. Thus the result presented here, of HRGMF model,
reproduces the lattice result. Deviation from the Skellam behaviour can be attributed to 
the repulsive interactions 
between baryons, and again, one cannot neglect its contribution in the conserved charge fluctuation studies. 

In the case of electric charge susceptibilities, the mesons are the dominant contributors and hence Boltzmann approximation is
not a good approximation.
Thus, simple expression similar to Eq.(\ref{chiBoltz}) cannot be obtained. Further, in the HRG model only baryons with 
baryon number $B = 1$ contribute to  $\chi_{B}^{n}$, while in case of $\chi_{Q}^n$ multiple charged hadrons contribute. 
In fact, as one considers higher order fluctuations, these multiply charged hadrons provide larger weightage, where both meson as well as 
baryons contribute. As a result, the characteristic deviation of $\frac{\chi_{Q}^4}{\chi_Q^2}$ from the baryonic sector is seen in Fig.\ref{chiBQ_ratios}(b). 
The effect of repulsive interactions is to suppress the number density at high temperature.  This can be seen in 
Fig.\ref{chiBQ_ratios}(b) as the ratio  $\frac{\chi_{Q}^4}{\chi_Q^2}$ in HRGMF deviate from HRG model results. 

The ratio of strangeness susceptibilities show somewhat similar picture. Like the electric charge scenario,
in the case of strangeness also, the Boltzmann approximation is not valid. Furthermore,
the particles with multiple strangeness contribute in this case which give larger weightage to higher order susceptibilities. The ratio increases monotonically with temperature for HRG.
Once the interaction is switched on, the ratio is suppressed. Furthermore,
the ratio becomes non-monotonic at high temperatures. The repulsive interaction seems to have a significant contribution in this quantity. 

\begin{figure}[h]
	\begin{center}
		\begin{tabular}{c c}
			\includegraphics[scale=0.55]{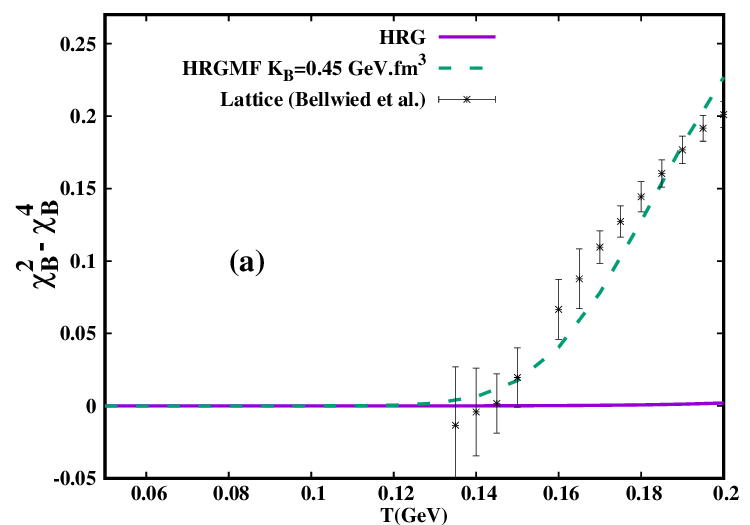}
			\includegraphics[scale=0.55]{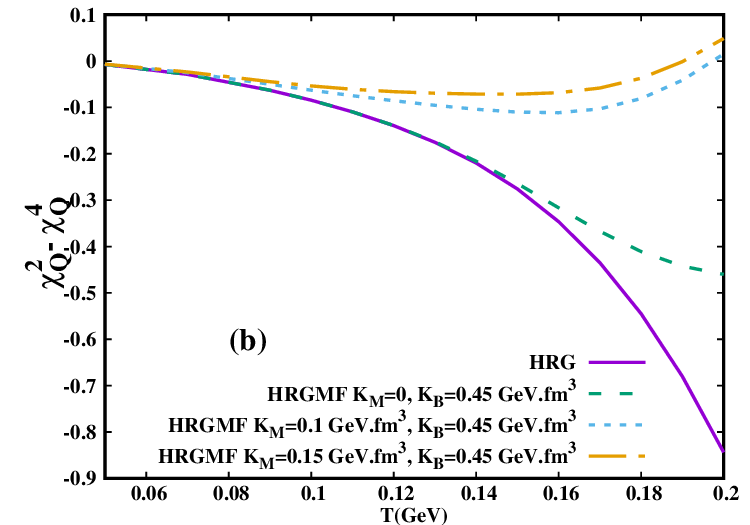}\\
			\includegraphics[scale=0.55]{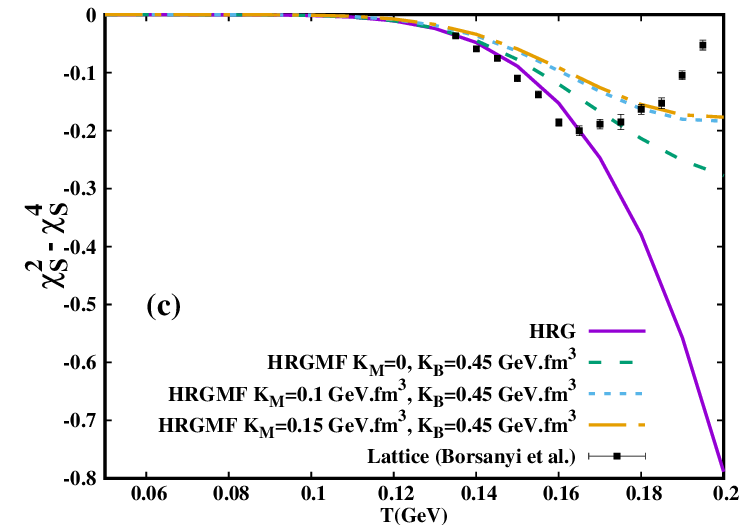}
		\end{tabular}
		\caption{Differences of second order and fourth order susceptibilities in HRGMF model. } 
		\label{chiBQ_diff}
	\end{center}
\end{figure}
It may not be easy to separate the effect of repulsive interactions from medium effects for the case of ratios of susceptibilities, like in-medium mass modification or the widening of spectral width. Differences of susceptibilities do not depend on the mass spectrum included in the HRG model. Differences of 2nd and 4th order susceptibilities are shown in  Fig.\ref{chiBQ_diff}. The difference, $\chi^2_B-\chi^4_B$ is zero in HRG model. But  this difference increases with temperature if one includes the repulsive interactions using mean-field approach. This behaviour is in agreement  with LQCD results~\cite{ratti1}. The difference increases with temperature as has been found in the HRGMF model and lattice results. The results of charge and strangeness sectors in HRG model, show different behaviour as opposed to baryon sector. 
The differences of these susceptibilities decrease with increase in temperature. The reason for such behaviour may be because of the fact that 
multiple charged hadrons, which contribute more to higher order susceptibilities, appear at high temperatures. The presence of repulsive interactions suppress the  density of charged hadrons at high temperature. Hence one observes less steeper decrease in HRGMF as compared to HRG.

So far, the results at zero baryon chemical potential have been discussed. Let us now discuss the case of non-vanishing
values for the chemical potentials $\mu_B$, $\mu_Q$ and $\mu_S$. Ref.~\cite{LQCD20} shows that, the susceptibilities
at finite chemical potential, in general, can be expanded in powers of $\mu_i/T$ $(i=B,Q,S)$ with  generalised susceptibilities being the coefficients
that can be evaluated at vanishing chemical potentials. The constraints of fixed strangeness and electric charge to 
baryon number fixes the relation between the electric charge chemical potential and the strangeness chemical potential 
to the baryon chemical potential at a given temperature~\cite{LQCD17}. As in Ref.~\cite{LQCD20}, a strangeness neutral system
i.e. $n_S=0$ and $n_Q/n_B=0.4$ has been considered here which are the conditions met in heavy ion collision experiments with gold or uranium 
nuclei. Thus 
one has the relation with $\hat\mu_i=\mu_i/T$
$$\hat \mu_Q\simeq q_1(T)\hat\mu_B+q_3(T)\hat\mu_B^3$$
\begin{equation}
\hat \mu_S\simeq s_1(T)\hat\mu_B+s_3(T)\hat\mu_B^3
\label{muqs}
\end{equation}

The coefficient functions $q_i$ and $s_i$ are given in terms of the susceptibilities at zero chemical potential and the
ratio $r=n_Q/n_B$ as in Ref.~\cite{LQCD172}. Different susceptibilities are then estimated using Eq.(\ref{chin}) for  a given
$\mu_B$, $\mu_Q$ and $\mu_S$.
\begin{figure}[h]
	\begin{center}
		\begin{tabular}{c c}
			\includegraphics[scale=0.55]{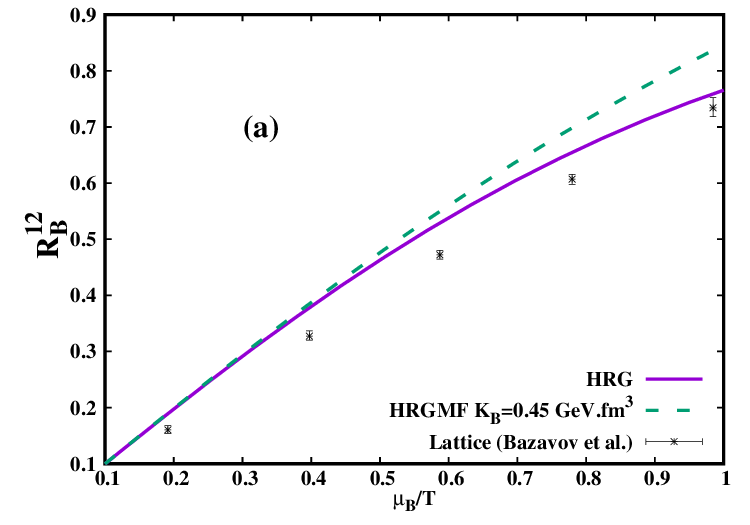}
			\includegraphics[scale=0.55]{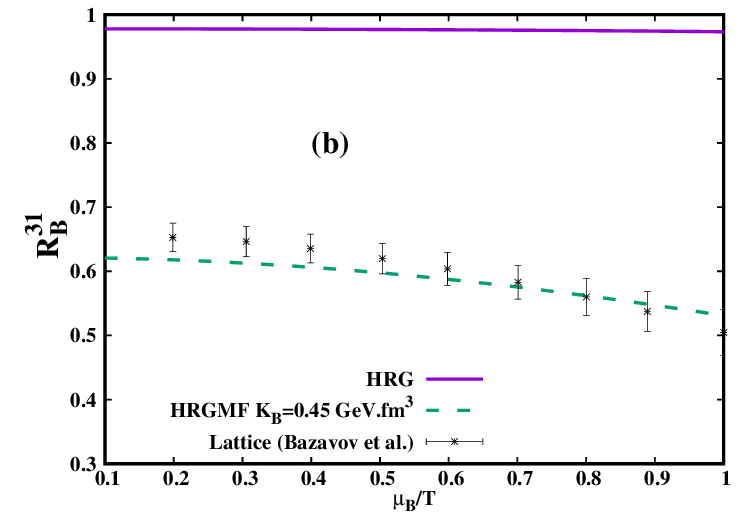}\\
			\includegraphics[scale=0.55]{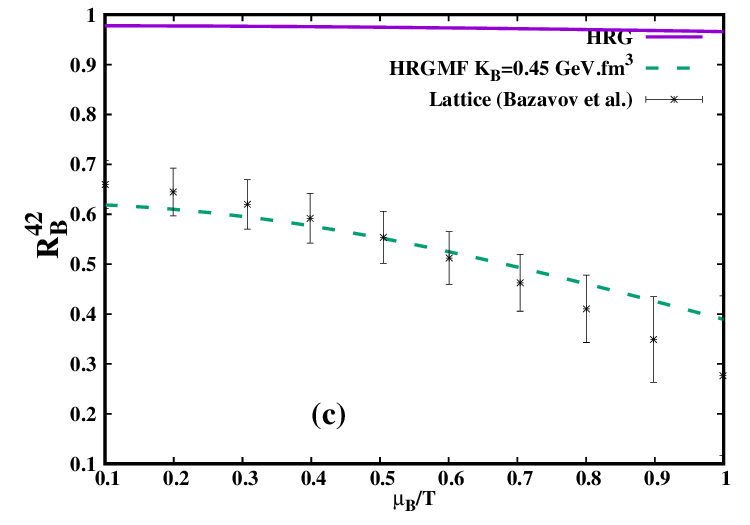}
		\end{tabular}
		\caption{$R_B^{12}$, $R_B^{31}$ and $R_B^{42}$ (defined in the text) as 
			functions of $\mu_B/T$ in HRGMF model. The lattice data is taken from Ref.~\cite{LQCD20}. } 
		\label{rb1}
	\end{center}
\end{figure}

Fig.\ref{rb1} shows the ratios of the cumulants for the net baryon number fluctuations as functions 
of $\mu_B/T$. The ratios of culumants considered here are:

\begin{eqnarray}
R^{12}_B=\chi^1_B(T,\mu_B)/ \chi^2_B(T,\mu_B)\nonumber\\
 R^{31}_B(T,\mu_B)=\frac{\chi^3_B(T,\mu_B)}{\chi^1_B(T,\mu_B)}\nonumber\\
  R^{42}_B=\frac{\chi^4_B(T,\mu_B)}{\chi^2_B(T,\mu_B)}
\end{eqnarray}
The temperature has been taken as $T=0.158$ GeV as in Ref.~\cite{LQCD20} corresponding to 
the upper end of the error band of the pseudocritical temperature for vanishing chemical potential.
On the top left panel $R^{12}_B$ has been plotted, as a function of $\mu_B/T$. On the top right panel  $R^{31}_B(T,\mu_B)$ has been plotted. In the bottom panel, the kurtosis 
ratio i.e. $R^{42}_B$ has been plotted as a function of $\mu_B/T$. 
As may be observed, the lower cumulant i.e. $R^{12}_B$ from lattice simulation
are in good agreement with the HRG model estimation which takes the hadrons to be point like and noninteracting. 
However, the simple HRG model is inadequate to describe the behaviour of  the higher order cumulants 
$R_{31}^B$ and $R_{42}^B$ obtained from lattice simulations. In fact, for simple HRG model, the values of both 
these higher order cumulant ratios are unity. 
When one includes the repulsive interactions among hadrons within HRGMF model, the results seem to be in agreement with the lattice QCD results even at finite chemical potential. 

	\newpage
\rhead{Hadronic matter in Magnetic field}
\chapter{\label{chap:thermodynamics} Hadronic matter in Magnetic field}
It is expected that off-central relativistic heavy-ion collisions produce very large magnetic field $B\sim m_{\pi}^2$ ($\sim 10^{18}$ G) due to the relativistic motion of electrically charged particles. Strong magnetic field is also expected to exist in dense neutron stars~\cite{Duncan:1992hi,dey}. This extra parameter $B$ can significantly change the equation of state (EoS) and the structure of the phase diagram of the strongly interacting matter produced in such experiments. The presence of magnetic field alters the energy levels of the particles of QCD matter significantly. The magnetic field can also affect the time evolution of the matter created in heavy-ion collision experiments. The presence of magnetic field can introduce interesting phenomena on strongly interacting matter, like, chiral magnetic effect~\cite{Fukushima:2008xe}, magnetic catalysis~\cite{Shovkovy:2012zn}, inverse magnetic catalysis~\cite{Bali:2011qj,Preis:2012fh}, etc. For these reasons, investigation of QCD matter in presence of magnetic field is very important for us.
\section{Hadron resonance gas model in magnetic field}
Let us now discuss some of the thermodynamic relations in presence of magnetic field. 
The free energy density is written in terms of partition function as 
\begin{equation}
F=-T\:\text{ln}\mathcal{Z}=F_{vac}+F_{th}
\end{equation}
Where $F_{vac}$ is the vacuum part and $F_{th}$ is the thermal part.
In the presence of constant external magnetic field, the free energy is written as
\begin{equation}
F=\mathcal{E}-T\mathcal{S}-B\mathcal{M_B}
\end{equation}
where $\mathcal{E}$ is the total energy, $\mathcal{S}$ is the total entropy, $B$ is the constant external magnetic field and $\mathcal{M_B}$ is the magnetization of the system. $F$ is related to the other quantities by the following differential equations:
\begin{equation}
\frac{\partial F}{\partial T}=-\mathcal{S}, \ \ \ \ \frac{\partial F}{\partial B}=-\mathcal{M_B}, \ \ \ \ \frac{\partial F}{\partial V}=-P
\end{equation}
with energy density $\epsilon=\mathcal{E}/V$, and entropy density $s=\mathcal{S}/V$\\
In the thermodynamic limit $V\rightarrow \infty$,
\begin{equation}
P=-\frac{F}{V}
\end{equation}
The free energy density of ideal HRG model in presence of constant external magnetic field is written as 

\begin{eqnarray}
F_{c}&=&\mp\sum_{i}\sum_{S_z}\sum_{n=0}^{\infty}\frac{eB}{(2\pi)^2}\int dp_{z}\bigg(E_{i,c}(p_{z},n,S_{z})\nonumber\\
&+&T\:\text{ln}(1\pm e^{-(E_{i,c}-\mu_i)/T})\bigg); e_i\neq 0
\label{idFen}
\end{eqnarray}

\begin{eqnarray}
F_{n}&=&\mp\sum_{i}\sum_{S_z}\int\frac{d^3p}{(2\pi)^3}\:\bigg(E_{i,n}\nonumber\\
&+&T\:\text{log}\bigg[1\pm \text{exp}\bigg(-\frac{(E_{i,n}-\mu_i)}{T}\bigg)\bigg]\bigg); e_i=0
\end{eqnarray}
\noindent
Where $F_c$ is the free energy density for electrically charged particles and $F_n$ is the free energy density for neutral particles. Here $e_i$ is the electric charge of $i^{\text{th}}$ hadronic species,
$E_{i,c/n}$ is the single-particle
energy for charged/neutral particle, $m_i$ is the mass, $T$ is the temperature and $\mu_i=B_i\mu_B+\mathcal{S}_i\mu_\mathcal{S}+Q_i\mu_Q$ is the
chemical potential. In the expression of previous line, $B_i$, $\mathcal{S}_i$, $Q_i$, are
respectively, the baryon number, strangeness and electric charge of the particle,
$\mu_i^,$s are corresponding chemical potentials. The upper and lower
signs correspond to fermions and bosons respectively.

The single particle
energy levels for neutral and charged particles in presence of a constant magnetic field $B$ are
given by~\cite{Endrodi:2013cs}

\begin{equation}
E_{i,n}=\sqrt{p^2+m_i^2}
\label{disp}
\end{equation}

\begin{equation}
E_{i,c}(p_z,n,S_z)=\sqrt{p_{z}^2+m_i^2+2e_{i}B(n+1/2-S_z)}
\label{dispmagn}
\end{equation}

\noindent
where, $n$ is any positive integer
corresponding to allowed Landau levels, $S_z$ is the component of spin
$S$ in the direction of magnetic field.
For a given $S$, there are $2S+1$ possible values of $S_z$. The
gyromagnetic ratios are taken as $g_i=2|e_i/e|$ ($e$ being elementary unit of electric charge) for all charged hadrons. It is to be noted that these values of gyromagnetic ratios are based on universal tree-level argument~\cite{Ferrara:1992yc}. Main reason for using this value is that the experimental value of $g$ is known only for few hadrons and the uncertainties in their values are also very large. The pressure of ideal hadron resonance gas  is $P^{\text{id}}=-F$ and it can be written as

\begin{eqnarray}
P_{c}^{\text{id}}&=&\pm\sum_{i}\sum_{S_z}\sum_{n=0}^{\infty}\frac{eB}{(2\pi)^2}\int dp_{z}\bigg(E_{i,c}(p_{z},n,S_{z})\nonumber\\
&+&T\:\text{ln}(1\pm e^{-(E_{i,c}-\mu_i)/T})\bigg); e_i\neq 0
\label{idprechrg}
\end{eqnarray}

\begin{eqnarray}
P_{n}^{\text{id}}&=&\pm\sum_{i}\sum_{S_z}\int\frac{d^3p}{(2\pi)^3}\:\bigg(E_{i,n}\nonumber\\
&+&T\:\text{log}\bigg[1\pm \text{exp}\bigg(-\frac{(E_{i,n}-\mu_i)}{T}\bigg)\bigg]\bigg); e_i=0
\label{idpreneutral}
\end{eqnarray}

Note that  the thermal part of the pressure is naturally convergent in the $UV$ limit but the vacuum part is divergent and needs to be regularized.

\subsection{Regularization of vacuum pressure}
The vacuum pressure in the above expression is divergent and it needs to be properly regularized first.  One should be careful in choosing appropriate regularization scheme to avoid certain unphysical results. For example, some studies have found the oscillations in the magnetization while others have found imaginary meson masses~\cite{Fayazbakhsh:2013cha}. Both the findings are unphysical and  can be considered to be the result of an inappropriate regularization choices. These unphysical results arise especially in case of magnetic field dependent regularization schemes. Thus, it is absolutely necessary to separate the magnetic field dependent and independent parts clearly through appropriate regularization scheme. Magnetic field independent regularization (MFIR) has recently been introduced to achieve this goal~\cite{Ebert:2003yk,Menezes:2008qt}. To obtain the regularized vacuum pressure for spin $\frac{1}{2}$ particles, MFIR method has been used in this work.

The vacuum part of the  pressure for a charged spin-$\frac{1}{2}$ particle in magnetic field  is 

\begin{equation}
P_{\text{vac}}(S=1/2,B)=\sum_{n=0}^{\infty}g_{n}\frac{eB}{2\pi}\int_{-\infty}^{\infty}\frac{dp_z}{2\pi}E_{p,n}(B)
\label{vacphalf}
\end{equation}
where $g_{n}=2-\delta_{n0}$ is the degeneracy of n$^{\text{th}}$ Landau level. Adding and subtracting  contribution from the lowest Landau level (i.e. $n=0$) from the above equation one gets
\begin{equation}
P_{\text{vac}}(S=1/2,B)=\sum_{n=0}^{\infty}2\frac{eB}{2\pi}\int_{-\infty}^{\infty}\frac{dp_z}{2\pi}\bigg(E_{p,n}(B)-\frac{E_{p,0}(B)}{2}\bigg)
\label{vacphalf1}
\end{equation}

Dimensional regularization~\cite{Peskin:1995} to regularize the divergence has been used here. In $d-\epsilon$ dimension Eq. (\ref{vacphalf1}) can be written as
\begin{eqnarray}
P_{\text{vac}}(S=1/2,B)&=&\sum_{n=0}^{\infty}\frac{eB}{\pi}\mu^{\epsilon}\int_{-\infty}^{\infty}\frac{d^{1-\epsilon}p_z}{(2\pi)^{1-\epsilon}}\nonumber\\
&\times& \bigg(\sqrt{p_z^2+m^2-2eBn}-\sqrt{p_z^2+m^2}\bigg)
\label{vacphalf2}
\end{eqnarray}
where the dimension of the above expression is fixed to 1 by the introduction of $\mu$.  The integration can be done using standard $d-$dimensional formula
\begin{equation}
\int_{-\infty}^{\infty}\frac{d^dp}{(2\pi)^d}\:(p^2+m^2)^{-A}=\frac{\Gamma(A-\frac{d}{2})}{(4\pi)^{d/2}\Gamma(A)(M^2)^{(A-\frac{d}{2})}}
\label{dDimint}
\end{equation}
Integration of the first term in  Eq.(\ref{vacphalf2}) gives
\begin{eqnarray}
I_1=\sum_{n=0}^{\infty}\frac{eB}{\pi}\mu^{\epsilon}\int_{-\infty}^{\infty}\frac{d^{1-\epsilon}p_z}{(2\pi)^{1-\epsilon}}(p_z^2+m^2-2eBn)^{\frac{1}{2}}\nonumber\\
=-\frac{(eB)^2}{2\pi^2}\bigg(\frac{2eB}{4\pi\mu}\bigg)^{-\frac{\epsilon}{2}}\Gamma\bigg(-1+\frac{\epsilon}{2}\bigg)\zeta\bigg(-1+\frac{\epsilon}{2},x\bigg)
\label{I1}
\end{eqnarray}
where $x\equiv\frac{m^2}{2eB}$. The Landau infinite sum has been expressed in terms of  Riemann-Hurwitz $\zeta-$function
\begin{equation}
\zeta(z,x)=\sum_{n=0}^{\infty}\frac{1}{(x+n)^z}
\label{RHdef}
\end{equation}

with the expansion
\begin{equation}
\zeta\bigg(-1+\frac{\epsilon}{2},x\bigg)\approx-\frac{1}{12}-\frac{x^2}{2}+\frac{x}{2}+\frac{\epsilon}{2}\zeta^{'}(-1,x)+\mathcal{O}(\epsilon^2)
\label{zetaexp}
\end{equation}
The expansion of $\Gamma$-function is
\begin{equation}
\Gamma\bigg(-1+\frac{\epsilon}{2}\bigg)=-\frac{2}{\epsilon}+\gamma-1+\mathcal{O}(\epsilon)
\label{gammaexp1}
\end{equation}
where $\gamma$ is the Euler constant. Eq.(\ref{I1}) can be written as
\begin{eqnarray}
I_1&=&-\frac{(eB)^2}{2\pi^2}\bigg(-\frac{2}{\epsilon}+\gamma-1+\text{ln}\bigg(\frac{2eB}{4\pi\mu^2}\bigg)\bigg)\bigg(-\frac{1}{12}\nonumber\\
&-&\frac{x^2}{2}+\frac{x}{2}+\frac{\epsilon}{2}\zeta^{'}(-1,x)+\mathcal{O}(\epsilon^2)\bigg)
\end{eqnarray}

Integration of the second term in  Eq.(\ref{vacphalf2}) can be written as
\begin{eqnarray}
I_2&=&\sum_{n=0}^{\infty}\frac{eB}{\pi}\mu^{\epsilon}\int_{-\infty}^{\infty}\frac{d^{1-\epsilon}p_z}{(2\pi)^{1-\epsilon}}(p_z^2+m^2)^{\frac{1}{2}}\nonumber\\
&=&\frac{(eB)^2}{2\pi^2}\bigg(-\frac{x}{\epsilon}-\frac{(1-\gamma)}{2}x+\frac{x}{2}\text{ln}\bigg(\frac{2eB}{4\pi\mu^2}\bigg)+\frac{x}{2}\text{ln}(x)\bigg)
\end{eqnarray}

Thus the vacuum pressure in presence of constant external magnetic field becomes
\begin{eqnarray}
P_{\text{vac}}(S=1/2,B)&=&\frac{(eB)^2}{2\pi^2}\bigg(\zeta^{'}(-1,x)-\frac{2}{12\epsilon}-\frac{(1-\gamma)}{12}-\frac{x^2}{\epsilon}-\frac{(1-\gamma)}{2}x^2\nonumber\\
&+&\frac{x}{2}\text{ln}(x)+\frac{x^2}{2}\text{ln}\bigg(\frac{2eB}{4\pi\mu^2}\bigg)+\frac{1}{12}\text{ln}\bigg(\frac{2eB}{4\pi\mu^2}\bigg)\bigg)
\label{vacphalfB}
\end{eqnarray}

The above expression is still divergent. Let us remove the $B=0$ contribution from it. This zero field vacuum pressure in $d=3-\epsilon$ dimension is
\begin{eqnarray}
P_{\text{vac}}(S=1/2,B=0)&=& 2\mu^{\epsilon}\int\frac{d^{3-\epsilon}p}{(2\pi)^{3-\epsilon}}\:(p^2+m^2)^{\frac{1}{2}}\nonumber\\
&=&\frac{(eB)^2}{2\pi^2}\bigg(\frac{2eB}{4\pi\mu^2}\bigg)^{-\frac{\epsilon}{2}}\Gamma\bigg(-2+\frac{\epsilon}{2}\bigg)x^{2-\frac{\epsilon}{2}}
\end{eqnarray}
Above equation can be further simplified to
\begin{eqnarray}
P_{\text{vac}}(S=1/2,B=0)&=&-\frac{(eB)^2}{2\pi^2}x^2\bigg(\frac{1}{\epsilon}+\frac{3}{4}-\frac{\gamma}{2}\nonumber\\
&-&\frac{1}{2}\text{ln}\bigg(\frac{2eB}{4\pi\mu^2}\bigg)-\frac{1}{2}\text{ln}(x)\bigg)
\label{vachalfB0}
\end{eqnarray}
where the use of the $\Gamma$-function expansion has been done.

One can write
\begin{eqnarray}
P_{\text{vac}}(S=1/2,B)&=&P_{\text{vac}}(1/2,B=0)+\Delta P_{\text{vac}}(1/2,B)
\label{pvacB}
\end{eqnarray}
where the first term on the r.h.s is independent of magnetic field and the second term is dependent on magnetic field. The second term in the r.h.s. is
\begin{eqnarray}
\Delta P_{\text{vac}}(S=1/2,B)&=&\frac{(eB)^2}{2\pi^2}\bigg(-\frac{2}{12\epsilon}+\frac{\gamma}{12}+\frac{1}{12}\text{ln}\bigg(\frac{m^2}{4\pi\mu^2}\bigg)+\frac{x}{2}\text{ln}(x)\nonumber\\
&-&\frac{x^2}{2}\text{ln}(x)+\frac{x^2}{4}-\frac{\text{ln}(x)+1}{12}+\zeta^{'}(-1,x)\bigg)
\label{delpvacB}
\end{eqnarray}

The field contribution given by (\ref{delpvacB}) is still divergent due to the presence of pure magnetic field dependent term $\propto \frac{B^2}{\epsilon}$~\cite{Schwinger:1951nm,Elmfors:1993bm,Andersen:2011ip}. This divergence can be cancelled by redefining field dependent pressure contribution by including magnetic field contribution in it as
\begin{equation}
\Delta P_{\text{vac}}^{r}=\Delta P_{\text{vac}}(B)-\frac{B^2}{2}
\end{equation}

The divergences are absorbed into the renormalization of the electric charge and magnetic field strength~\cite{Endrodi:2013cs},
\begin{equation}
B^2=Z_{e}B_r^2; \hspace{0.5cm} e^2=Z_e^{-1}e_r^2; \hspace{0.5cm}  e_rB_r=eB
\end{equation}
where the electric charge renormalization constant is
\begin{equation}
Z_{e}\bigg(S=\frac{1}{2}\bigg)=1+\frac{1}{2}e_r^2\bigg(-\frac{2}{12\epsilon}+\frac{\gamma}{12}+\frac{1}{12}\text{ln}\bigg(\frac{M_{*}}{4\pi\mu^2}\bigg)\bigg)
\end{equation}

Here, $M_{*}=m$ has been taken, i.e the physical mass of the particle has been used. Thus the renormalized field dependent pressure (without pure magnetic field contribution) is
\begin{eqnarray}
\Delta P_{\text{vac}}^r(S=1/2,B)&=&\frac{(eB)^2}{2\pi^2}\bigg(\zeta^{'}(-1,x)+\frac{x}{2}\text{ln}(x)\nonumber\\
&-&\frac{x^2}{2}\text{ln}(x)+\frac{x^2}{4}-\frac{\text{ln}(x)+1}{12}\bigg)
\label{delpvacBfin}
\end{eqnarray}

The renormalized $B$ dependent pressure for spin zero and spin one can be obtained using similar method and are given by
\begin{itemize}

	\item Spin-0 particle:
	\begin{eqnarray}
	\Delta P_{\text{vac}}^r(s=0,B)&=&-\frac{(eB)^2}{4\pi^2}\bigg(\zeta^{'}(-1,x+1/2)\nonumber\\
	&-&\frac{x^2}{2}\text{ln}(x)+\frac{x^2}{4}+\frac{\text{ln}(x)+1}{24}\bigg)
	\label{delpvacBfin0}
	\end{eqnarray}
	
	\item Spin-1 particle:
	\begin{eqnarray}
	\Delta P_{\text{vac}}^r(s=1,B)&=&-\frac{3}{4\pi^2}(eB)^2\bigg(\zeta^{'}(-1,x-1/2)+\frac{(x+1/2)}{3}\text{ln}(x+1/2)\nonumber\\
	&+&\frac{2}{3}(x-1/2)\text{ln}(x-1/2)
	-\frac{x^2}{2}\text{ln}(x)+\frac{x^2}{4}\nonumber\\
	&-&\frac{7}{24}(\text{ln}(x)+1)\bigg)
	\label{delpvacBfin1}
	\end{eqnarray}
	
\end{itemize}
The vacuum pressure depends on the quantum numbers like spin, mass, charge etc. Thus the total vacuum  pressure of hadron gas is evaluated by adding the vacuum pressure of all the hadronic species included in HRG model.
\section{Thermodynamic quantities}
Table \ref{tab:hadrons} shows the hadrons and resonances included in the model with magnetic field. For the stability reasons, as discussed in Ref.~\cite{Endrodi:2013cs}, only those hadrons having spin $S<\frac{3}{2}$ has been considered. 

\begin{table}[h]
	\centering
	\begin{tabular}{|l|c|c|c|c||l|c|c|c|c|}
\hline
hadron & $m $ & $|e|$ & Spin & $\textmd{deg.}$ & hadron & $m $ & $|e|$ & Spin & $\textmd{deg.}$ \\ 
 & $(\textmd{GeV})$ &  &  & &  & $(\textmd{GeV})$ &  &  &  \\ 
\hline
\hline
$\pi^{\pm}$ & 0.135 & 1 & 0 & 2  & $p$ & 0.938 & 1 & 1/2 & 2  \\
$\pi^0$ & 0.135 & 0 & 0 & 1 & $n$ & 0.938 & 0 & 1/2 & 2  \\
$K^{\pm}$ & 0.495 & 1 & 0 & 2 & $\eta'$ & 0.958 & 0 & 0 & 1  \\
$K^0$ & 0.495 & 0 & 0 & 2 & $f_0$ & 0.980 & 0 & 0 & 1  \\
$\eta$ & 0.548 & 0 & 0 & 1 & $a_0$ & 0.980 & 0 & 1 & 1  \\ 
$\rho^{\pm}$ & 0.776 & 1 & 1 & 2 & $\phi$ & 1.020 & 0 & 1 & 1  \\
$\rho$ & 0.776 & 0 & 1 & 1 & $\Lambda$ & 1.116 & 0 & 1/2 & 1 \\  
$\omega$ & 0.782 & 0 & 1 & 1 & $h_1$ & 1.170 & 0 & 1 & 1 \\ 
$K_{*}^{\pm}$ & 0.892 & 1 & 1 & 2 & $\Sigma^\pm$ & 1.189 & 1 & 1/2 & 2 \\ 
$K_*^0$ & 0.892 & 0 & 1 & 2 &  $\Sigma^0$ & 1.189 & 0 & 1/2 & 1 \\ 
\hline
\end{tabular}

	\caption{\label{tab:hadrons} Hadrons and resonances  included in the hadron resonance gas model with magnetic field. Particle data is taken from Ref.~\cite{Beringer:1900zz}.
	}
\end{table}
Let us now discuss the thermodynamic properties of hadronic matter in presence of magnetic field. Here, type 2 EVHRG model has been used. For the investigation of the model with magnetic field, where it has been assumed that there is repulsive interaction among all the mesons and baryons. This type of meson-baryon and baryon-antibaryon repulsive interactions were included in studies of earliest versions of EVHRG model. Here the sizes of hadrons are taken as  $R_\pi$ (pion radius) $= 0$ fm, $R_K$ (kaon radius) $= 0.35$ fm, $R_m$ (all other meson radii) 
$= 0.3$ fm and $R_b$ (baryon radii) $= 0.5$ fm.

\begin{figure}[h]
	\vspace{-0.4cm}
	\begin{center}
		\begin{tabular}{c c}
			\includegraphics[width=6.5cm,height=6cm]{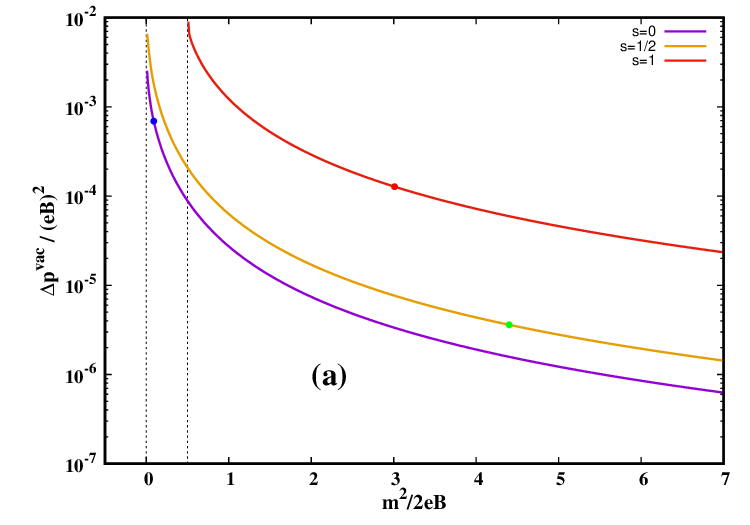}&
			\includegraphics[width=6.5cm,height=6cm]{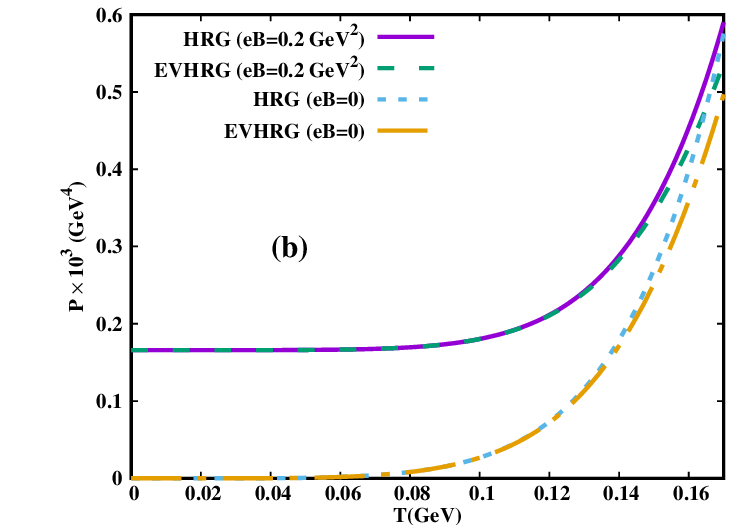}
		\end{tabular}
		\caption{Left panel shows vacuum pressure of charged particles computed using MFIR scheme plotted against dimensionless variable $x=\frac{m^2}{2eB}$.
			Right panel shows the total pressure as a function of temperature in presence of magnetic field for $\mu_B=0$.   } 
		\label{pressure}
	\end{center}
\end{figure}

Fig.\ref{pressure}(a) shows the variation of the vacuum  pressure as a function of the dimensionless variable $x=\frac{m^2}{2eB}$. The vacuum pressure is positive for spin-0, spin $\frac{1}{2}$ and spin-1 hadrons for wide range of magnetic fields and hence it can safely be assumed that the HRG description is valid.  Fig.\ref{pressure}(b) shows the variation of the total pressure with temperature both in presence and in absence of magnetic field for HRG and EVHRG models at zero $\mu_B$. For HRG and EVHRG models, it is to be noted that the pressure in presence of magnetic field has magnetic field dependent vacuum contribution. Such contribution is zero in absence of magnetic field. Hence the HRG pressure in absence of magnetic field vanishes at $T=0$ GeV while it is non-zero at finite magnetic field. This is the reason why the pressure instead of scaled pressure ($\frac{P}{T^4}$) has been plotted.  It can be seen that the pressure increases with temperature in all cases considered. The reason is that, the contribution to pressure of a 
particular species, with mass $m$,  is proportional to the Boltzmann factor $e^{-m/T}$. For ideal HRG, the thermal part of the pressure in presence of magnetic field is smaller than that in absence of magnetic field. However, it is the other way round for the  vacuum part of the pressure.  This can be explained  by comparing  the dispersion relations in presence of  magnetic field (Eq.(\ref{dispmagn})) and in the absence of magnetic field (Eq.(\ref{disp})). The effective mass of the charged particles in presence of magnetic field is $m_{eff}^2=m^2+2eB(\frac{1}{2}-S)$. Thus the mass of a spin-0 particle increases  whereas that of spin-1 particle decreases if $eB\neq 0$ or when magnetic field is present. For spin-$\frac{1}{2}$ particles, mass remains unchanged. Since the particles with spin-1 are heavier (lightest spin-1 particle $\rho$ weighs 0.776 GeV which is more than five times heavier than pion), their contributions to pressure  are much smaller compared to spin-0  particles and show up only at higher temperature. Since pions are dominant particles in the hadronic matter at low temperature, and since their effective mass increases in presence of magnetic field, the thermal part of the pressure, for HRG model, in presence of magnetic field is smaller than that for without magnetic field.  

Fig.\ref{pressure}(b) further shows the effect of repulsive interaction on the pressure. If the repulsive interactions are included through excluded volume correction to ideal gas partition function, then  the pressure is smaller for both $eB=0$ and $eB\neq0$ cases. If finite size of hadrons is taken into account, the available free space for hadrons decreases with increasing temperature. This decreases the number density and hence the pressure of hadrons compared to free HRG case. From Fig.\ref{pressure} (b) it is seen that  pressures for HRG and EVHRG models are almost identical upto $T\sim0.1 $ GeV  for $eB=0$ case and up to $T\sim0.12$ GeV  for $eB=0.2\ \text{GeV}^2$ case respectively. Above these temperatures, there is a notable decrease in  pressure for EVHRG case.

\begin{figure}[h]
	\vspace{-0.4cm}
	\begin{center}
		\begin{tabular}{c c}
			\includegraphics[scale=0.5]{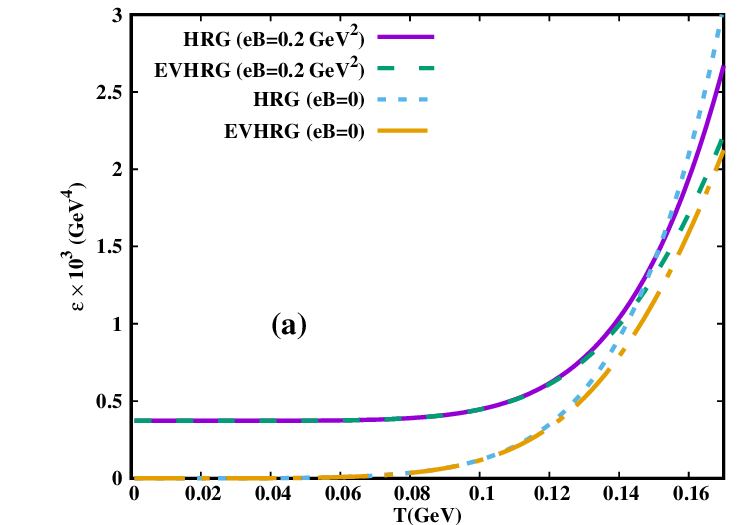}&
			\includegraphics[scale=0.5]{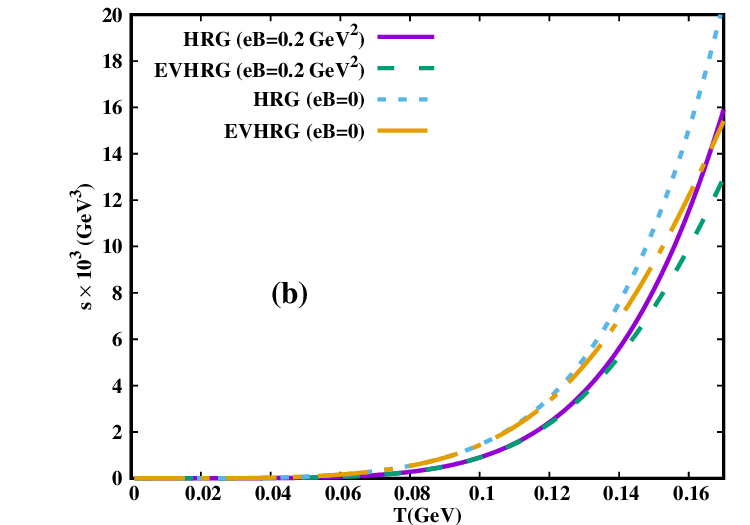}
		\end{tabular}
		\caption{Energy density (left panel) and  entropy density (right panel) calculated in HRG and EVHRG models with and without  magnetic field for $\mu_B=0$. } 
		\label{energy}
	\end{center}
\end{figure}

Fig.\ref{energy} shows the effect of magnetic field on the energy density and entropy density for both HRG and EVHRG models at zero $\mu_B$. The influence of magnetic field and hardcore repulsive interactions on energy density  is similar to that of pressure. The thermal part of the energy density for $eB\ne 0$ is smaller than that for $eB=0$ case. When repulsive interactions are turned on, it further suppresses the energy density.  The effect of magnetic field as well as the repulsive interactions on the entropy density is quite interesting. In ideal HRG model, the entropy density increases rapidly as the temperature increases due to rapid production of hadrons. The effective mass of a spin-0 particle increases in presence of magnetic field. Thus the thermal production of spin-0 particles is suppressed by Boltzmann factor $e^{-m_{eff}/T}$ and hence, the entropy is suppressed in presence of magnetic field. The excluded volume repulsive interaction also suppresses the entropy productions since the finite size of hadrons reduces the number density at high temperature.

\begin{figure}[h]
	\vspace{-0.15cm}
	\begin{center}
		\begin{tabular}{c c}
			\includegraphics[scale=0.5]{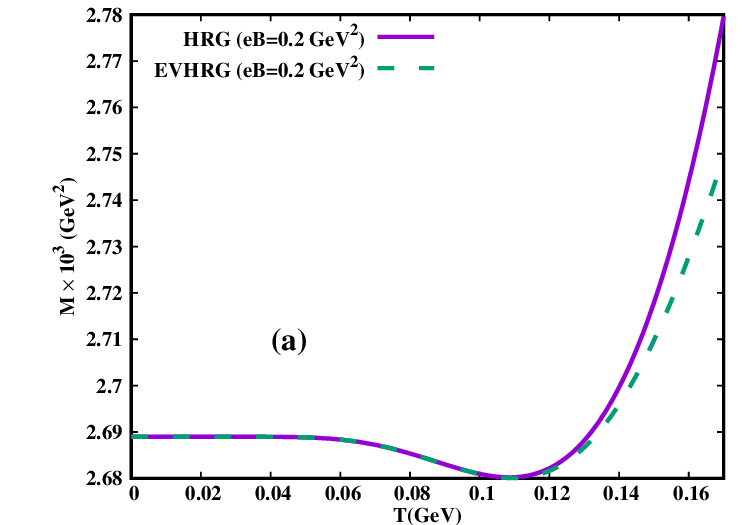}&
			\includegraphics[scale=0.5]{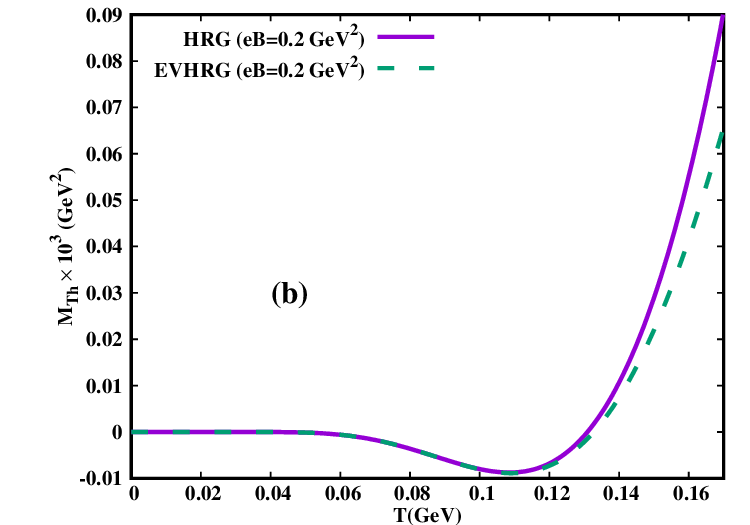}
		\end{tabular}
		\caption{Magnetization of hadronic matter estimated in HRG and EVHRG models in presence of magnetic field. Left panel shows the total (vacuum+thermal)
			magnetization and the right panel shows only the thermal part of magnetization. Calculations are performed at  $\mu_B=0$} 
		\label{magnetization}
	\end{center}
\end{figure}

Fig.\ref{magnetization} shows the magnetization as a function of temperature for zero $\mu_B$. Fig.\ref{magnetization}(a) shows the variation of total magnetization with temperature.  The total magnetization is positive which indicates that the hadronic matter is paramagnetic. Fig.\ref{magnetization}(b) shows only thermal contribution to the magnetization. The thermal part of magnetization is practically zero at very low temperature due to the absence of charged hadrons. The pions, which are lightest hadronic species, are thermally excited at $T\sim 0.06$ GeV. As they are scalar bosons, their magnetization is negative (diamagnetic) and hence magnetization decreases with increase in temperature. It becomes positive only when lightest spin-1 particle, $\rho$-meson, starts populating the hadronic matter and gives positive contribution to the magnetization. Thereafter, it rises rapidly as spin-$\frac{1}{2}$ particles also start to make positive contribution to the magnetization. Even though the thermal part of magnetization becomes negative for certain range of temperatures, the total magnetization including vacuum part is always positive. It turns out that the sign of magnetization of low temperature hadronic matter is a fundamental characteristic of thermal QCD vacuum.

\begin{figure}[t]
	\vspace{-0.4cm}
	\begin{center}
		\begin{tabular}{c c}
			\includegraphics[scale=0.5]{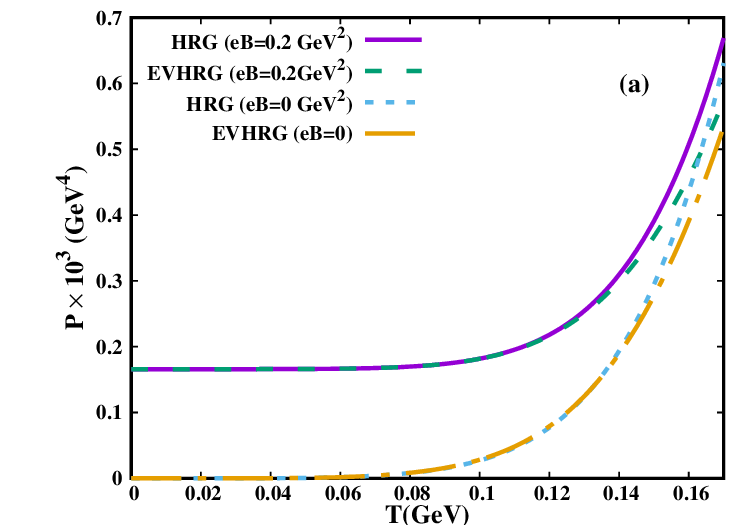}
			\includegraphics[scale=0.5]{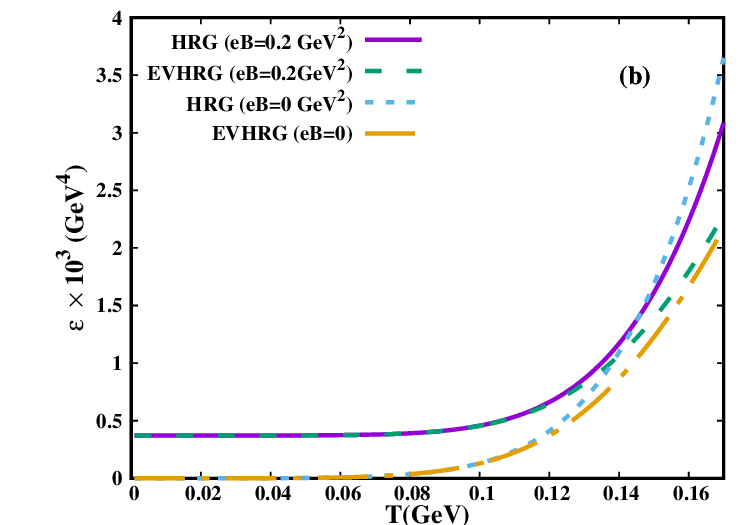} \\ 
			\includegraphics[scale=0.5]{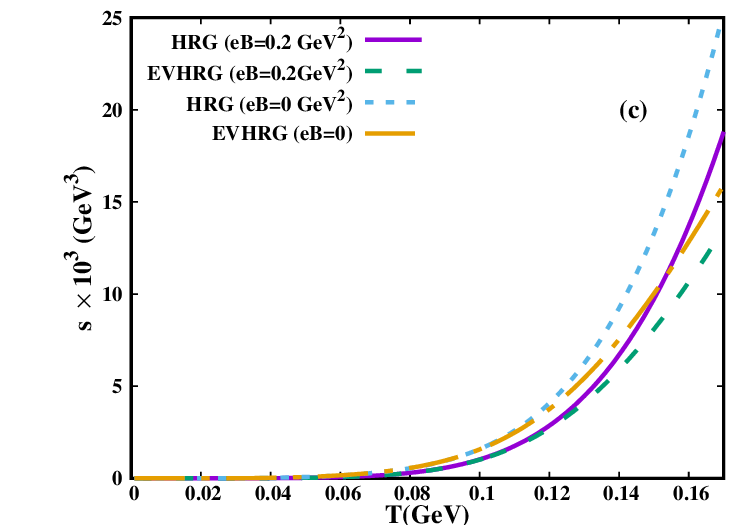}
			\includegraphics[scale=0.5]{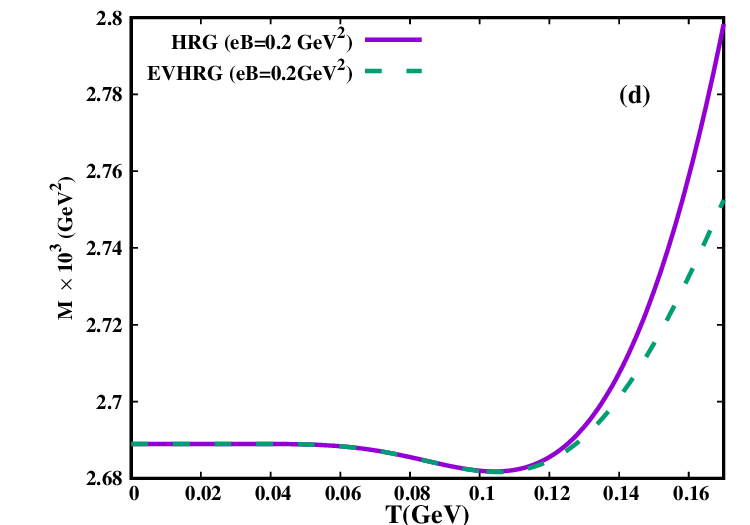} 
		\end{tabular}
		\caption{Thermodynamic variables at $\mu_B=300 MeV$ in   HRG and EVHRG models with and without  magnetic field.
			a) Pressure, b) Energy Density, c) Entropy density and d) Magnetization. } 
		\label{figmu}
	\end{center}
\end{figure}
Fig.\ref{figmu} shows the pressure, energy density, entropy density and magnetization at non zero $\mu_B$. The reason for choosing  $\mu_B$ = 300 MeV is
because it is close to CEP. In all the plots of this figure, it can be seen that all the qualitative behaviour of thermodynamic quantities are similar to that obtained 
in absence of magnetic field. The pressure has been plotted in Fig.\ref{figmu}(a). When this plot is compared with Fig~\ref{pressure}(b), one finds that, at high temperatures and in absence of magnetic field,  
the pressure at finite $\mu_B$ is greater than that in absence of magnetic field. This increase for the case of HRG is about $5\%$ whereas for EVHRG, there is no noticeable 
change in the pressure. At finite $\mu_B$, in presence of magnetic field, the  increase in pressure, for HRG and EVHRG, are  $16\%$ 
and $9\%$ respectively. A similar situation is revealed when one compares energy density (Fig.\ref{energy}(a) and Fig.\ref{figmu}(b)), at zero and finite $\mu_B$.  It is seen that 
the magnitude of change, in the energy density, at finite $\mu_B$, is larger compared to that in pressure. 
In absence of magnetic field, the energy density increases at finite $\mu_B$ compared to zero $\mu_B$ case. This increase for HRG is about $25\%$ whereas for EVHRG, it remains unchanged. For $eB=0.2\ \text{GeV}^2$, the corresponding changes are  $19\%$ 
and $8\%$ respectively. The changes in entropy density for HRG, are similar as to that of energy density whereas for EVHRG, there is almost no change. No appreciable effect of finite $\mu_B$ on magnetization was found in this study. It is found that the effect of finite $\mu_B$ 
is much less, at times negligible, for EVHRG. This is because of the fact that as $\mu_B$ increases, the suppression due to excluded volume also increases.
\section{Susceptibilities}
The susceptibilities of conserved charges in the model with magnetic field are discussed below.
\begin{figure}[H]
	\begin{center}
		\begin{tabular}{c c}
			\includegraphics[scale=0.5]{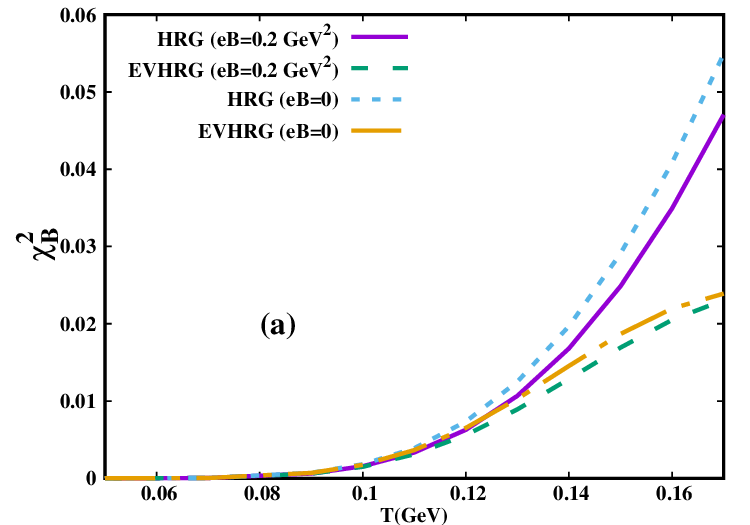}
			\includegraphics[scale=0.5]{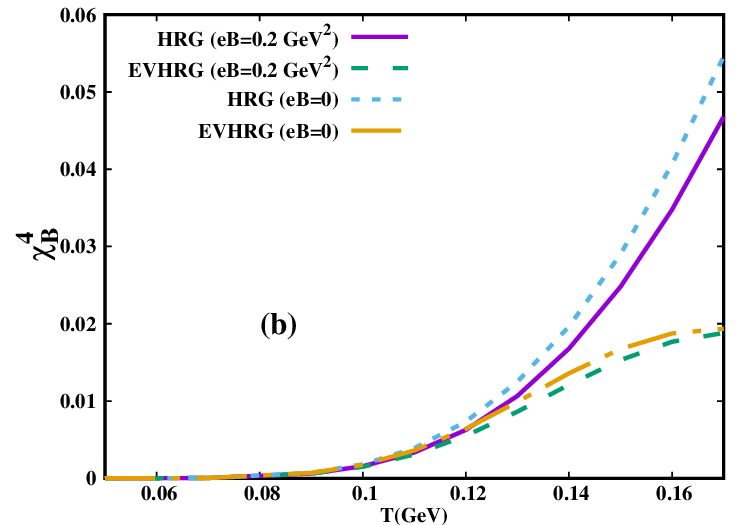}\\
			\includegraphics[scale=0.5]{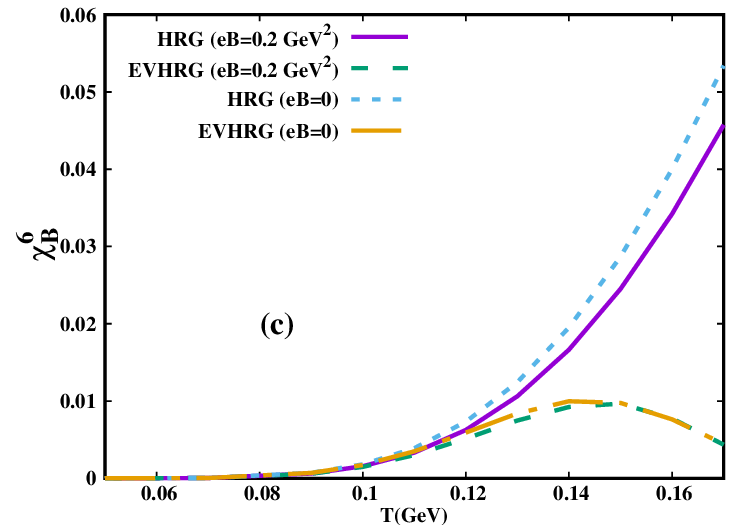}
		\end{tabular}
		\caption{Baryon susceptibilities of the hadronic matter estimated in HRG and EVHRG models with and without magnetic field. } 
		\label{chiB_mag}
	\end{center}
\end{figure}

Fig.\ref{chiB_mag} shows baryon number susceptibilities calculated within HRG and EVHRG models in presence and in absence of magnetic field. For HRG model, 
$\chi_B^{n}$s ($n=2,4,6$) increase rapidly when the temperature is high. It is to be noted that at a given temperature, the susceptibilities in presence of magnetic field, is less than that for $eB=0$.  The reason is that, the presence of magnetic field decreases the contribution to pressure from spin-$\frac{1}{2}$ charged particles as compared to the case of zero magnetic field. 
(Any baryon with spin greater than $\frac{1}{2}$ has not been considered here).  At low temperatures, nucleons, namely protons and neutrons which carry baryon number $1$, are the dominating contributors to $\chi_{B}$. It is to be noted that the mesons, which are predominant degrees of freedom at all temperatures, do not contribute to $\chi_{B}$ directly since they do not carry baryon number.   The probability that a baryon with mass $m$ populates at a given temperature $T$ is proportional to the Boltzmann factor $e^{-m/T}$. Thus as one goes to higher temperatures, other heavier baryons are thermally excited  and start to contribute to the pressure and hence to $\chi_{B}$. Here all the baryons considered, have baryon number $\pm1$. For such a case, there is almost no 
difference in magnitude for  susceptibilities of different orders for $\chi_B$ in case of HRG model.

The effects of repulsive interactions incorporated through excluded volume corrections can also be seen in Fig.\ref{chiB_mag}. It is seen that the numerical values of $\chi_B^n$ with repulsive interactions are smaller as compared to those estimated in ideal HRG model. The effect of repulsive interaction is very small at low temperatures since the number of various particles is small at low temperature and hence there is enough room for particles in the system. The inclusion of repulsive interaction suppresses the pressure as the available space for additional number of particles decreases. This suppression is traded into the $\chi_B^n$. Thus, the presence of repulsive interactions  affect the baryon number susceptibilities. Unlike the ideal HRG, this suppression effect becomes stronger for higher order susceptibilities i.e. $\chi_B^n$s of different orders are not of same magnitude when interaction is present. The $\chi_B^6$, at low temperatures, increases with temperature and then decreases for both $eB=0$ and $eB=0.2\ \text{GeV}^2$.

\begin{figure}[H]
	\begin{center}
		\begin{tabular}{c c}
			\includegraphics[scale=0.5]{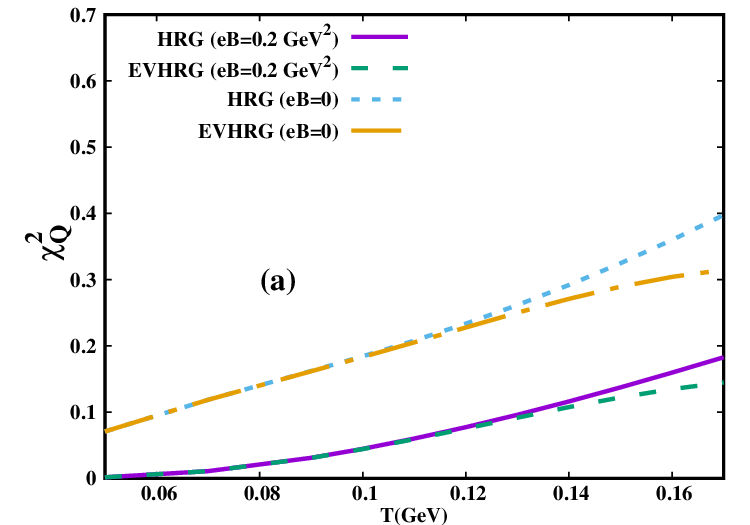}
			\includegraphics[scale=0.5]{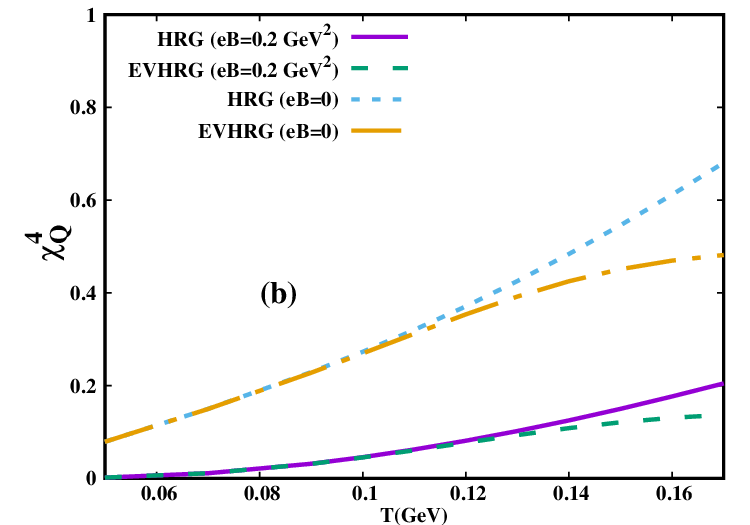}\\
			\includegraphics[scale=0.5]{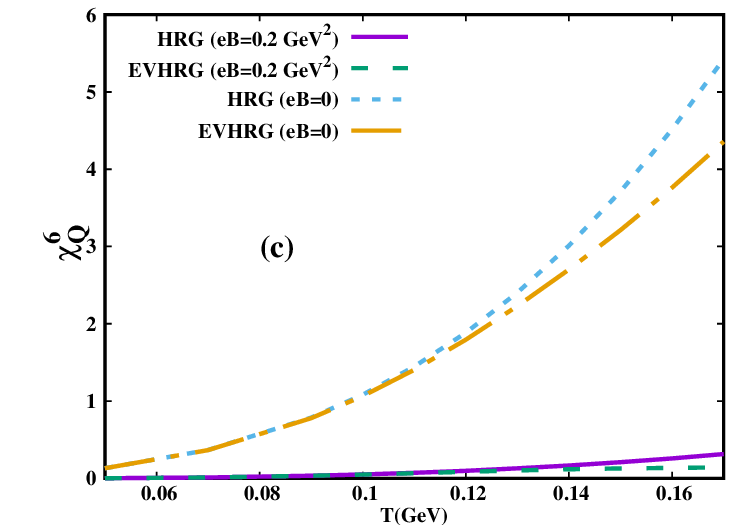}
		\end{tabular}
		\caption{Electric charge susceptibilities of the hadronic matter estimated in HRG and EVHRG models with and without magnetic field. } 
		\label{chiE_mag}
	\end{center}
\end{figure}

Fig.\ref{chiE_mag} shows electric charge susceptibilities of second, fourth and sixth order calculated within HRG and EVHRG models both in presence and in absence of magnetic field. Electric charge susceptibilities of all orders, in HRG model, are larger in magnitude than baryon susceptibilities. This is expected as in the particle spectrum, which has been considered here, has electrically charged mesons having very low mass, for example charged pions, which contribute significantly to $\chi^{n}_{Q}$ unlike $\chi^{n}_{B}$. It is also seen that higher order electric charge susceptibilities increase at a faster rate with increasing temperature than lower ones. This is because of lower masses of electrically charged hadrons. Dominant contribution to $\chi_Q^n$s, at low temperatures, come from charged pions. With the increase of temperature, heavier hadrons start to contribute to pressure and hence to susceptibilities.  Here, the effect of Van der Waals excluded volume repulsive interaction have also been taken into account.
The hardcore repulsive interaction reduces the pressure and hence $\chi_Q^n$ as compared to HRG Model. The effect of repulsive interaction is not significant at low temperatures ($T\sim 0.13$ GeV). This is because, the excluded volume radius $r_\pi=0$ is chosen for pions, which are dominant degrees of freedom and the population of hadrons with non-zero hard-core radius is low. Thus the effect of repulsive interaction starts to appear only after $T\sim 0.13 $ GeV when hadrons heavier than pions start to populate significantly. It is also to be noted that the higher order susceptibilities are suppressed more. There is a significant difference in magnitude between electric charge susceptibilities for $eB=0$  case and $eB=0.2\ \text{GeV}^2$ case even at low temperatures, which is not the case with baryon susceptibilities. This observation can again be attributed to the fact that the lower mass electrically charged particles  are suppressed in presence of magnetic field as compared to $eB=0$ case.

This behaviour can be understood by calculating the susceptibilities analytically in pure HRG model. The second and fourth order susceptibilities 
for pure HRG, with unit charge, can be written as

\begin{equation*}
\chi_i^2=\pm \sum_j\frac{g_j}{2\pi^2T^3}\int_{0}^{\infty}p^2 dp\frac{\exp{[(E_j-\mu_i)/T]}}{\{ \exp{[(E_j-\mu_i)/T]} \pm 1\}^2}
\end{equation*}

\begin{eqnarray}
\chi_i^4&=\pm\sum_j \frac{g_j}{2\pi^2T^3}\int_{0}^{\infty}p^2 dp\left[\frac{\exp{[(E_j-\mu_i)/T]}}{\{ \exp{[(E_j-\mu_i)/T]} \pm 1\}^2} \right . \nonumber \\
&\left .- 6  \frac {(\exp{[E_j-\mu_i)/T])})^2} {\{ \exp{[(E_j-\mu_i)/T]} \pm 1\}^3} 
+  6  \frac{(\exp{[E_j-\mu_i)/T])})^3}{\{ \exp{[(E_j-\mu_i)/T]} \pm 1\}^4} 
\right]
\end{eqnarray}
where $E_j$ is the energy of a particle of species $j$ and $\mu_i$ is chemical potential corresponding to some conserved charge.

The above expressions for $\chi^4_i$ contains some extra terms compared to that for $\chi^2_i$. These extra terms are negligible at low temperatures if the mass of the 
particle is quite heavy compared to the temperature. The $\chi_B^2$ and $\chi_B^4$  have almost same values for the case of baryon, as the baryon charge 
is dominantly carried by protons which has a mass much higher than the temperature.  On the other hand, for electric charge, the leading order contribution comes from pion, the 
mass of which is comparable to the temperature. So all the terms in the expression of $\chi^4_Q$ contribute significantly and as a result,  electric charge 
susceptibilities of different order have significantly different values.

\begin{figure}[h]
	\begin{center}
		\begin{tabular}{c c}
			\includegraphics[scale=0.5]{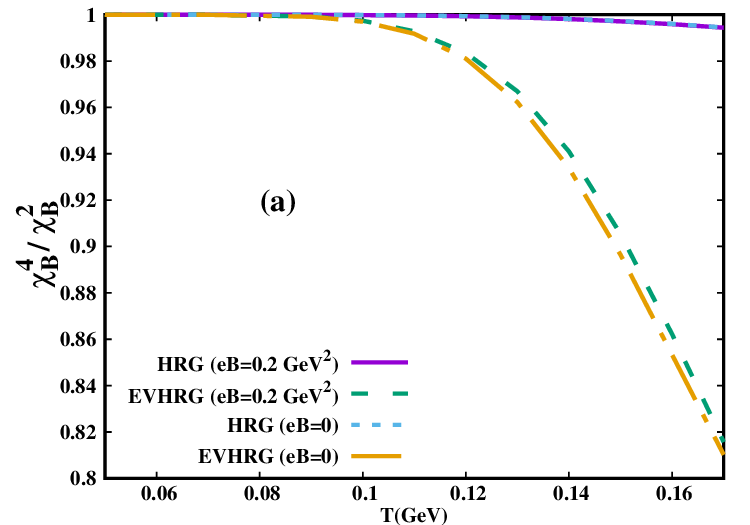}
			\includegraphics[scale=0.5]{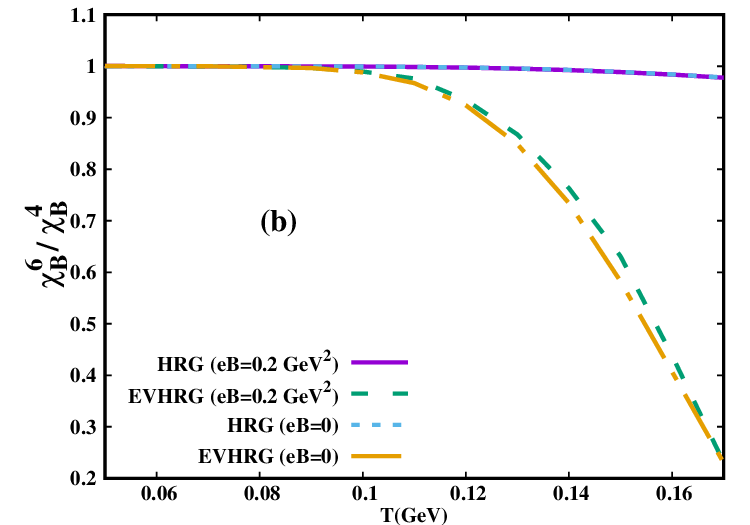}
		\end{tabular}
		\caption{Ratios of baryon susceptibilities of the hadronic matter estimated in HRG and EVHRG models with and without magnetic field. } 
		\label{chiB_ratios_mag}
	\end{center}
\end{figure}

\begin{figure}[H]
	\begin{center}
		\begin{tabular}{c c}
			\includegraphics[scale=0.5]{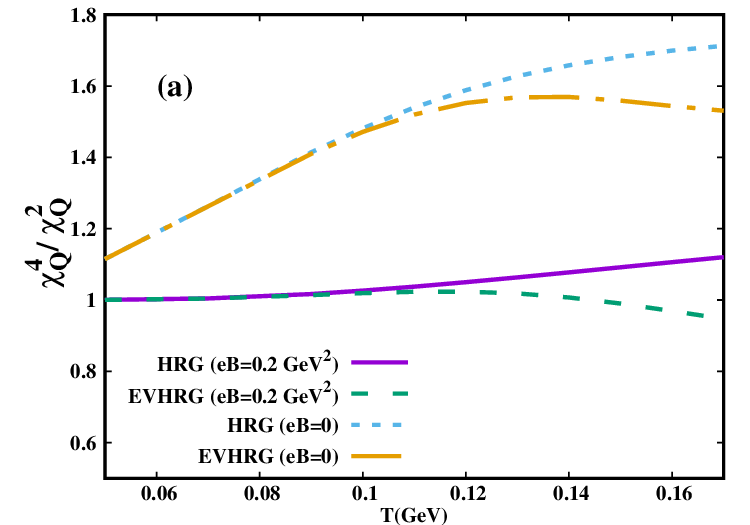}
			\includegraphics[scale=0.5]{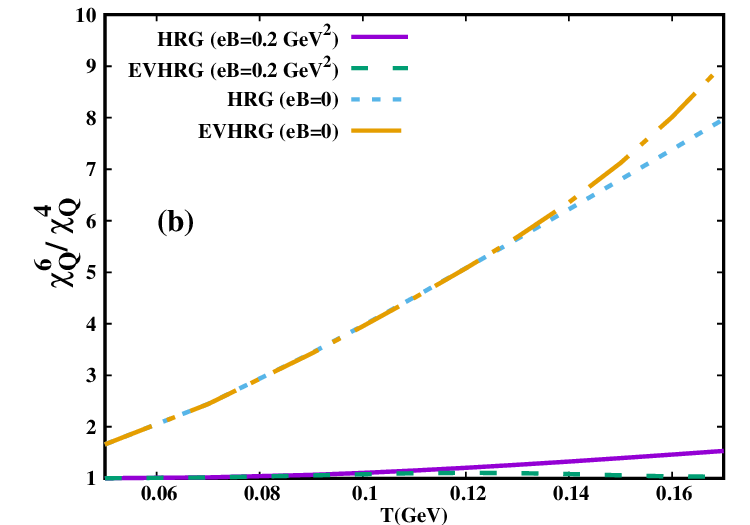}
		\end{tabular}
		\caption{Ratios of electric charge susceptibilities of the hadronic matter estimated in HRG and EVHRG models with and without magnetic field.} 
		\label{chiQ_ratios_mag}
	\end{center}
\end{figure}

Figures \ref{chiB_ratios_mag} and \ref{chiQ_ratios_mag} show the ratios of susceptibilities. The ratio $\chi_B^4/\chi_B^2$ is almost unity throughout the temperature 
range for pure HRG model. When repulsive interaction is included, this ratio decreases sharply by about $20\%$ at a temperature of 0.17 GeV.  The ratio 
$\chi_B^6/\chi_B^4$ also stays close to unity throughout the temperature range for pure HRG model.   
For interacting  scenario, it decreases even faster than $\chi_B^4/\chi_B^2$. At a temperature of 0.17 GeV its value becomes almost $0.2$. 
The ratios of electric charge susceptibilities have a more complicated behaviour with temperature.  The ratio $\chi_Q^4/\chi_Q^2$ is close to unity at low 
temperature. For pure HRG, in absence of magnetic field, it monotonically increases with temperature. For EVHRG without magnetic field, it first increases with temperature 
and then starts decreasing. For the cases with magnetic field, the changes are relatively small. At high temperature the change is about $10\%$. The ratio $\chi_Q^6/\chi_Q^4$ 
increases almost linearly with temperature for both HRG and EVHRG model with zero magnetic field. 
For constant magnetic field the magnitude of this ratio remains almost unity throughout the temperature range. 
	\newpage
\rhead{Summary and Conclusion}
\chapter{\label{chap:Summary and Conclusion} Summary and Conclusion}
The main purpose of this work is to investigate the properties of strongly interacting matter under extreme conditions of high temperature and/or density. Although QCD, the theory of strong interactions should be used to study the properties of such matter, the non-perturbative nature of the physics involved in the temperature and density regime we are concerned, restricts one to use QCD directly. Lattice QCD is a first principle method to circumvent this problem. It discretizes the space-time and uses large computing power to give reliable results at high temperature and low density regime. In the high density region, lattice QCD suffers from the sign problem, as discussed in the introduction section. Another approach to look at the strongly interacting matter is the use of effective models. In this work, exploration of the properties of strongly interacting matter at high temperatures and densities in the effective model approach has been performed. This approach is convenient from the numerical perspective as compared to lattice QCD and also provides important information in the region of the phase diagram where LQCD fails.

In chapter 2, the Hadron Resonance Gas (HRG) model and its extended version with short range hard-core repulsive interaction, namely, Excluded Volume Hadron Resonance Gas (EVHRG) model have been reviewed. The Modified Excluded Volume Hadron Resonance Gas (MEVHRG) model, which takes into account the effect of unequal excluded volume radii of different hadron species, has also been introduced.  Effect of unequal radii in EVHRG model has been studied in Refs.~\cite{Alba:2017bbr,Vovchenko:2016ebv} and effect of Lorentz contraction has been considered in Ref.~\cite{Oliinychenko:2012hj}. But these approaches are somewhat different than the one used here. Then, the effect of Lorentz contraction in the MEVHRG model in the rest frame of the system has been included and the model has been named as Lorentz Contracted Modified Excluded Volume Hadron Resonance Gas (LMEVHRG) model. Another way of including the repulsive interactions among hadrons is by the consideration of a mean-field potential. This approach is called the HRG mean-field (HRGMF) model. In this approach, the single particle energy levels are shifted by a density dependent term and ultimately, mimicking the inclusion of repulsive interaction.

In chapter 3, the thermodynamic properties of hadronic matter have been studied. The pressure in HRG model, increases monotonically with temperature. The results of ideal HRG model gives the best agreement with lattice data for pressure. When repulsive interaction through excluded volumes is included, the pressure drops significantly, especially at high temperatures. The effect of unequal radii incorporated in MEVHRG model becomes more prominent as one considers larger variety of radii. Effect of Lorentz contraction on pressure is to increase it and its effect is significant at higher temperatures. It is found that, for ideal HRG model scenario, the specific heat increases with increasing temperature. It is expected because as temperature increases, heavier hadrons populate the system gradually. Nonetheless, HRG model does not show critical behaviour due to lack of exponential rise in the spectral density needed to show such critical behaviour. When repulsive interaction is turned on, the magnitude of $C_V$ gets reduced.  Including the repulsive interactions in HRG model by mean-field approach leads to satisfactory qualitative reproduction of the overall behaviour of the lattice data for $C_V$. It is also seen that the effect of repulsive interactions on $\kappa_T$ is very mild, while  its role in case of $C_s^2$  is very nontrivial. With the inclusion of repulsive interactions modeled through mean-field approach, $C_S^2$ shows good qualitative agreement with the LQCD data over wide range of temperatures and baryon chemical potentials. So, the repulsive interactions have a very strong bearing on $C_V$ and $C_s^2$. The inclusion of repulsive interactions have significant implications on the results in the context of heavy-ion collision experiments. It was found that at low $\sqrt{s}$, both $C_V$ and $\kappa_T$ are large. Then these quantities drop sharply as the collision energy increases, and finally saturate at constant value at 
large $\sqrt{s}$. On the other hand, behaviour of $C_s^2$ is opposite in nature. It is small at low $\sqrt{s}$, then rises very rapidly as the collision energy increases, and finally saturates to a constant value at high $\sqrt{s}$. Thus, in the context of HIC experiments, all these quantities are found to be strongly sensitive to the strength of repulsive interactions.

Chapter 4 is devoted to studying the susceptibilities of conserved charges. Although ideal HRG model can give satisfactory agreement with LQCD data for the scaled pressure, in order to explain the susceptibilities, repulsive models are absolutely essential. Second order strangeness susceptibility in HRG is smaller than that in lattice data. The pure HRG result for fourth order strangeness susceptibility matches with lattice data till 0.165 \ GeV of temperature. However, at higher temperature, the LMEVHRG results are in good agreement with LQCD. It is argued that the strangeness sector gives incomplete picture about the particle spectrum and hence on the thermodynamic quantities when only the hadronic states listed in Particle Data Group are considered~\cite{Alba:2017bbr,Alba:2017mqu}. The effect of unequal excluded volumes is very mild for the choice of excluded volumes in this work. Inclusion of Lorentz contraction has significant impact on the susceptibilities. A study of the effect of repulsive interaction on the susceptibilities of conserved charges using the HRGMF model has also been presented here. The results of ideal HRG model for baryonic susceptibilities fails to explain the LQCD data. 
Inclusion of repulsive interaction through HRGMF model performs well in explaining the data, especially the higher order susceptibilities. The HRGMF model explains the lattice data for the ratios of baryon susceptibilities at non zero baryon chemical potentials quite well.

In chapter 5, the hot and dense hadronic matter, in presence of a magnetic field, has been studied. The energy levels of electrically charged particles in presence of magnetic field are given by the Landau levels. The thermodynamic properties of hadronic matter are significantly suppressed in presence of magnetic field with magnitude of $eB=0.2\ \text{GeV}^2$, which was chosen as the representative value of the magnetic field in this work. In the presence of magnetic field, the effective masses of the hadrons are modified, which, leads to the modification of the population of different electrically charged hadron species while that of electrically neutral particles remain unchanged. It is found that the thermal parts of all the thermodynamic quantities considered in this work, are smaller in presence of magnetic field. The total magnetization of hadronic matter is positive. The susceptibilities of conserved charges are also significantly suppressed by the magnetic field both in HRG and EVHRG models. The ratios of susceptibilities of different orders also show significant dependence on excluded volume and magnetic field.

In the following, we list the significant conclusions of this work.
\begin{enumerate}
	\item The repulsive interaction, modelled through excluded volume or mean-field potential, have significant impact on the pressure, energy density, entropy density, specific heat and speed of sound.
	\item Ideal HRG model, although shows good agreement with lattice results for the pressure, fails to reproduce the lattice data for the susceptibilities. Repulsive interaction among the hadrons are necessary to describe the susceptibilities.
	\item Models with repulsive interactions are found to yield unsatisfactory results for the strangeness susceptibilities. Inclusion of strange resonances, in addition to those listed in Particle Data Group, which are predicted by some models, may lead to good agreement with the lattice data for the strange sector.
    \item The effect of unequal excluded volume radii in MEVHRG model becomes stronger if one considers different excluded volumes for each particle species.
    \item Lorentz contraction of excluded volume has significant impact on the quantities studied here.
    \item There is still no consensus regarding the exact size of the excluded volume of the hadrons. 
    \item The HRGMF model describes the ratios of baryon number susceptibilities considered here at non zero baryon chemical potential quite satisfactorily.
    \item The presence of magnetic field influences the results of both the HRG and EVHRG models significantly. It is found in this study that the total magnetization of hadronic matter is positive. The magnetic field has stronger effect on the electric charge susceptibilities as compared to baryon susceptibilities.
\end{enumerate}
	\newpage
\rhead{Bibliography}

\end{document}